\pdfoutput=1 \pdfsuppresswarningpagegroup=1
\documentclass[
 aps,%
 prb,
 reprint,%
 groupedaddress,% 
 superscriptaddress,%
 showpacs,%
 letterpaper,%
 footinbib,%
 noeprint%
]{revtex4-1}
\usepackage{bm}
\usepackage{graphicx}
\usepackage{graphics}
\usepackage{amsmath}
\usepackage{amsfonts}
\usepackage{amssymb}
\usepackage{makeidx}
\usepackage[colorlinks=true,citecolor=blue,linkcolor=blue,linktocpage=true,pagebackref=false]{hyperref}
\hypersetup{colorlinks=true,citecolor=blue,linkcolor=blue,filecolor=blue,urlcolor=blue}
\usepackage{color,soul} 
\usepackage{float}
\usepackage{subfigure}
\usepackage[colorinlistoftodos,prependcaption,textsize=small]{todonotes}
\usepackage{soul}
\usepackage[normalem]{ulem}

\renewcommand{\eqref}[1]{(\ref{#1})}

\newcommand{\be}{\begin{equation}}
\newcommand{\ee}{  \end{equation}}
\newcommand{\ba}{\begin{eqnarray}}
\newcommand{\ea}{  \end{eqnarray}}
\newcommand{\ve}{\varepsilon}

\graphicspath{ {./Figs/} } 

\begin{document}

%\title{Impurity solvers: a case study applied to quantum dots}
%Suggestion for the title
\title{Quantitative comparison of Anderson impurity solvers applied to transport in quantum dots}
\author{Bruno Max de Souza Melo}
\address{Instituto de F\'{\i}sica, Universidade Federal
  Fluminense, 24210-346 Niter\'oi, RJ, Brazil}
\author{Luis G.~G.~V. Dias da Silva}
\affiliation{Instituto de F\'{\i}sica, Universidade de S\~{a}o Paulo,
C.P.\ 66318, 05315--970 S\~{a}o Paulo, SP, Brazil}
\author{Alexandre Reily Rocha}
\address{Instituto de F\'{\i}sica Te\'orica, S\~ao Paulo State University (UNESP), S\~ao Paulo SP, Brazil}
\author{Caio Lewenkopf}
\address{Instituto de F\'{\i}sica, Universidade Federal
  Fluminense, 24210-346 Niter\'oi, RJ, Brazil}
  
%%%%%%%%%%%%%%%%%%%%%%%%%%%%%%%%%%%%%%%%
\begin{abstract}
We study the single impurity Anderson model (SIAM) using the equations of motion method (EOM), 
the non-crossing approximation (NCA), the one-crossing approximation (OCA), and Wilson's numerical 
renormalization group (NRG).
We calculate the density of states and the linear conductance focusing on their dependence on the 
chemical potential and on the temperature paying special attention to the Kondo and Coulomb blockade 
regimes for a large range of model parameters. We report that some standard approximations based on
the EOM technique display a rather unexpected poor behavior in the Coulomb blockade regime
even at high temperatures.  
Our study offers a critical comparison between the different methods as well as a detailed compilation of 
the shortcomings and limitations due the approximations involved in each technique, thus allowing for 
a cost-benefit analysis of the different solvers that considers both numerical precision and computational 
performance.
\end{abstract}
\maketitle

%%%%%%%%%%%%%%%%%%%%%%%%%%%%%%%%%%%%%%%%
\section{Introduction}
\label{sec:introduction}
%%%%%%%%%%%%%%%%%%%%%%%%%%%%%%%%%%%%%%%%

The single impurity Anderson model (SIAM) is one of the most recurrent models in condensed matter physics 
playing an important role on the study of strongly correlated systems. \cite{Tsvelick1983,Hewson1997,Georges1996}
% \todo{Se vamos user este tipo de rerefer\^encias, tem que passar tudo pra depois da pontua\c{c}\~ao} 
Originally put forward to describe the physics of doped metals with magnetic impurities \cite{Anderson1961,Krishna-Murthy1975,Krishna-murthy1980a,Krishna-murthy1980b}, the model has reached a remarkable protagonism being used to study the mixed valence regime in rare earth compounds \cite{Newns1987,Coleman1984,Hewson1997} as well as the Coulomb blockade and the Kondo effect in quantum dots. \cite{Meir1991,Pustilnik2004,Meir1992a,Ng1988,Costi1994,Goldhaber-Gordon1998Nature, Goldhaber-Gordon1998PRL, 
Pustilnik2001,Glazman2003} In the context of the dynamical mean field theory (DMFT) \cite{Georges1996} the model has gained new traction, and contributed to describe the properties of a huge variety of strongly correlated materials like transition metal oxides \cite{Kugler2019}, heavy-fermion systems, high temperature superconductors, etc.. \cite{Kotliar2006,Georges1996} In the field of molecular electronics \cite{Liang2002,Park2002,Thoss2018} the numerical solution of the SIAM is an important step to study the transport properties of strongly correlated molecular devices through an approach that combines the density functional theory (DFT), the dynamical mean field theory (DMFT) and the non-equilibrium Green's function formalism (NEGF). \cite{Droghetti2017,Chioncel2015,Appelt2018,DavidJacob2015,Jacob2009,Jacob2010}
 
 The large variety of scenarios in which the model can be applied demands the calculation of spectral densities, thermodynamical quantities like the free energy and entropy as well as transport characteristics. In order to compute this variety of physical quantities, several impurity solvers have been developed for the SIAM over the years such as the numerical renormalization group (NRG) \cite{Krishna-murthy1980b,Krishna-murthy1980a}, the equations of motion (EOM) \cite{Theumann1969,Lacroix1981,Lacroix1982}, slave boson mean field approximation \cite{Coleman1984}, non-crossing (NCA)\cite{bickers1987} and one-crossing approximations (OCA) \cite{Pruschke1989},  
exact diagonalization method \cite{Georges1996}, quantum 
Monte-Carlo algorithms \cite{Hirsch1986,Kotliar2006}, iterative perturbation theory \cite{Yamada1975,Yosida1970,Yosida1970a}, to name a few.
 
These impurity solvers differ mainly by the computational cost involved and by the range of the model parameters where the used method is reliable. Indeed, the search for different methods is mainly a consequence of the fact that no method can provide reliable results over the whole physically relevant parameter space \cite{Georges1996,Kotliar2006}. In many cases it is necessary to employ more than one method to better understand a specific problem \cite{Meir1992a}.
 
In view of the variety of available methods and scenarios in which the Anderson model can be applied the task 
of benchmarking impurity solvers in each context is very important and far from being exhausted. The choice of 
a specific solver involves the knowledge of the aspects that make it more competitive in some range of parameters 
than others.

In this work we carry out a systematic comparison of the accuracy of the standard impurity solvers in the literature, namely, 
the equations of motion method \cite{Kashcheyevs2006}, the non crossing approximation \cite{bickers1987}, 
the one-crossing approximation \cite{Pruschke1989,Haule2010}, and the numerical renormalization group (NRG)\cite{Bulla2008}. 
We compute the Green's function of a quantum dot using the different methods and obtain the density of states and 
the linear conductance of the dot as well as its dependence on the chemical potential and the temperature.
We discuss the strengths and weaknesses, as well as the computational efficiency of the above mentioned impurity solvers for a 
broad range of parameters of physical interest.

This paper is organized as follows. In Sec.~\ref{sec:model} we briefly present the model 
system and the impurity solvers assessed in this study. In Sec.~\ref{sec:results} we compare their
accuracy and numerical efficiency by computing the density of states and the linear conductance as a function of charging energy, chemical potential, temperature, etc., covering all the standard regimes of the model. Finally, in Sec.~\ref{sec:conclusion} we 
present a summary of our findings and our conclusions.

%%%%%%%%%%%%%%%%%%%%%%%%%%%%%%%%%%%%%%%%%%%
\section{Model and methods}
\label{sec:model}
%%%%%%%%%%%%%%%%%%%%%%%%%%%%%%%%%%%%%%%%%%%

% \starr{We begin by presenting the model Hamiltonian and the main expression for the 
% calculation of the conductance in the the linear response regime. Next, we give a brief
% description of the many-body solvers we analyze, namely, equations of motion (EOM)
% \cite{Haug2008}, slave-boson approximation (SBA) \cite{Coleman1984}, and numerical 
% renormalization group (NRG) \Caio{citation needed}.
% Since most of this material is well known, we present only the elements necessary for 
% the discussion of the results that follow, referring the interested reader to the original literature.} 
%

The model Hamiltonian we use to benchmark the impurity solvers is the single-impurity Anderson 
model \cite{Anderson1961}. We write the SIAM as
\begin{equation}
H = H_{\rm imp} + H_{\rm C} + H_{\rm B}.
\label{eq:SIAMHamiltonian}
\end{equation}
Here, $H_{\rm imp}$ reads
\be
H_{\rm imp} = \sum_\sigma \ve_\sigma f^\dagger_\sigma f^{}_\sigma + U n_\uparrow n_\downarrow,
\ee 
where $f^\dagger_{ \sigma}$  $(f^{}_{ \sigma})$ creates (annihilates) an electron with spin projection 
$ \sigma $ at the impurity site, $n_{\sigma} = f^\dagger_{ \sigma} f^{}_{ \sigma} $ is the occupation 
number operator, and $U$ is the Coulomb charging energy, the energy cost for double occupancy 
of the impurity site.

We address a two-probe setup, where the system of interest is connected to electronic reservoirs in thermal 
and chemical equilibrium by considering two semi-infinite electrodes. The corresponding  
Hamiltonian reads
\begin{equation}
H_{\text{B}} = \sum_{\alpha,\mathbf{k}, \sigma} \ve^{}_{\alpha \mathbf{k} \sigma} 
c^{\dagger}_{\alpha \mathbf{k} \sigma} c^{}_{\alpha \mathbf{k} \sigma} ,
\label{eq:LeadHamiltonian}
\end{equation}
where $\alpha = \left\{L,R\right\}$ labels the leads, $c^\dagger_{\alpha \mathbf{k} \sigma}$ $(c^{}_{\alpha \mathbf{k} \sigma })$ 
creates (annihilates) an electron of wave number $\bf k$ and spin projection $\sigma$ at the $\alpha$ lead, and 
$\ve_{\alpha \mathbf{k} \sigma}$ is the corresponding single-particle energy.

Finally,
\begin{equation}
H_{\text{C}}  = \sum_{\alpha, \mathbf{k}, \sigma} \left( V^{}_{\alpha \mathbf{k} \sigma } c_{\alpha \mathbf{k}\sigma}^{\dagger} f^{}_{ \sigma} +
 {\rm H.c.}\right),
\end{equation}
represents the coupling between the impurity and the electrodes, where $ V_{\alpha \mathbf{k} \sigma} $ is 
the so-called hybridization matrix element.

The nonequilibrium Green's function theory allows one to write the linear conductance $\mathcal{G}$ of the system as 
 \cite{MeirWingreen1992,Wingreen1994}
\begin{equation}
\mathcal{G} = - \frac{e^2}{h} \sum_\sigma \int_{-\infty}^{\infty} \! d \omega \left(-\frac{\partial n_{\rm F}}{\partial \omega} \right) 
\Gamma_\sigma (\omega) \,{\rm Im} [G_\sigma^r(\omega)],
\label{eq:G_linear_response} 
\end{equation}
where the retarded Green's function $G_{\sigma}^{r} (\omega)$ is the Fourier transform of
\begin{equation}
G_{\sigma}^{r}(t) = -i \theta(t) \langle \{ f_{\sigma}(t) f_{\sigma}^{\dagger} (0) \} \rangle,
\label{eq:impurityGF}
\end{equation}
$n_{\rm F}(\omega) = 1/(1+e^{\beta \omega})$
is the Fermi distribution, with $\beta = (k_B T)^{-1}$. Throughout this work the chemical potential is set to $\mu = 0$
and $k_B = 1$. Finally, $\Gamma_{\sigma}$ stands for
\begin{equation} 
\Gamma_{\sigma} (\omega) = \frac{\Gamma_{\sigma}^{L} (\omega) \Gamma_{\sigma}^{R} (\omega)}
{ \Gamma_{\sigma}^{L}(\omega) + \Gamma_{\sigma}^{R}(\omega)}.
\end{equation}
The decay widths $\Gamma_\sigma^\alpha(\omega)$  
\footnote{In this paper we use $\Gamma$, the typical scattering theory notation used in quantum dots and molecular electronics, 
instead of the hybridization function $\Delta$, more familiar to the strongly correlated systems community. Note that 
$\Gamma = 2 \Delta$.}
are given by
\begin{equation}
\Gamma^{\alpha}_{\sigma} (\omega) = 2 \pi \sum_{\mathbf{k} } |V_{ \alpha \mathbf k \sigma }|^2 \delta(\omega - \varepsilon_{\alpha \mathbf k \sigma})
\end{equation}
and are proportional to the imaginary part of the embedding self-energy $\Sigma^\alpha_\sigma (\omega)$  due to 
the scattering processes between the impurity and the leads \cite{Haug2008}
\begin{equation}
\Sigma^\alpha_\sigma (\omega) = \sum_{{\bf k} }V_{\alpha \mathbf{k} \sigma} \frac{1}{\omega -\ve^{}_{\alpha \mathbf{k} \sigma} } 
V_{\alpha \mathbf{k} \sigma}^*.
\end{equation} 
As usual, by making the substitution $\omega \to \omega \pm i \eta$ with $\eta \to 0^{+}$, one obtains the 
retarded and advanced counterparts of the Green's functions and self-energies presented in this paper. 

The standard derivation of Eq.~\eqref{eq:G_linear_response} relies on assuming proportional coupling, namely, 
$\Gamma^R (\omega)= \lambda \Gamma^L (\omega)$ \cite{MeirWingreen1992}. It has been recently shown 
\cite{DiasdaSilva2017} that, within the linear response regime, Eq.~\eqref{eq:G_linear_response} is valid as long 
as $\Gamma_{\sigma} (\omega)$ varies slowly on the scale of $kT$. 

It is sometimes useful to take the wide-band limit and consider $V_{ \alpha \mathbf k \sigma }$ to be independent 
of $\mathbf k$ as well as taking the (non-interacting) density of states of the electrons to be constant within the band-width:
\begin{equation}
\rho^{\alpha}_{\sigma}(\omega) = \sum_{\mathbf{k} } \delta(\omega - \varepsilon_{\alpha \mathbf k \sigma}) \equiv \rho^{\alpha}_{\sigma}(0) \mbox{ for } -D \leq \omega \leq D \; ,
\label{eq:LeadDOS}
\end{equation}
where $D$ is the half-bandwidth and $\omega\!=\!0$ is taken to be the Fermi energy in the leads. In this case, $\Gamma^{\alpha}_{\sigma} \equiv 2 \pi |V_{ \alpha \sigma }|^2 \rho^{\alpha}_{\sigma}(0)$ becomes independent of the energy $\omega$.

Next we briefly review the methods we use to calculate the Green's functions $G_{\sigma} ^{r}$, namely, 
equations of motion (EOM) \cite{Haug2008}, slave-boson approximations (SBA) \cite{bickers1987, Pruschke1989}, 
and numerical renormalization group (NRG) \cite{Bulla2008}.
Since most of this material is well known, we present only the elements necessary for 
the discussion of the results that follow, referring the interested reader to the original literature.

%================================================================
\subsection{Equations of motion (EOM) method}
\label{sec:EOM}

{The most straightforward technique} to calculate equilibrium and nonequilibrium Green's functions is the
equations of motion method \cite{Haug2008}. 
This technique consists in taking time-derivatives of the Green's functions to generate a set of coupled equations
of motion. 
For bilinear Hamiltonians the system of equations can be solved exactly \cite{Haug2008,Hernandez2007}. 
For many-body Hamiltonians the hierarchy of equations is infinite and it is 
necessary to truncate the equations by means of physical arguments, yielding different approximate solutions
 to the problem. 
 
The EOM simplest approximation for the SIAM is to obtain the exact Green's function of the uncoupled impurity Hamiltonian and to
include an embedding self-energy by hand \cite{Haug2008}, namely
\begin{equation}
G_{\sigma} (\omega) = \frac{1 - \langle n_{\bar{\sigma}}\rangle}{\omega - \varepsilon_\sigma - \Sigma_0 (\omega)} + 
                                      \frac{ \langle n_{\bar{\sigma}}\rangle} {\omega - \varepsilon_\sigma - U -  \Sigma_0 (\omega)},
\label{eq:GF_EOM0_approx}
\end{equation}
where 
\begin{equation}
\Sigma_0 (\omega) = \Sigma^L_\sigma\left(\omega\right) +  \Sigma^R_\sigma\left(\omega\right) = \sum_{\alpha\mathbf{k}} \frac{\left|V_{\alpha\mathbf{k}}\right|^2}{\omega-\epsilon_{\alpha\mathbf{k}}},
\end{equation}
and the occupation is 
\begin{equation}
 \langle n_{\bar{\sigma}} \rangle =   \int_{-\infty}^{\infty} d \omega n_F(\omega) \rho_{\sigma}({\omega}),
 \label{eq:occupation}
\end{equation}
with the impurity density of states given by
\begin{equation}
\rho_{\sigma}({\omega}) = - \frac{1}{\pi} \text{Im}[G_{\sigma}^{r}(\omega)]~.
\label{eq:impurity_DOS}
\end{equation}
Equations \eqref{eq:GF_EOM0_approx} and \eqref{eq:occupation} are solved self-consistently. This approach,
hereafter called EOM0, is frequently regarded as a good approximation to the SIAM in the weak coupling 
regime \cite{Haug2008}.

A more consistent approach is to write the EOM for the impurity Green's function, Eq.~\eqref{eq:impurityGF}, using 
the full Hamiltonian. The simplest truncation of the resulting hierarchy of coupled equations is to consider the two-particle
Green's functions at the Hartree-Fock level \cite{Haug2008}. This approximation, that we call EOM1, is formally equivalent to the Hubbard I approximation
\cite{Hubbard1963,Gebhard1997}. The resulting Green's function is
\begin{widetext}
\begin{equation}
G_{\sigma} (\omega)  = \frac{\omega - \varepsilon_\sigma - U (1 - \langle n_{\bar{\sigma}}\rangle )}
     {(\omega - \varepsilon_\sigma) (\omega - \varepsilon_\sigma - U) - 
     \Sigma_0(\omega) \left[\omega - \varepsilon_\sigma - U(1 - \langle n_{\bar{\sigma}}\rangle)\right]}.
\label{eq:GF_EOM0}
\end{equation}
It is often stated that Eq.~\eqref{eq:GF_EOM0_approx} is a good approximation to \eqref{eq:GF_EOM0} \cite{Haug2008}. As we discuss in the next section, the agreement is at most qualitative. 

The next level of complexity is to neglect spin correlations in the two-particle Green's functions \cite{Meir1991} to write
\begin{equation}
G_\sigma (\omega) = \frac{\ve - \ve_\sigma - U(1- \langle n_{\bar \sigma}\rangle) - \Sigma_0(\ve) -\Sigma_3(\omega) }
{[\omega - \ve_\sigma - \Sigma_0(\omega)][\omega - \ve_\sigma - U(1- \langle n_{\bar \sigma}\rangle) - \Sigma_0(\omega) -\Sigma_3(\omega) ] + 
U \Sigma_1(\omega) }
\label{eq:GF-EOM1},
\end{equation}
where the self-energies $\Sigma_{1,3}$ are given by \cite{Meir1991}:
\begin{align}
\label{eq:EOM1_self-energies}
\Sigma_i (\omega) = \sum_{\mathbf q \beta} A^{(i)}_{\mathbf q \beta \sigma} | V_{\mathbf q \beta \sigma}|^2 & \Bigg( \frac{1}{\omega + \ve_{\mathbf q \beta \sigma} - \ve_{\overline\sigma} - \ve_\sigma - U}
% + \nonumber \\ &
+ \frac{1}{\omega - \ve_{\mathbf q \beta \sigma} + \ve_{\overline\sigma} - \ve_\sigma} \Bigg)
\end{align}
\end{widetext}
with $A^{(1)}_{\mathbf q \beta \sigma} = n_{\rm F}(\ve_{\mathbf q \beta \sigma})$ and $A^{(3)}_{\mathbf q \beta \sigma}=1$. 
We call this truncation scheme EOM2. We note that it is formally equivalent to the Hubbard III approximation \cite{Hubbard1964}. 

A more sophisticated truncation scheme that includes spin correlations has been introduced in Ref.~\onlinecite{Kashcheyevs2006}. 
In the absence of a magnetic field, the impurity Green's function reads
\begin{widetext}
\begin{equation}
G_{\sigma} (\omega) = \frac{u(\omega) - \langle n_{\bar{\sigma}} \rangle - P_{\bar{\sigma}}(\omega) - P_{\bar{\sigma}} (\omega_{2})    }{u(\omega) [\omega - \epsilon_{\bar{\sigma}} - \Sigma_{0} (\omega) ] + [P_{\bar{\sigma}}(\omega) + P_{\bar{\sigma}} (\omega_{2})]\Sigma_{0} (\omega) - Q_{\bar{\sigma}}(\omega) + Q_{\bar{\sigma}} (\omega_{2}) }~,
\label{eq:GreenFunctionAharony_main}
\end{equation}
\end{widetext}
where $u(\omega) \equiv U^{-1}[U-\omega+\epsilon_{\sigma} + 2 \Sigma_{0} (\omega) - \Sigma_{0} (\omega_{2})]$ and $\omega_{2} = -\omega + \epsilon_{\sigma} + \varepsilon_{\bar{\sigma}} + U$. The functions $P_{\sigma}(\omega)$ and $Q_{\sigma}(\omega)$ are given by:
\begin{align}
P_{\sigma}(\omega)\equiv & \frac{i}{2 \pi}\oint_C n_{\rm F}(z)
G_\sigma(z) \frac{\Sigma_{0}(z)-\Sigma_{0}(\omega)}{\omega-z} d z \nonumber \\
Q_{\sigma}(\omega)\equiv & \frac{i}{2 \pi}\oint_C n_{\rm F}(z)
[1+ \Sigma_{0}(z)G_\sigma(z)] \frac{\Sigma_{0}(z)-\Sigma_{0}(\omega)}{\omega-z} d z
\, . \label{eq:P(z)_Q(z)_main}
\end{align}
In this approximation, called hereafter EOM3, $G_{\sigma} (\omega)$ is obtained by solving equations \eqref{eq:GreenFunctionAharony_main}, \eqref{eq:P(z)_Q(z)_main}, and \eqref{eq:occupation} self-consistently. 

Let us briefly address the particular case of particle-hole symmetry in which 
$\varepsilon_0 = - U/2$ and $ \langle n_{\bar{\sigma}} \rangle = 1/2$ for the 
approximation schemes presented above. 
In this  case, the EOM0 Green's function, Eq~ \eqref{eq:GF_EOM0_approx}, reduces to
\begin{equation}
[G^{\rm sym}_\sigma (\omega)]^{-1} = \omega -  \Sigma_0  (\omega)  - \frac{  U^2   }{4 [ \omega -  \Sigma_0 (\omega) ] }~,
\end{equation}
Interestingly, the EOM1 approximation, Eq. (\ref{eq:GF_EOM0}), simplifies to
\begin{equation}
[G^{\rm sym}_\sigma (\omega)]^{-1} = \omega -  \Sigma_0  (\omega)  - \frac{  U^2   }{4  \omega }~,
\end{equation}
Finally, the EOM2 and EOM3 Green's functions,  Eqs.~\eqref{eq:GF-EOM1} and \eqref{eq:GreenFunctionAharony_main}, respectively, coincide and can be written as
\begin{equation}
[G^{\rm sym}_\sigma (\omega)]^{-1} = \omega -  \Sigma_0  (\omega)  - \frac{  U^2   }{4 [ \omega - 3 \Sigma_0 (\omega) ] }~.
\end{equation}

It is surprising and somewhat unexpected that the EOM0 result is closer to the Green's function of the more involved EOM2 and EOM3 schemes 
than the EOM1 one since Eq.~\eqref{eq:GF_EOM0_approx} is frequently regarded as an approximation to the exact form in Eq.~\eqref{eq:GF_EOM0}.

Clearly these Green's functions are temperature independent and hence the EOM solutions can not describe Kondo physics in the particle-hole symmetric point \cite{Theumann1969,Kashcheyevs2006}.

%================================================================
\subsection{Slave boson approximation (SBA)}
\label{sec:NCA_OCA}

Let us now discuss the slave-boson method for solving the SIAM Hamiltonian \cite{Coleman1984} in 
both the non-crossing (NCA) \cite{Pruschke1989,Gerace2002,Aguado2003}
and the one-crossing approximations (OCA)  \cite{Haule2001,Sposetti2016}. 

In some contexts, the interaction part of the SIAM is one of the largest energy scales of the system. 
Hence, a perturbation expansion in $U$ may be problematic. Alternatively, noting that the hybridization
terms are small compared to $U$ and to the kinetic part of the Hamiltonian \cite{bickers1987}, one can 
treat them as the perturbation term of the problem. However, we can not use the standard machinery of 
perturbation theory, since the SIAM Hamiltonian entails an unperturbed term that is not bilinear and, 
in this case, Wick's theorem  does not apply \cite{Barnes1976}.

It is possible to circumvent this problem by employing another representation for the impurity 
operators, the so-called slave boson approximation (SBA).
In the SBA one writes the impurity operator as  \cite{Sposetti2016,Barnes1976,Barnes1977}
\begin{equation}
f^{\dagger}_{\sigma} = s_{\sigma}^{\dagger} b + d^{\dagger}_{\bar{\sigma} \sigma } s_{\bar{\sigma}}~,
\label{eq:pseudo_representation} 
\end{equation}
where $b^{\dagger}$, $s_{\sigma}^{\dagger}$ and $d^{\dagger}_{\bar{\sigma} \sigma }$ are auxiliary  or pseudoparticle (PP)
operators which create, over the vacuum $| vac \rangle$, the states $| 0 \rangle$, $| \sigma \rangle$ and 
$| \bar{\sigma} \sigma \rangle$, respectively.  $b^{\dagger}$ and $d^{\dagger}_{\bar{\sigma} \sigma }$ are 
bosonic operators, while $s_{\sigma}^{\dagger}$ is a fermionic operator.

In this representation the SIAM Hamiltonian reads
\begin{align}
H  = & \sum_{\mathbf k  \sigma}\varepsilon_{\mathbf k  \sigma} c_{\mathbf k  \sigma}^{\dagger}  c_{\mathbf k  \sigma}+ \sum_{\sigma}\varepsilon_{0} s^{\dagger}_{\sigma}s_{\sigma}
+(2 \varepsilon_{0}+U) d^{\dagger}_{ \bar{\sigma} \sigma}d_{\bar{\sigma} \sigma } + \nonumber \\ 
& + \sum_{\mathbf k  \sigma}\left(V_{\mathbf k  \sigma} s^{\dagger}_{\sigma}b \;c_{\mathbf k  \sigma }+ {\rm H.c.}\right) \nonumber \\ 
& +\sum_{\mathbf k  \sigma}\left(V_{\mathbf k } d^{\dagger}_{ \bar{\sigma} \sigma }s_{\bar{\sigma} }c_{\mathbf k  \sigma}+ {\rm H.c.}\right) ~.
\label{eq:SIAMBarnesrepresentation}
\end{align}
Here, for simplicity, we consider {the case where} $\varepsilon_0 = \varepsilon_{\uparrow} = \varepsilon_{\downarrow}$.  
The subspace of impurity states is formed by four states, namely, $| 0 \rangle$, $\mid\uparrow \rangle$, 
$\mid\downarrow \rangle$, and $\mid\uparrow \downarrow \rangle$ representing an empty, singly-occupied 
with spin up/down, and doubly-occupied impurity states, respectively. This implies that, at any given time \cite{Aguado2003,Haule2001,Haule2010,Jacob2010}
\begin{equation}
Q \equiv b^{\dagger} b + \sum_{\sigma} s_{\sigma}^{\dagger} s_{\sigma} + d_{\bar\sigma \sigma}^{\dagger} = 1 ~,
\end{equation}
where $Q$ is called pseudo-particle charge.
The constraint $Q=1$ has to be enforced in the calculation of the expectation value of any given observable $\langle O \rangle$. 
In practice, one first calculates an unrestricted expectation value, for instance, using diagrammatic techniques in the grand canonical ensemble to obtain $\langle O \rangle_{G}$. 
In this ensemble a chemical potential $- \lambda$ is associated to the pseudo-particle charge $Q$. 
Next, one projects the result into the  $Q=1$ canonical subspace to find $\langle O \rangle_{C}$ using 
the so-called Abrikosov's trick, namely \cite{Coleman1984,Kroha1998,Abrikosov1965,Hewson1997,Wingreen1994}
\begin{equation}
\langle O \rangle_{C} = \lim_{\lambda \to \infty} \frac{ \langle O \rangle_{G} }{\langle Q \rangle_{G}}.
\end{equation}

The impurity Green's function is written in terms of the pseudo-particle Green's functions \cite{Kroha1998}
\begin{align}
\label{eq:G_OCA}
G_{b} (\omega) & = [\omega - \lambda - \Sigma_b (\omega)]^{-1} \nonumber \\
G_{s_\sigma} (\omega) & =  [\omega - \lambda - \varepsilon_0 - \Sigma_{s_{\sigma}} (\omega)]^{-1} \nonumber \\
G_{d} (\omega) & =  [\omega -\lambda - 2\varepsilon_0 - U - \Sigma_{d} (\omega)]^{-1},
\end{align}
whose self-energies are obtained by analytic continuation and projection into the $Q=1$ subspace, namely \cite{Sposetti2016}
\begin{eqnarray}
\label{eq:self-energy_OCA}
\begin{split}
 \Sigma_b(\omega) & = \int_{-\infty}^{\infty} \frac{d\epsilon}{\pi}n_{\rm F}(\epsilon)\sum_{\sigma}\Delta_{\sigma}(\epsilon)G_{s_{\sigma}}(\epsilon+\omega)
 \Lambda_{\sigma}^{(0)}(\omega,\epsilon),\\
 \Sigma_{s_\sigma}(\omega) & =\int_{-\infty}^{\infty} \frac{d\epsilon}{\pi}n_{\rm F}(\epsilon)\Big[\Delta_{\sigma}(-\epsilon)G_{b}(\epsilon+\omega) \times \\
 & \times \Lambda_{\sigma}^{(0)}(\epsilon+\omega,-\epsilon) +\Delta_{\bar{\sigma}}(\epsilon) G_{d_{\sigma \bar{\sigma}}}(\epsilon+\omega) \times \\ 
 & \times \Lambda_{\sigma \bar{\sigma}}^{(2)}(\epsilon+\omega,\epsilon) \Big],\\
 \Sigma_{d_{\sigma \bar{\sigma}}}(\omega) & =\int_{-\infty}^{\infty} \frac{d\epsilon}{\pi}n_{\rm F}(\epsilon) \left[ \Delta_{\sigma}(-\epsilon)G_{s_{\bar{\sigma}}}(\epsilon+\omega)
      \Lambda_{\sigma \bar{\sigma}}^{(2)}(\omega,-\epsilon) + \right. \\
 & + \left. \Delta_{\bar{\sigma}}(-\epsilon)G_{s_{\sigma}}(\epsilon+\omega)\Lambda_{\bar{\sigma} \sigma}^{(2)}(\omega,-\epsilon)\right],\\
 \end{split}
\end{eqnarray}
where $\Delta_\sigma(\ve) \equiv \sum_\alpha \Gamma_\sigma^{\alpha}(\ve)/2$.
Equations \eqref{eq:G_OCA} and \eqref{eq:self-energy_OCA} are solved self-consistently. 

In the simplest diagrammatic perturbation theory approximation, which neglects vertex corrections, $\Lambda_{\bar{\sigma} \sigma}^{(2)} (\omega,\omega') = \Lambda_{\sigma}^{(0)} (\omega,\omega') = 1$. This simplification corresponds to the so-called non-crossing approximation (NCA) \cite{bickers1987}. By including higher order diagrams, one obtains
\begin{eqnarray}\label{vertices_oca}
\begin{split}
\Lambda_{\sigma}^{(0)}(\omega,\omega')&=  1+ \int_{-\infty}^{\infty}\frac{d\epsilon}{\pi}n_{\rm F}(\epsilon) \Delta_{\bar{\sigma}}(\epsilon) G_{s_{\bar{\sigma}}}(\omega+\epsilon) \\ 
&~~~~~~~~~~~~~~~~~~ \times 
G_{d_{\bar{\sigma}} \sigma }(\omega+\omega'+\epsilon) \\
\Lambda_{\sigma \bar{\sigma}}^{(2)}(\omega,\omega')&=1+ \int_{-\infty}^{\infty}\frac{d\epsilon}{\pi}n_{\rm F}(-\epsilon) \Delta_{\sigma}(\epsilon) \times \\
&~~~~~~~~~~~~~~~~~~ \times G_{s_{\bar{\sigma}}}(\omega-\epsilon)G_{b}(\omega-\omega'-\epsilon).
\end{split}
\end{eqnarray}
which contain vertex corrections at the order of one-crossing approximation (OCA) \cite{Haule2001,Sposetti2016}.

The impurity density of states, Eq.  \eqref{eq:impurity_DOS}, is given in terms of the pseudo-particle spectral functions, namely \cite{Sposetti2016, Haule2001, Pruschke1989}
\begin{align}
\label{eq:pseudoparticleSpec}
\rho_{\sigma}(\omega)= \frac{Z_C (0)}{Z_C (1) }\int_{-\infty}^{\infty} d\epsilon ~e^{-\beta\epsilon}  \Big[ & A_b(\epsilon)
A_{ s_{\sigma} }(\epsilon+\omega) ~+ \nonumber \\
& A_{ s_{\sigma} }(\epsilon)  A_d(\epsilon + \omega) \Big]
\end{align}
where $Z_C (1)$ is the canonical partition function for a system with one impurity and $Z_C (0)$ is the canonical partition 
function of the conduction electrons (no impurity). The ratio $Z_C (1)/Z_C (0)$, which appears from the projection to the 
$Q=1$ subspace, can be written as
\begin{equation}
\label{eq:Z1Z0}
\frac{Z_C (1)}{Z_C (0) } = \int_{-\infty}^{\infty} d \epsilon \, e^{- \beta \epsilon} \Big[ A_b(\epsilon) + \sum_{\sigma}
A_{ s_{\sigma} }(\epsilon) + A_d(\epsilon ) \Big] .
\end{equation}
These elements allow us to compute the conductance $\mathcal{G}$ given by Eq.~\eqref{eq:G_linear_response}.

The energy integrals in Eqs.~\eqref{eq:pseudoparticleSpec} and \eqref{eq:Z1Z0} have to be evaluated carefully because of the exponential divergences of the statistical factors appearing in both expressions. 
Refs.~\onlinecite{Hettler1998,Costi1996} discussed how to deal with large negative values of $\beta \varepsilon$ in the integrand.
The exponential factor is compensated by the threshold behavior of the auxiliary spectral functions, vastly reported in the literature. \cite{Hettler1998, Costi1996, Kroha1998} The singular threshold structure poses serious difficulties 
to converge the NCA and OCA equations.\cite{Hettler1998}. The trick to achieve convergence exploits the fact that theses equations 
are invariant under a frequency argument shift of the PP spectral functions, namely, $\omega \to \omega + \lambda_0$. \cite{Hettler1998,Costi1996,Kroha1998}
%In this case, the PP chemical potential $\lambda$ is shiftedcan assume an arbitrary value. 
In the implementation used here \cite{Haule2010} $\lambda_0$ can be calculated in two ways:
(i) $\lambda_0$ is fixed and $Q$ is calculated \footnote{See, for example, the Appendix of Ref. \onlinecite{Sposetti2016} }and (ii) $\lambda_0$ is calculated and $Q$ is fixed \footnote{See, for example, the Appendix A of Ref. \onlinecite{Hettler1998} }.  In both methods, the parameter $\lambda_0$ 
shifts the peak of the main \footnote{The main pseudo-particle is the one with the lowest energy. Note that $\varepsilon_b = 0$, $\varepsilon_{s_{\sigma}}=\varepsilon_0$, and $\varepsilon_d = 2\varepsilon_0 + U$}  PP to $\omega = 0$. The $\lambda_0$ calculation combined with a non-uniform frequency mesh with a high-density of points around $\omega = 0$ greatly improves the convergence of the equations. However, it is worth to emphasize that the sharp peaks in the PP spectral functions can still lead to instabilities in the calculation, in particular for very low temperatures and 
away from the particle-hole symmetry point.   

Despite its success in capturing important low temperature features of the SIAM, the SBA has several known problems. \cite{Tosi2011,Grewe2008, Anders1995,Schmitt2009,Vildosola2015} The NCA, for instance, fails to give the correct Kondo scale $T_K$ at finite $U$ \cite{Haule2001}. The inclusion of vertex corrections solve this pathology to some extent. \cite{Vildosola2015,Grewe2008} However for sufficiently low temperatures ($T\lesssim 0.1 T_K$) the height of the Kondo resonance of the OCA spectral functions slightly overestimates the limit imposed by Friedel's sum rule. \cite{Haule2010,Tosi2011,Schmitt2009,Vildosola2015} As pointed out in Ref. \onlinecite{Ruegg2013} 
the OCA also violates the sum rules for the coefficients of the high-frequency expansion of the electronic self-energy, especially in the case of multi-orbital Anderson models.

%================================================================
\subsection{Numerical renormalization group (NRG)}
\label{sec:NRG}

Last, we employ Wilson's numerical renormalization group (NRG) method \cite{Bulla2008} to the SIAM in order to calculate the impurity spectral function  $\rho_{\sigma}(\omega)$ as well as the conductance $\mathcal{G}$, given by Eq.\ (\ref{eq:G_linear_response}).

We can summarize the main ingredients of the NRG scheme as follows. The first step (i) is to perform a logarithmic discretization (in energy) of the non-interacting conduction electron Hamiltonian [Eq.\ (\ref{eq:LeadHamiltonian})] by considering discrete energy intervals $D \Lambda^{-(m+1)} \leq \omega \leq D \Lambda^{-m}$ where  $\Lambda > 1$ is a discretization parameter.  Step (ii) involves mapping the impurity+discretized band to an effective 1D tight-binding chain (usually dubbed the ``Wilson chain"), whose first site is coupled to the interacting impurity. Due to the logarithmic discretization in the original band, the mapping produces couplings $t_n$ between sites $n$ and $n+1$ which decay as $t_n \sim \Lambda^{-n/2}$. This feature also defines a characteristic energy scale for a given Wilson chain length $N$ as $D_N=\frac{1}{2}\left(1+\Lambda^{-1}\right)\Lambda^{-(N-1)/2}D$  where $D$ is the half-bandwidth of the metallic band appearing in Eq.\ (\ref{eq:LeadDOS}). 

Finally, step (iii) amounts to an iterative numerical diagonalization of the resulting Hamiltonian $H(N)$ of the impurity plus a Wilson chain of length $N$, and the subsequent mapping to a system with one extra site $H(N+1)$, given by the renormalization condition: 
\begin{equation}
H(N+1) = \sqrt{\Lambda} H(N) + \xi_N \sum_{\sigma}( f^{\dagger}_{N+1, \sigma}f_{N, \sigma} + f^{\dagger}_{N, \sigma}f_{N+1, \sigma}) ,
\end{equation}
where $f^{\dagger}_{N, \sigma}$ creates an electron with spin $\sigma$ on site $N$ of the Wilson chain (sometimes referred to as ``shell $N$'') and $\xi_N \propto t_N \Lambda^{N/2} \sim 1$ is the renormalized coupling. The final approximation is to use the 1000--2000 lowest-energy states of $H(N)$ (the ``kept'' states) to generate a basis for $H(N+1)$. The process is then repeated for $N+2, N+3,\cdots, N_{\rm max}$  until the desired lowest energy scale $D_{N_{\rm max}}$ is achieved. 

The key result out of the NRG algorithm is the calculation of the (many-body) energy spectrum $\{ | r \rangle_N \}$ for each $H(N)$ in the form $H(N) | r \rangle_N \!=\! E^{N}_{r} | r \rangle_N$ along with matrix elements $\langle r | \hat{\mathcal{O}} | r^{\prime} \rangle_N$ for local operators $\hat{\mathcal{O}}$ within the approximations involved (logarithmic discretization of the conduction band and truncation of the spectrum at each $N$).

%---------------------------------------------------------------------------------------------------
\subsubsection{Impurity spectral density}
\label{sec:SpectralDensityNRG}

With the many-body spectrum at hand, one can formally write the leading contribution of the NRG spectrum to the impurity spectral function at a frequency $\omega \sim D_N$ in the Lehmann representation as
\begin{equation}
\rho^{N}_{\sigma}(\omega) \! = \! \sum_{r, r^{\prime}} | \mathcal{A}^{N}_{r r^{\prime}}|^2 \frac{e^{-\beta E^{N}_{r}}+e^{-\beta E^{N}_{r^{\prime}}}}{Z_N} \delta(\omega - E^{N}_{r^{\prime}} + E^{N}_{r}) \; ,
\label{eq:LehmannRep} 
\end{equation}
where $Z_N = \sum_{r} e^{-\beta E^{N}_{r}}$ and $\mathcal{A}^{N}_{r r^{\prime}} =  \langle r | f_{\sigma} | r^{\prime} \rangle_N$ is the matrix element of the impurity operator $f_{\sigma}$. 

As it stands, Eq.\ (\ref{eq:LehmannRep}) yields a sum of delta functions centered at the excitation energies $\Delta E^{N}_{r,r^{\prime}} \equiv E^{N}_{r^{\prime}} - E^{N}_{r}$ in the spectrum for $N$-sites in the Wilson chain. In order to obtain the continuous spectral function $\rho_{\sigma}(\omega)$ at energy $\omega$, some additional approximations are needed. First, the delta functions in Eq.\ (\ref{eq:LehmannRep}) need to be broadened, {a procedure which might introduce overbroadening errors \cite{Zitko:Phys.Rev.B:085142:2011}. In order to minimize such errors,  Gaussian or log-Gaussian kernels with small broadening widths are typically used \cite{Bulla2008,Peters2006,Weichselbaum2007}}.

The second approximation is to collect the spectral contributions for different shells to the spectral function at a given energy $\omega$. Early NRG calculations \cite{Costi1994} employed the so-called ``single-shell approximation'', which amounts to take a single contribution at each shell $N$ in the form: $\rho_{\sigma}(\omega \! \approx \! D_N) \approx \rho^{N}_{\sigma}( \omega\!=\!a(\Lambda)D_N)$ where  and $a(\Lambda)\sim 2-3$ is a $\Lambda$-dependent numeric pre-factor. Modern NRG implementations \cite{Peters2006,Weichselbaum2007} use contributions from several shells, weighted by the reduced density-matrix at iteration $N$ and separating contributions from states ``kept'' and ``discarded'' during the truncation process at each $N$ (the ``complete Fock space'' (CFS) approach) in order to avoid double counting between contributions of different shells.

In this work, we employ the ``full-density-matrix NRG'' (FDM-NRG) approach \cite{Weichselbaum2007} to calculate the spectral functions at zero and finite temperatures. For zero-temperature calculations, both FDM-NRG and CFS approaches are equivalent \cite{Peters2006,Weichselbaum2007}. For finite temperatures, FDM-NRG is usually the method of choice, with a caveat: as it is true in all NRG implementations, the spectral function resolution in energy is limited for $\omega \ll T$ due to the appearance of spurious peaks related to errors introduced by the logarithmic discretization procedure \cite{Zitko:Phys.Rev.B:085142:2011}. As such, even the FDM-NRG procedure with ``z-trick averaging'' \cite{Yoshida:Phys.Rev.B:41:9403:1990} produces spurious peaks for $\omega \ll T$. Thus, good quality data is typically limited to energies $\omega \gtrsim T$.

\subsubsection{Conductance}
\label{sec:ConductanceNRG}

In order to obtain the conductance from NRG data, we can combine Eqs.\ (\ref{eq:G_linear_response}) and (\ref{eq:LehmannRep}) and use the delta functions to perform the integrals in energy for each shell. As a result, we obtain the contribution to the conductance from shell $N$ as: \cite{Zawadzki2018}  
%s
\begin{equation}
g_N(T) = \frac{\pi \beta}{Z_N} \sum_{r, r^{\prime}} \frac{| \mathcal{A}^{N}_{r r^{\prime}}|^2}{e^{\beta E^{N}_{r}}+e^{\beta E^{N}_{r^{\prime}}}} \; ,
\label{eq:gNT}
\end{equation}
and
\begin{equation}
{\mathcal G}(T) = \frac{2 e^2}{h} \Gamma \sum_{N} g_N(T) \; .
\label{eq:CondNRG}
\end{equation}

Within the NRG approximations, Eq.\ (\ref{eq:CondNRG}) gives ${\mathcal G}(T)$ down to arbitrarily low-temperatures of order $T \sim D_{N_{\rm max}}$, fully describing the regime $T \ll T_K$, where $T_K$ is the Kondo temperature. Notice that no broadening in the delta functions is necessary in using Eqs.\ (\ref{eq:gNT}) and (\ref{eq:CondNRG}). For both spectral function and conductance calculations, it is advisable to perform $z$-averaging in order to get rid of the spurious oscillations arising from the NRG discretization errors. In the calculations, we have averaged over $N_z\!=\!5-10$ values of $z$.

%%%%%%%%%%%%%%%%%%%%%%%%%%%%%%%%%%%%%%%%%%%
\section{Results}
\label{sec:results}
%%%%%%%%%%%%%%%%%%%%%%%%%%%%%%%%%%%%%%%%%%%

In this section we critically compare calculations of the density of states and the conductance for 
the SIAM obtained by the methods discussed above.
For the OCA we used the implementation of Haule and collaborators \cite{Haule2010}, while for 
the others we used in-house codes. 
All calculations were performed in the wide-band approximation and in the absence 
of a magnetic field, namely, $\ve_\uparrow = \ve_\downarrow=\ve_0$.

When studying the SIAM it is usual to separate two limits based on the characteristic Kondo temperature 
of the system: the high and the low temperature regime when, $T \gg T_K$ and $T\lesssim T_K$, respectively. 
We proceed accordingly using the Haldane estimate \cite{Haldane1978} for $T_K$, namely,
\begin{equation}
T_{K} \sim \sqrt{U \Gamma} \exp \!\left(\!-\pi\frac{|\varepsilon_0(\varepsilon_0 + U)|}{U \Gamma}  \right),
\label{eq:Haldane_expression}
\end{equation}
as a guideline. 

%===================================================
\subsection{Density of states}
\label{sec:DOS}

\subsubsection{High-temperature limit}
\label{sec:hightempDOS}

\begin{figure}[h]
\includegraphics[width=1.0\columnwidth]{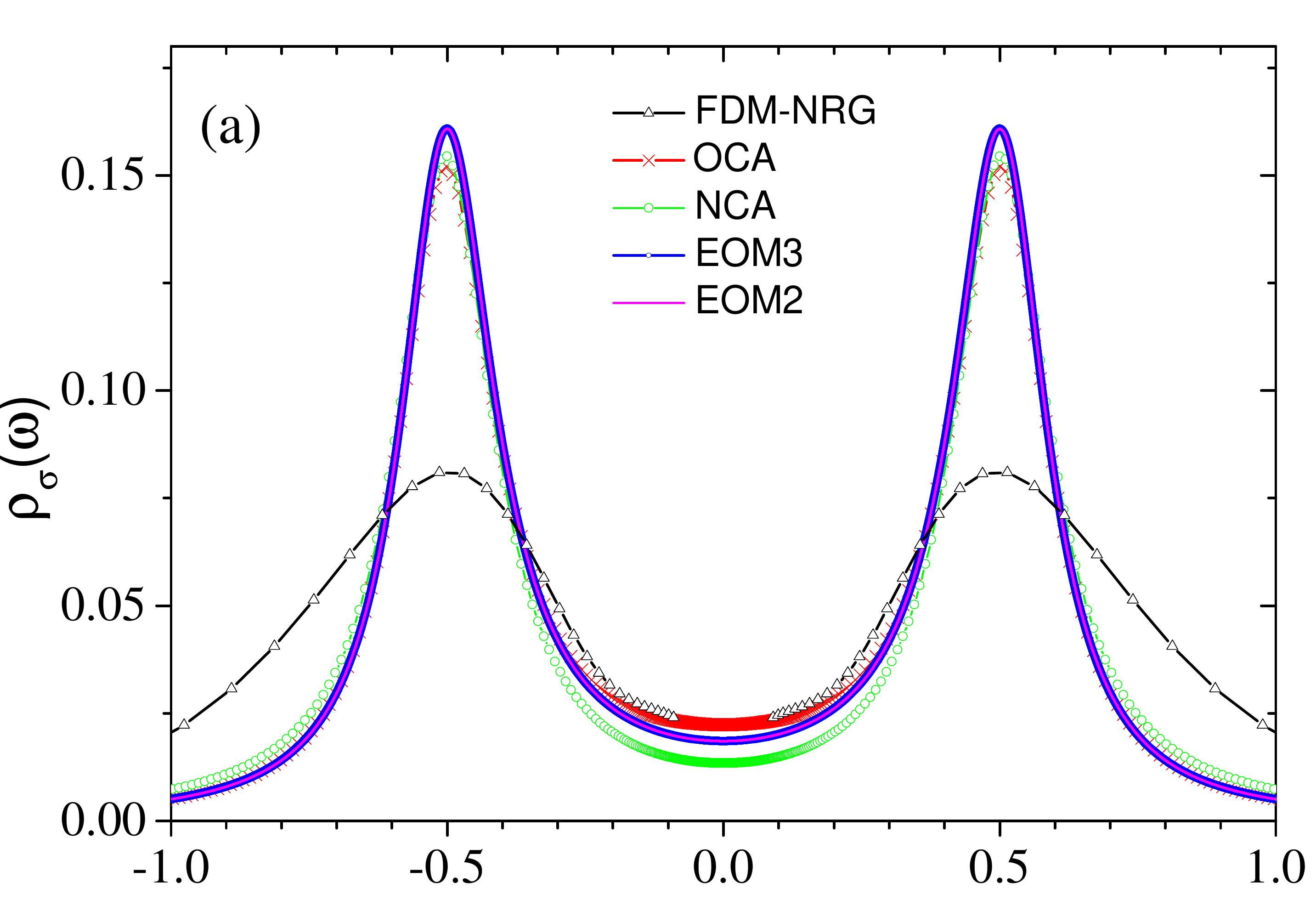}
\includegraphics[width=1.0\columnwidth]{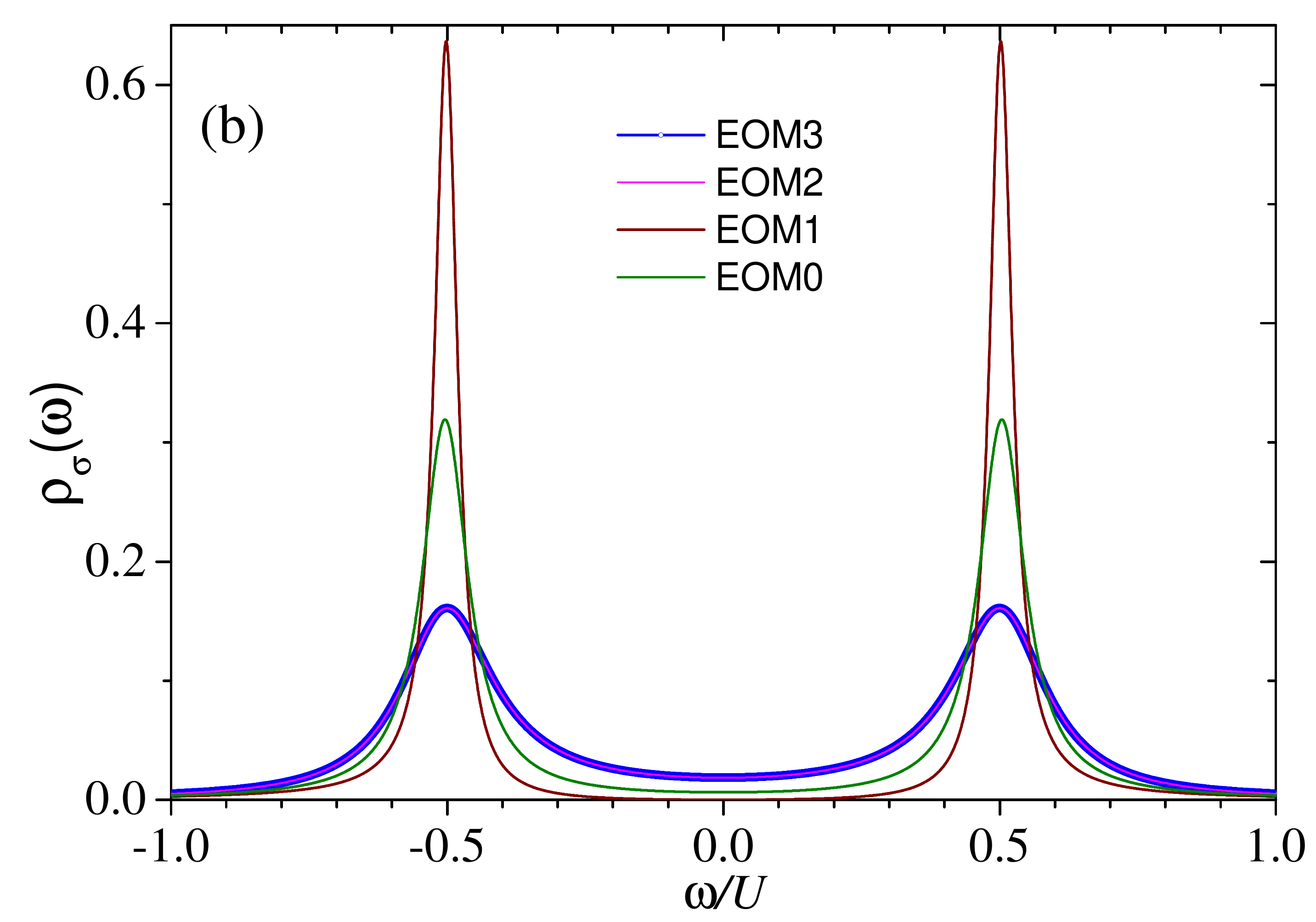}
\caption{(a) Density of states of the SIAM for different methods for $\varepsilon_0\!=\! -U/2$, $\Gamma=0.1U$, and $T=0.1U$ . 
%\Luis{Bruno, please use the new FDM-NRG data.} 
(b) Same parameters as above, comparison between different EOM schemes.
}
\label{fig:DOScomparison_all_symmetric}
\end{figure}

Let us begin with the high temperature limit, $T\gg T_K$. 
Figure \ref{fig:DOScomparison_all_symmetric} shows $\rho_\sigma(\omega)$ for the particle-hole symmetric case, $\ve_0=-U/2$.
We find that all methods, except NRG, give  similar qualitative results. The main feature is the double peak structure with resonances at $\omega = \varepsilon_0$ and $\omega = \varepsilon_0 + U $, both with a width $\Gamma$ characteristic of the Coulomb blockade regime in quantum dots. 

Despite the qualitative similarities, even at the high temperature limit, there are significant differences between the approximation schemes. We find that the EOM0 and EOM1 give very poor results as we discuss below. Fig. \ref{fig:DOScomparison_all_symmetric}b shows that both peaks have the same height, implying that $\langle n_\sigma \rangle = 0.5$.
Using this input, one can obtain an analytical estimate for the EOM DOS at $\omega=\pm U/2$ , namely, 
\begin{align}
%&\rho^{\rm EOM0} (\omega= \pm \frac{U}{2})  = \frac{1}{ \pi \Gamma} %L Not good: parenthesis too small.
%&\rho^{\rm EOM0} \left( \omega= \pm \frac{U}{2} \right)  = \frac{1}{ \pi \Gamma} %L Not good: parenthesis too large.
&\rho^{\rm EOM0} \left( \omega= \pm U/2 \right)  = \frac{1}{ \pi \Gamma} %L ok.
\nonumber \\
%&\rho^{\rm EOM1} (\omega=\pm \frac{U}{2}) = \frac{2}{ \pi \Gamma}
&\rho^{\rm EOM1} \left( \omega=\pm U/2 \right) = \frac{2}{ \pi \Gamma}
\nonumber \\
%\rho^{\rm EOM2} (\omega= &\pm \frac{U}{2})  = \rho^{\rm EOM3} (\omega=\pm \frac{U}{2}) = \frac{1}{2 \pi \Gamma}.
\rho^{\rm EOM2} (\omega=  \pm U/2 ) & = \rho^{\rm EOM3} (\omega=\pm U/2) = \frac{1}{2 \pi \Gamma}.
\end{align}
This indicates that, at odds with the claims found in the literature \cite{Haug2008}, the different EOM truncation 
schemes render very discrepant density of states, see Fig. \ref{fig:DOScomparison_all_symmetric}b.

\begin{figure}
\includegraphics[width=1.0\columnwidth]{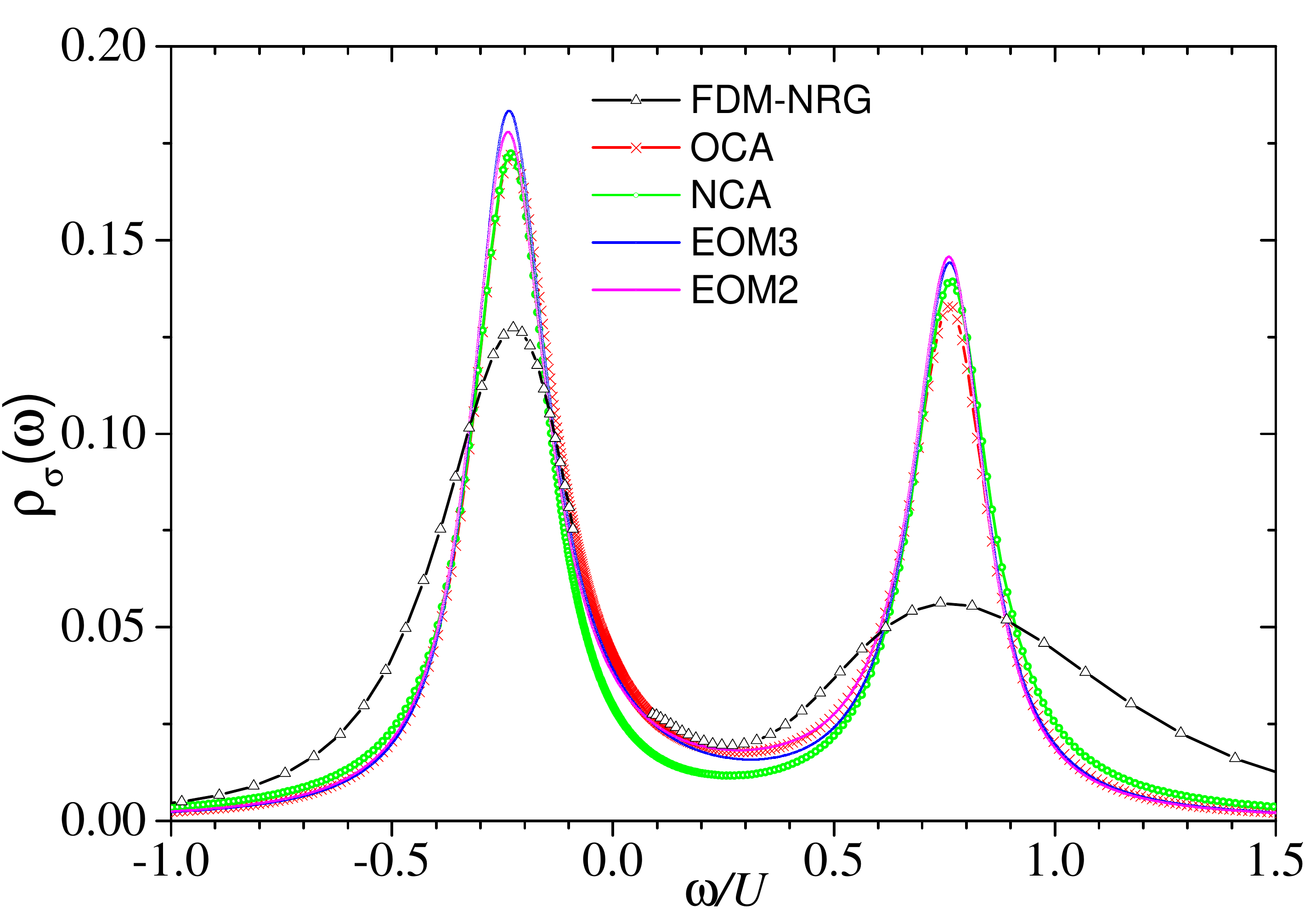}
\caption{Density of states of the SIAM for $\Gamma=0.1U$ and $T=0.1U$ for  $\varepsilon_0 = -U/4$.
}
%\vspace{-10pt}
\label{fig:DOScomparison_all}
\end{figure}

Figure \ref{fig:DOScomparison_all_symmetric} shows that EOM2, EOM3, NCA, and OCA give very similar results for $\rho(\omega)$, 
while NRG overbroadens the Hubbard peaks, a well-known limitation of the method \cite{Grewe2008,Georges1996}.
We point out that, as discussed in Sec.~\ref{sec:SpectralDensityNRG}, the FDM-NRG procedure for obtaining spectral functions suffers from spurious oscillations in the $\omega \ll T$ range \cite{Weichselbaum2007}. Thus, only NRG data for $ |\omega| \gtrsim T$ is shown in Figs. \ref{fig:DOScomparison_all_symmetric}a and \ref{fig:DOScomparison_all}.

%In Fig. \ref{fig:DOScomparison_all_symmetric}a we also note that the two resonances appearing in the density of %states are also not well described in the NRG result. The poor description of the high energy features of the
% spectral functions is a recurrent problem in NRG calculations \cite{Grewe2008,Georges1996}.
%Not even the FDM-NRG procedure is capable of overcoming the limitations of the method in the $T \gg T_K$ limit.

% ------------------------------ FIGURE 1 --------------------------------------
% ------------------------------ FIGURE 1 --------------------------------------

% ------------------------------ FIGURE 2 --------------------------------------
% --------------------------- END FIGURE 2 -------------------------------------

\begin{figure}[h]
\includegraphics[width=1.0\columnwidth]{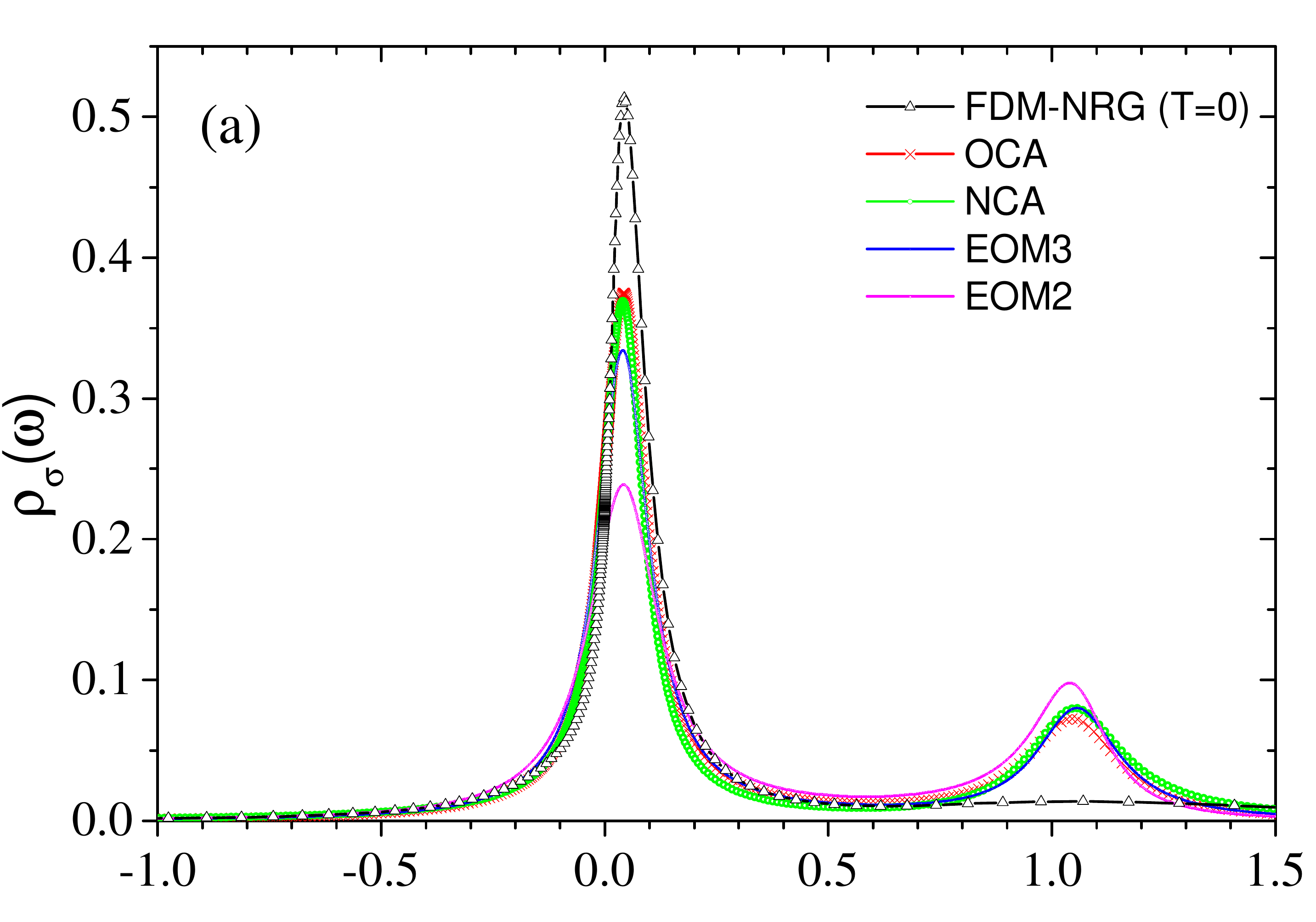}
\includegraphics[width=1.0\columnwidth]{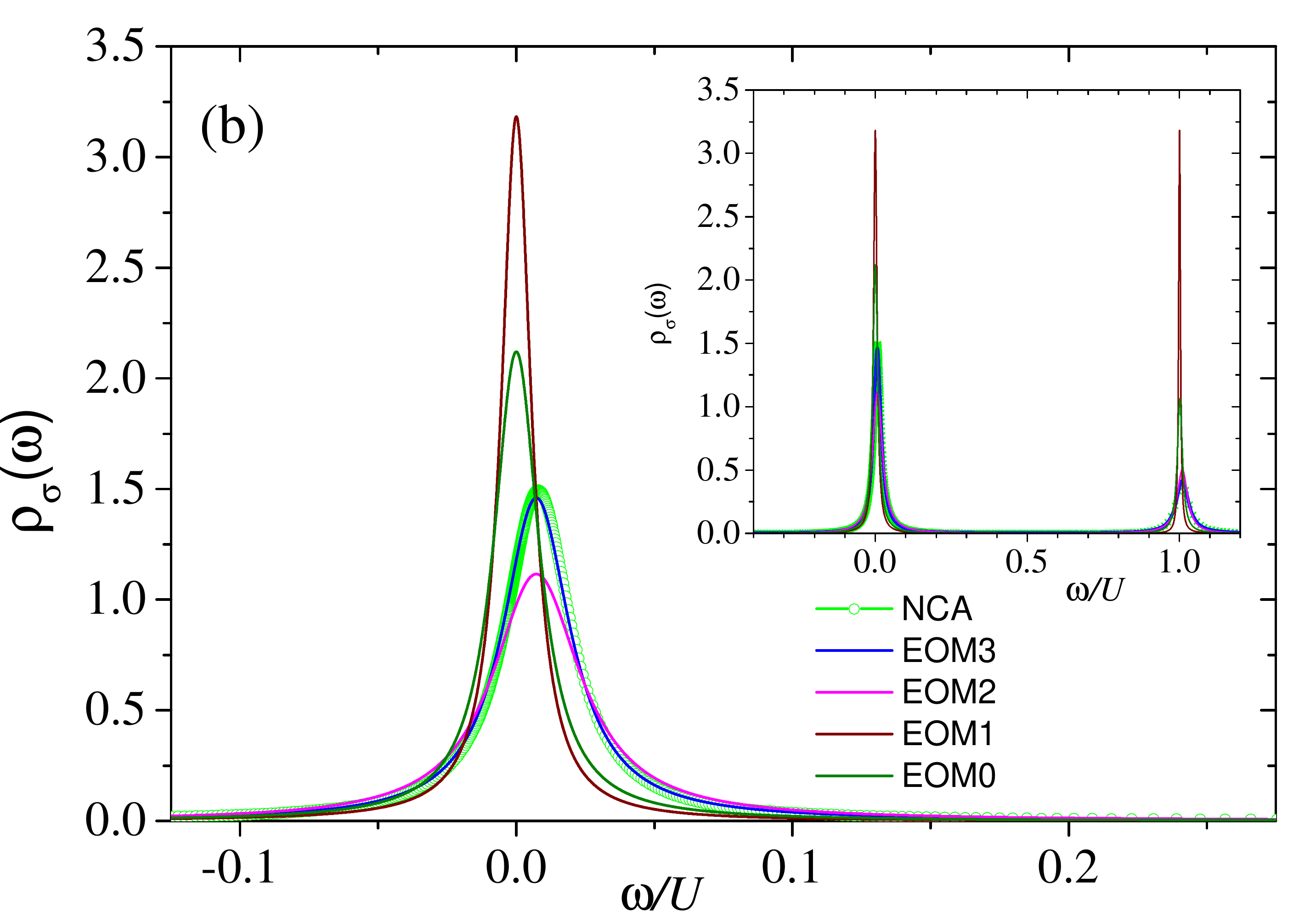}
\vspace{-10pt}
\caption{ Impurity density of states for $\varepsilon_0 = 0$, $T=0.1U$. In (a) we use $\Gamma=0.1U$. In (b) we set $\Gamma = 0.02U$ and compute the DOS near the Fermi level $\omega=0$ for different EOM schemes and also for the NCA. Here the estimated Kondo temperature is $T_K \sim 10^{-18}U$ for $\varepsilon_0 = -U/2$. In the inset we show the density of states in a more extended range of energies. }\vspace{-10pt}
\label{fig:DOSasymmetric-comparison}
\end{figure}

As $\ve_0$ is moved away from the particle-hole symmetry point, the DOS peaks become asymmetric. For $\varepsilon_0 = -U/4$ and keeping the same values for $\Gamma$ and $T$,
the different solvers still give very similar results, in line with the analysis we presented for the symmetric case. The results are shown in Fig.~\ref{fig:DOScomparison_all}. The differences between the approximation schemes become more pronounced for $\ve_0 =0$, see Fig. \ref{fig:DOSasymmetric-comparison}a). In particular, the EOM2 density of states exhibits a 
smaller peak height at the Fermi level as compared with the one obtained by the other methods.

% \Caio{Haldane predicts a rise in $T_K$ as the DOS peaks approach the $\mu=0$. How much can we quantitatively trust Haldane's estimate in such cases? At a qualitative level, a larger $T_K$, makes $T/T_K$ smaller and explains the discrepancies.} 
% \Luis{The $\mu=0$ limit is tricky since the dot occupation falls well below 1 (typically to less than 0.5) for $\mu \gtrsim 0$. If the dot is unoccupied, it makes no sense in talking about Kondo or $T_K$ since the Schrieffer-Wolff transformation breaks down. }  

Note that for $T \alt \Gamma$ the most relevant resonance configurations for the conductance calculations are
the asymmetric ones, where the DOS peak matches $\mu=0$. The failure of the EOM2 approximation in reproducing 
the DOS obtained by more accurate approaches for the asymmetric case indicates that, even 
at the high temperature regime, spin correlations cannot be entirely neglected.

Figure \ref{fig:DOSasymmetric-comparison}b sheds further light on the EOM results. It clearly shows that since the EOM0
self-energy is just the embedding $\Sigma_0(\omega)$, this approximation underestimates $\Sigma_\sigma(\omega)$ and,
hence, overestimates the DOS peak heights. An analysis of $\rho_\sigma (\omega)$ at $\omega = 0$ and $\omega =U$ reveals 
that Eq.~\eqref{eq:GF_EOM0} has another rather unphysical feature: at the peaks, $\rho^{\rm EOM1}_\sigma$ becomes 
independent of $\langle n_{\overline{\sigma}} \rangle$, that is, both DOS peaks have the same height. 
Curiously, the simplest heuristic approximation EOM0 gives superior results in this respect. 

% ------------------------------ FIGURE 3 --------------------------------------
% --------------------------- END FIGURE 3 -------------------------------------

\subsubsection{Low-temperature limit}
\label{sec:lowtempDOS}

\begin{figure}[h]
\centering
\vspace{-10pt}
\includegraphics[width=1.0\columnwidth]{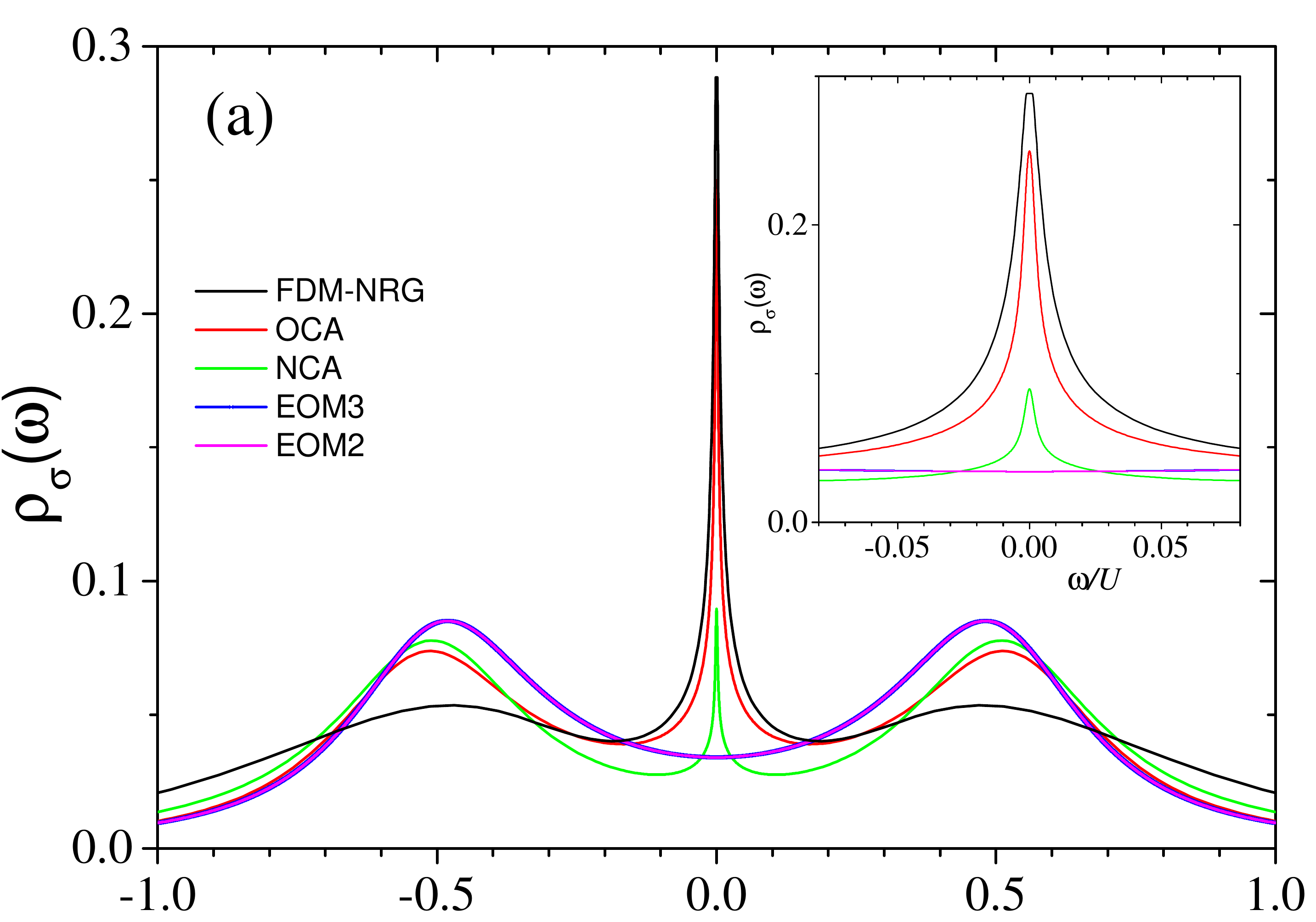}
\includegraphics[width=1.0\columnwidth]{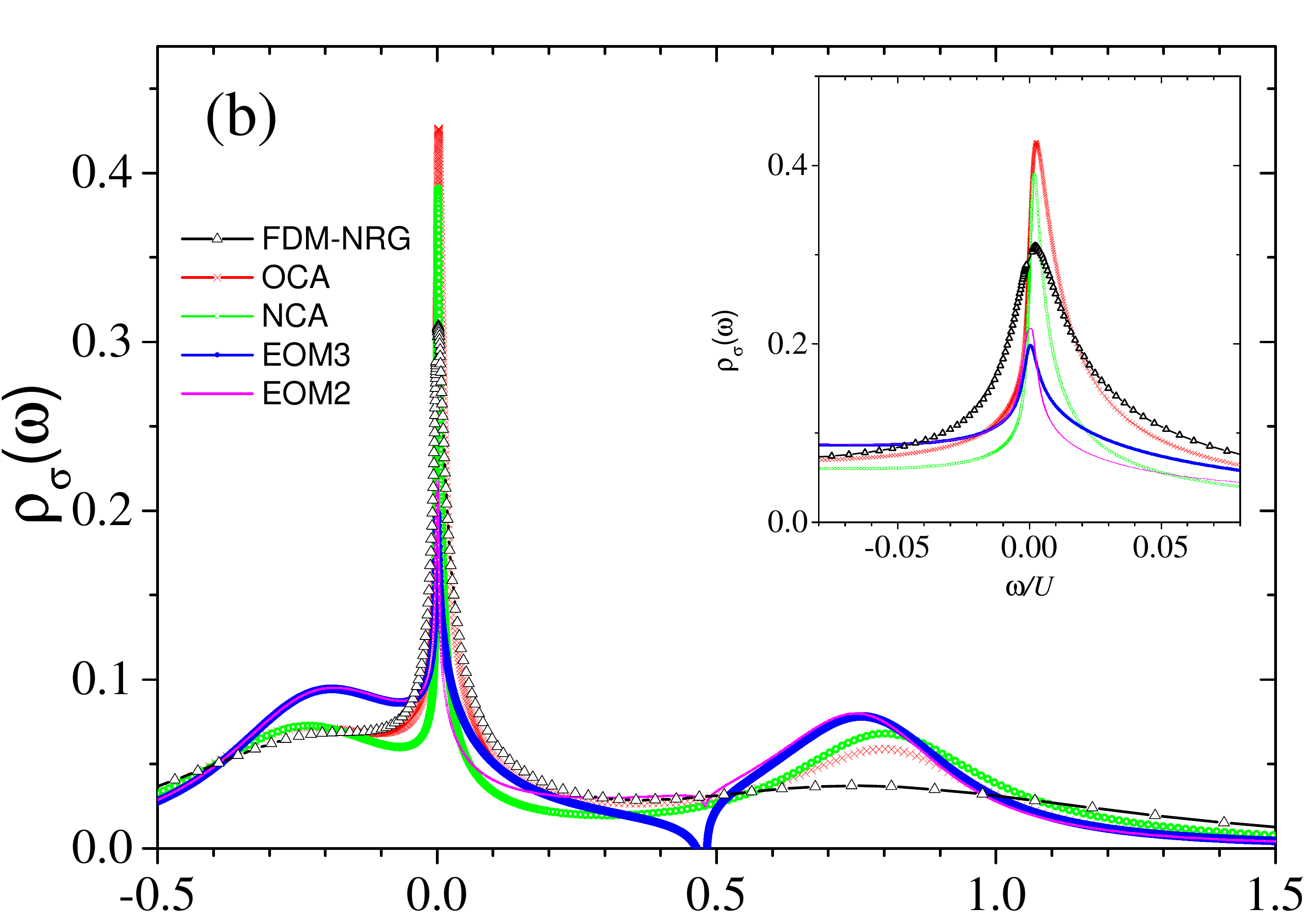}
\includegraphics[width=1.0\columnwidth]{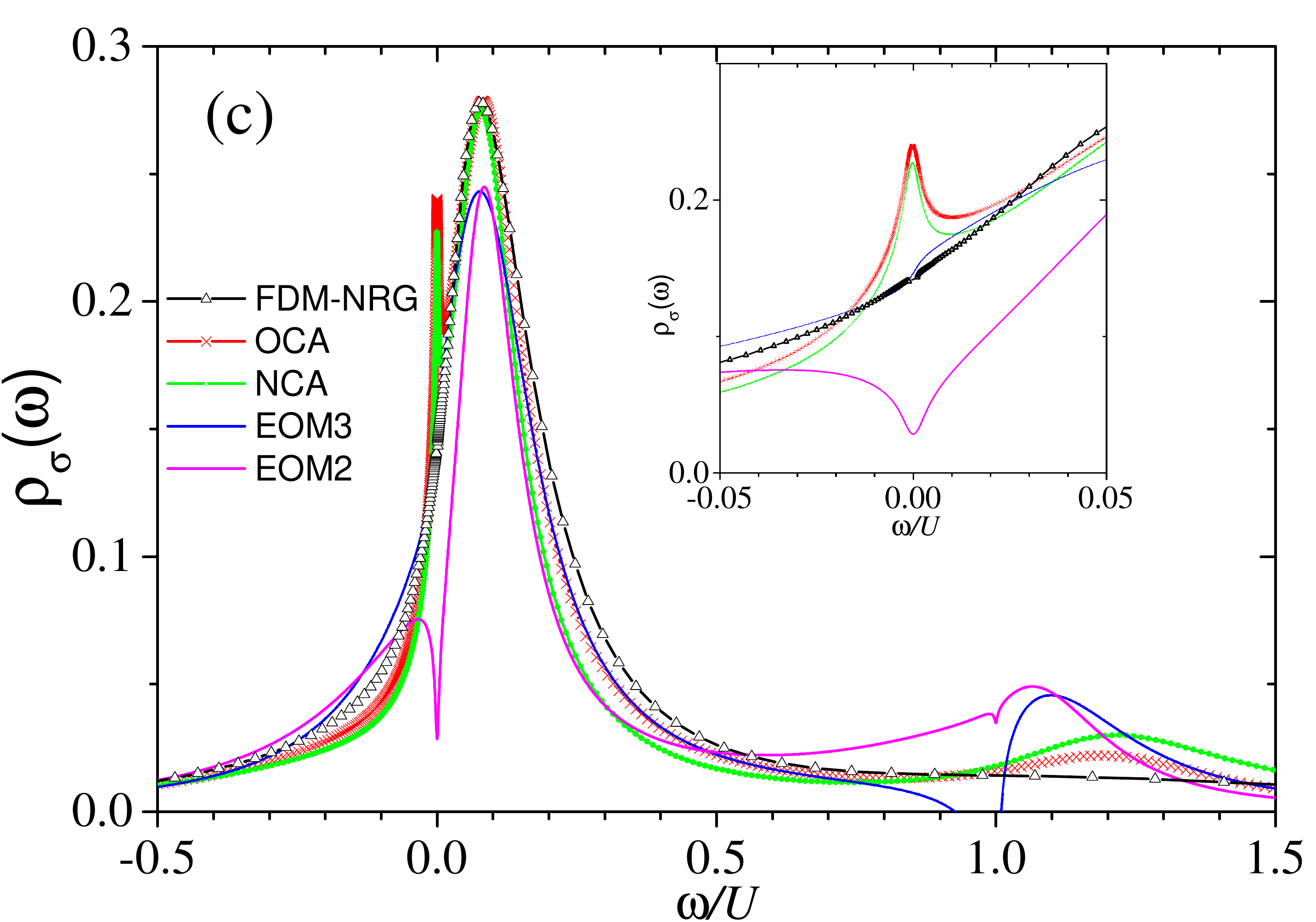}
\caption{Density of states of the SIAM for $\Gamma=0.2U$, $T=0.001U$, and (a) $\varepsilon_0=-U/2$,
(b) $\ve_0=-0.26U$, and (c) $\ve_0 = 0$. The insets show a zoom near the region $\omega=0$.}
\vspace{-10pt}
\label{fig:DOS-lowT}
%\todo[inline]{Dados do NRG perto do pico Kondo parece ter pontos faltando. Há ou não um pico condo em c)? O pico Kondo em b) e tao diferente para o OCA?}
\end{figure}

Now we turn to the low-temperature regime,  Fig. \ref{fig:DOS-lowT}. 
Besides the usual resonances at the energies $\omega \approx \varepsilon_0$ and $\omega \approx \varepsilon_0 + U$ 
the $\rho_\sigma(\omega)$ calculated using the SBA and NRG exhibit a pronounced and sharp peak at the Fermi energy, 
a clear signature of the Kondo effect. 
Interestingly,  while the slave-boson methods suffer from instabilities for $T \ll T_K$ (see, for instance, the discussion at the end of
Sec.~\ref{sec:NCA_OCA}), this is the regime where NRG works best.

% ------------------------------ FIGURE 4 --------------------------------------
% --------------------------- END FIGURE 4 -------------------------------------

Figure \ref{fig:DOS-lowT} shows the results we obtain for $\Gamma = 0.2 U$ and $T = 0.001U$. 
For the particle-hole symmetric case, $\varepsilon_0 = -U/2$, the Haldane estimate gives $T_{K} \approx 0.009U \gg T$.
This $T_K$ value is consistent with the NRG estimate of $T_{K} \approx 0.00208U$ obtained from the impurity magnetic susceptibility $\chi(T)$ (not shown) using the criteria\cite{Krishna-murthy1980a} $T_K \chi(T_K) = 0.0701$ 
The $T\! \ll \! T_K$ regime is precisely where the NRG is expected to work best, giving an accurate description of the $\omega\!=\! 0 $ peak all the way to $T/T_K \to 0$\cite{Tsvelick1983,Hewson1997,Grewe2008}. As such, we will use the $T\!=\!0$ NRG results as a benchmark for our $T \lesssim T_K$ analysis.

As shown in Fig.~\ref{fig:DOS-lowT}a, the OCA significantly improves the width and the height of the Kondo resonance in comparison with the NCA. In addition, it gives an improvement in the estimate of $T_K$ of at a least an order of magnitude, 
as previously reported \cite{DavidJacob2015,Haule2001,Kotliar2006,Haule2001}. 
The peak at $\omega = 0 $ is absent in the EOM solutions. It is well known  \cite{Kashcheyevs2006} that, in the truncation schemes (and the corresponding neglected correlation processes) discussed in this paper, due to the particle-hole symmetry the EOM Green's function cannot exhibit a resonance at $\omega = 0$ \cite{Kashcheyevs2006}. More recently, it has been shown \cite{VanRoermund2010,Lavagna2015} that one can overcome this shortcoming of the EOM method by considering processes beyond the EOM3 truncation scheme. 

We also compare the different methods in the particle-hole asymmetric case. 
In Fig. \ref{fig:DOS-lowT}b we plot the spectral functions for $\varepsilon_0 = -0.26U \approx -U/4$. 
In accordance with NRG results, both NCA and OCA show a Kondo peak at  $\omega \!\approx\! 0$ in addition to the usual 
single-particle peaks at   $\omega \!=\! \varepsilon_0 + U = 0.74U$ and $\omega \!=\! \varepsilon_0 = - 0.26U$. 
The latter resembles more of a ``shoulder'' due to the broadening and the overlap with the Kondo peak at the Fermi energy.
The EOM spectral functions also exhibit a resonance at the Fermi level. In addition, there is an unphysical peak at 
$\omega = 2 \varepsilon_0 + U = 0.48U $ which is enhanced in the EOM3 solution. This spurious singularity can be identified 
with the singularities of $\Sigma_{1}(\omega)$ in EOM2, Eq.~\eqref{eq:EOM1_self-energies} and with the ones of the 
functions $P(z)$ and $Q(z)$, Eq. \eqref{eq:P(z)_Q(z)_main}, in EOM3.
Spurious peaks in EOM have been already reported in a different context \cite{Czycholl1985}.

Figure \ref{fig:DOS-lowT}c shows the computed density of states for $\varepsilon_0 \!=\! 0$. In this ``mixed-valence'' regime, no clear Kondo effect is expected as the net magnetic moment in the impurity is suppressed. As such, the NRG $T\!=\!0$ spectral density shows only a broad, single-particle-like peak centered at $\omega \!=\! \varepsilon_0\!=\!0$. It is also in this regime where the differences between the methods at the energies $\omega$ very close to the Fermi level are quite remarkable. 

Both the NCA and OCA density of states  show a sharp peak near $\omega \!=\! 0$, while there is no such peak in the EOM3 and NRG results (see inset of Fig.~\ref{fig:DOS-lowT}c). The presence of these peaks in the NCA and OCA results strongly influence the conductance calculations, as discussed in the next section. The EOM2 density of states exhibits a somewhat unexpected result with an anti-peak (deep) at the Fermi level. This rather artificial feature severely compromises the performance of the EOM2 scheme in the conductance calculations, as it will be made more clear in the following section.

The results of this section can be summarized as follows. While NRG is the best method to describe the low-energy excitations of the spectrum at low temperatures, the SBA approaches (OCA and NCA) give a quite reasonable qualitative picture of the physics, with OCA remarkably improving over NCA results. The EOM schemes, by contrast, display very poor performance in capturing the low-energy properties. On the other hand, the most appropriate methods to describe the high-energy features of the spectrum are the SBA, followed by the EOM schemes, while NRG shows a characteristic poor performance in describing the spectral features in this limit.

Let us conclude this section with a discussion of the computational times involved in the spectral function calculations. 
The EOM solvers take few seconds to converge and run very well on just one core (processor Intel 
\textregistered Xeon \textregistered X5650 2.67GHz). 
The simple implementation and the low computational cost explain why the EOM methods, despite their limitations, 
are still widely used. 
The NCA solver also quickly converges and its performance is similar to the EOM one. Using the same processor 
and $N = 1060$ points in the
pseudo-particle frequency mesh NCA converges in $9s$ within 20 iterations. 
In contrast, OCA converges in 76 minutes within 15 iterations, that is, it consumes typically 400 times more CPU than NCA.
The main reason for this are the double convolutions in the OCA equations, which involve overlapping integrations over
functions with sharp peak structures \cite{Grewe2008}. 
The NRG calculations were performed in a cluster with a similar processor unit, namely, 
Intel(R) Xeon(R) CPU E5620 @ 2.40GHz with 8Gb RAM. 
The zero-temperature spectral functions with DM-NRG, which is a bit less precise but numerically much less expensive requires
 $\sim$ 30min per $z$ value $\times$ 5 values to average $\sim$ 2h30 per spectral curve. In turn, the zero-temperature spectral functions with CFS consumed $\sim$ 2h40 per $z$ value $\times$ 5 values to average $\sim$ 13h-14h per spectral curve. Finally, the finite-temperature FDM-NRG cost $\sim$ 5h per $z$ value $\times$ 5 values to average $\sim$ 25h per spectral curve. All CPU times refer to single-core calculations. Table \ref{tab:cpu_times} summarizes our findings. 
 
\begin{table}
\caption{\label{tab:cpu_times}Typical times taken by each method to deliver a converged spectral function. All CPU times refer to single-core calculations.}
\begin{ruledtabular}
\begin{tabular}{cc}
Method & CPU time \\ \hline
EOM & $< 1 $ min  \\ 
NCA & $< 1 $ min \\ 
OCA & $\sim 1$ h \\ 
CFS-NRG & $\sim 13-14$ h\\
FDM-NRG & $\sim 25$ h \\
\end{tabular}
\end{ruledtabular}
\end{table}

%===================================================
\subsection{Conductance}
\label{sec:conductance}

Let us now consider the calculation of two-point electrical conductance in the linear response regime.  Hereafter, 
we assume that the single-particle energy $\varepsilon_0$ can be tuned by an external applied gate voltage, 
$\varepsilon_0 = \varepsilon_0 (V_g)$ and set $\mu = 0$.

The linear conductance $\mathcal{G}$, given by Eq. \eqref{eq:G_linear_response}, is a convolution of the DOS and the derivative of the Fermi distribution. Hence, the features discussed in the previous section give insight on the behavior of $\mathcal{G}$. It is important to stress that the NRG computes the conductance directly from Eq. \eqref{eq:CondNRG}, avoiding the difficulties  of the method in calculating high-energy properties of the spectral function. Thus, the NRG results will serve as a benchmark to the conductance calculations in all regimes studied here.

In quantum dots, the high-temperature limit corresponds to the Coulomb blockade regime. Here, the conductance peaks are close to the resonances at $\varepsilon_0$ and $\varepsilon_0+U$. 
The Coulomb blockade conductance peak at $\varepsilon_0 \approx 0$ corresponds to a dot with either empty or single electron occupation, while the resonance at $\varepsilon_0 + U=0$ is associated with a single or double occupied configuration. The low conductance in the mid-valley region arises mainly due to cotunneling processes \cite{Aleiner2002,FoaTorres2003}, 
that are also nicely described by a real-time diagrammatic technique \cite{Schoeller1994}.  As the temperature is lowered, Kondo physics sets in. Its main manifestation is to increase the mid-valley conductance with decreasing $T/T_K$, reaching the unitary regime as $T/T_K \to 0$.

\begin{figure}[!ht]
\centering
\includegraphics[width=1.0\columnwidth]{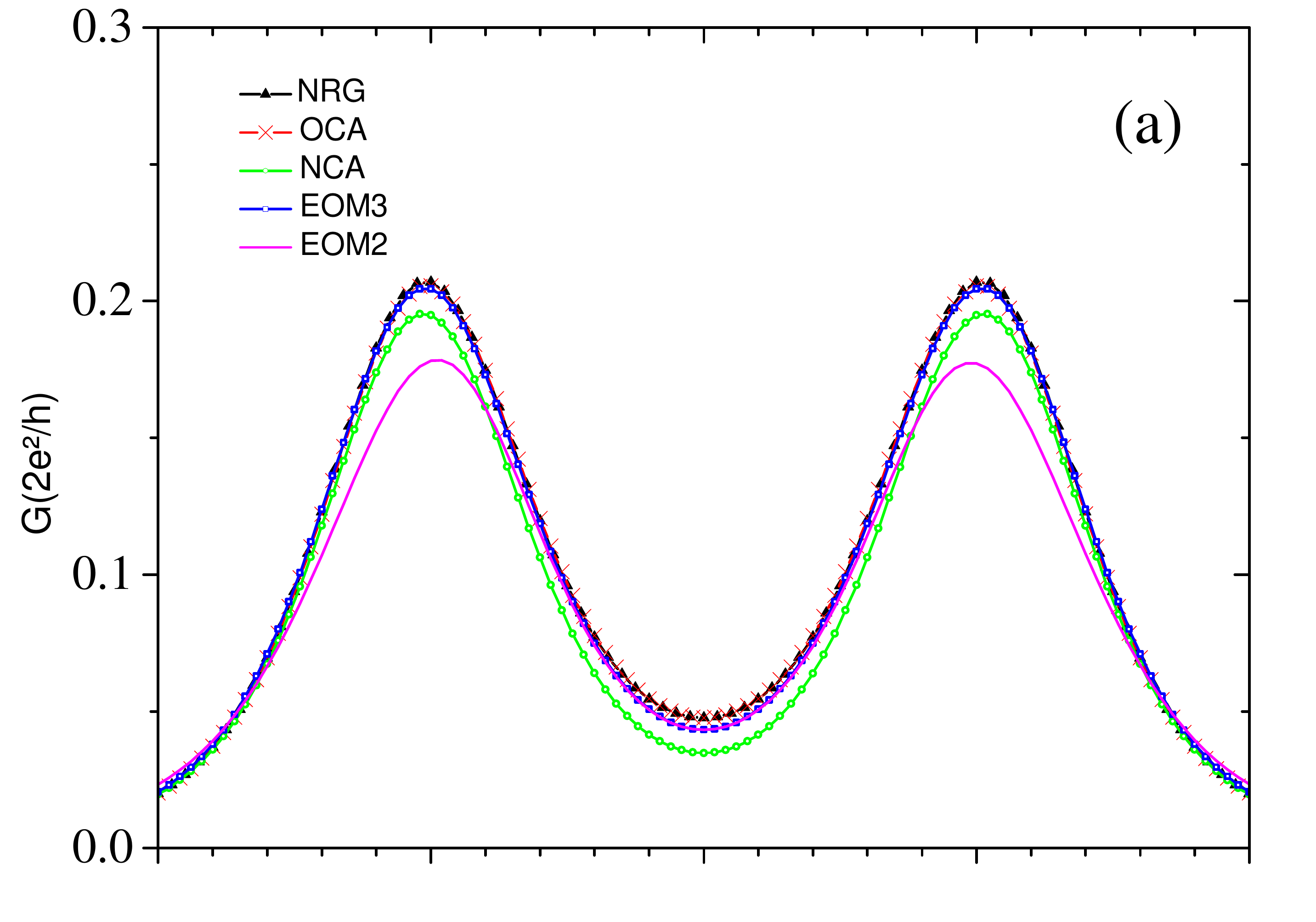}
\includegraphics[width=1.0\columnwidth]{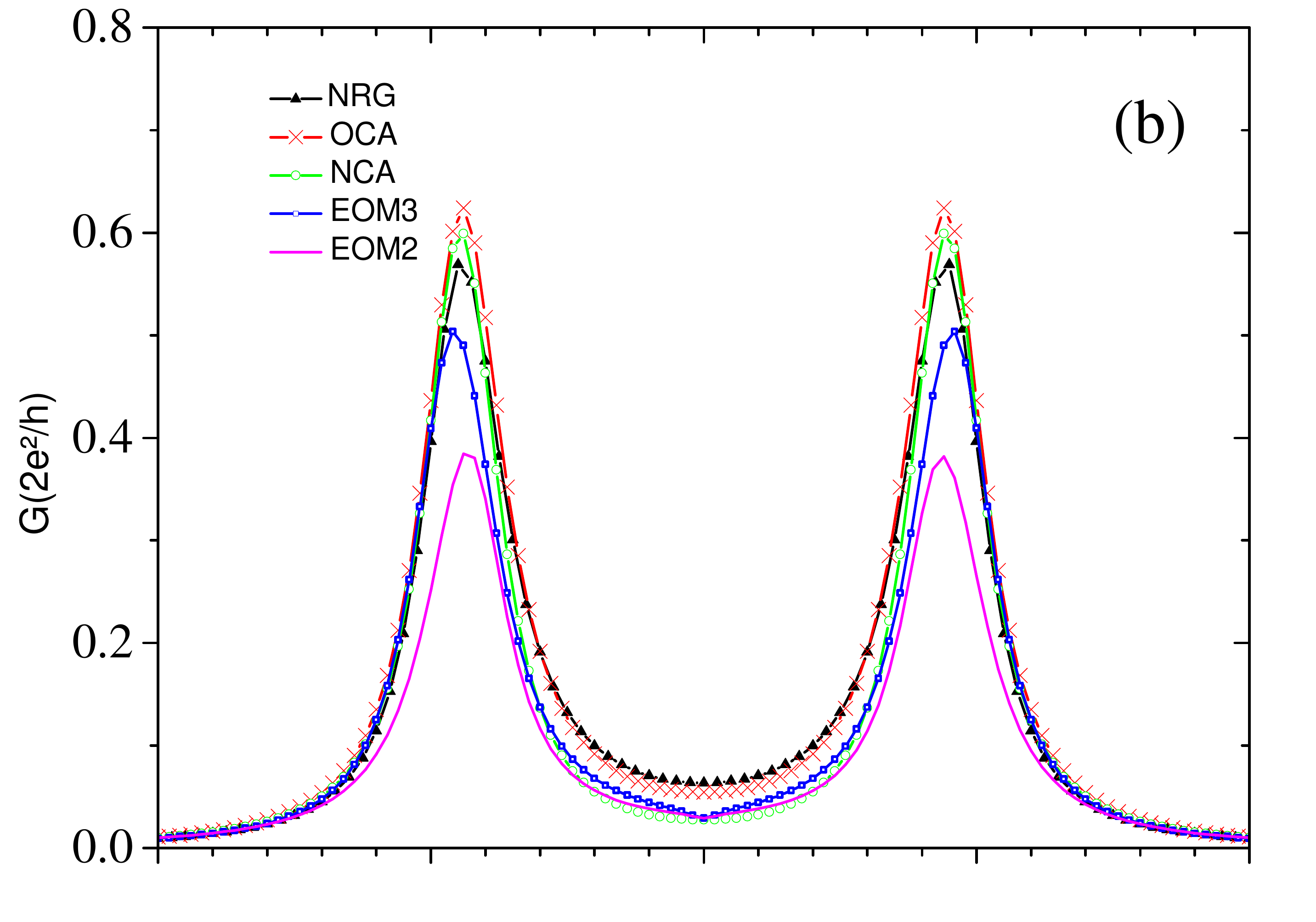}
\includegraphics[width=1.0\columnwidth]{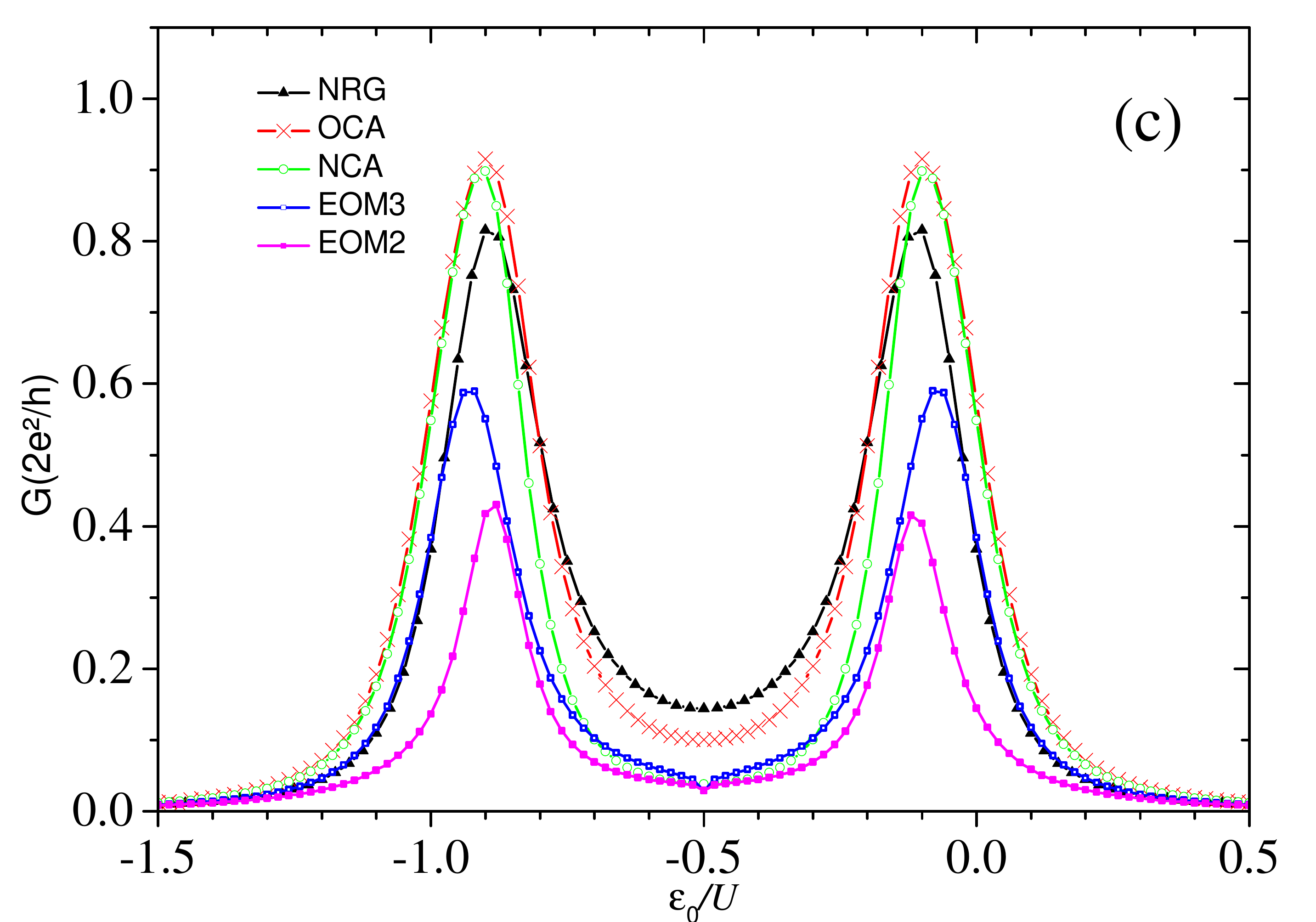}
\caption{Conductance $\mathcal{G}$  in units of $2e^2/h$ as a function of the single particle energy $\varepsilon_0$ 
in units of $U$ for $\mu =0$, $\Gamma = 0.1U$, and different temperature values (a) $T=0.1U$, (b) $T=0.01U$, 
and (c) $T=0.001U$. }
\label{fig:conductancexTxe0_Gamma1}
\end{figure}

Let us begin analyzing the high-temperature limit. All panels of Fig. \ref{fig:conductancexTxe0_Gamma1} exhibit the main features of the Coulomb blockade regime: Two peaks separated by an energy $\sim U$ with low mid valley conductance. 
Figure~\ref{fig:conductancexTxe0_Gamma1}a shows the conductance obtained by the methods presented in Sec.~\ref{sec:model} 
for $T=0.1U$. 
All approximations show a good qualitative agreement, except for the EOM2 result, which presents pronounced discrepancies, 
particularly near the charge fluctuation regime. This rather unexpected behavior of the EOM2 result is a consequence of the 
deviation in the density of states compared to other methods, as discussed in the previous section.

% ------------------------------ FIGURE 5 --------------------------------------
% --------------------------- END FIGURE 5 -------------------------------------

As we decrease the temperature towards the Kondo regime, the discrepancies between the results delivered by 
the different methods become more pronounced. The conductance peak heights given by 
the EOM are consistently smaller than those obtained by NRG, while SBA tends to overestimate them. 

As expected, EOM fails to describe the rise of the mid-valley conductance with decreasing $T/T_K$, while within 
the SBA methods, only OCA gives good results compared with NRG. We stress that the EOM3 gives better results 
than the SBA for the empty  ($\varepsilon_0 > 0$) and double occupation ($\varepsilon_0 < - U$) 
quantum dot configurations.

\begin{figure}[!ht]
\centering
%\vspace{-10pt}
\includegraphics[width=1.0\columnwidth]{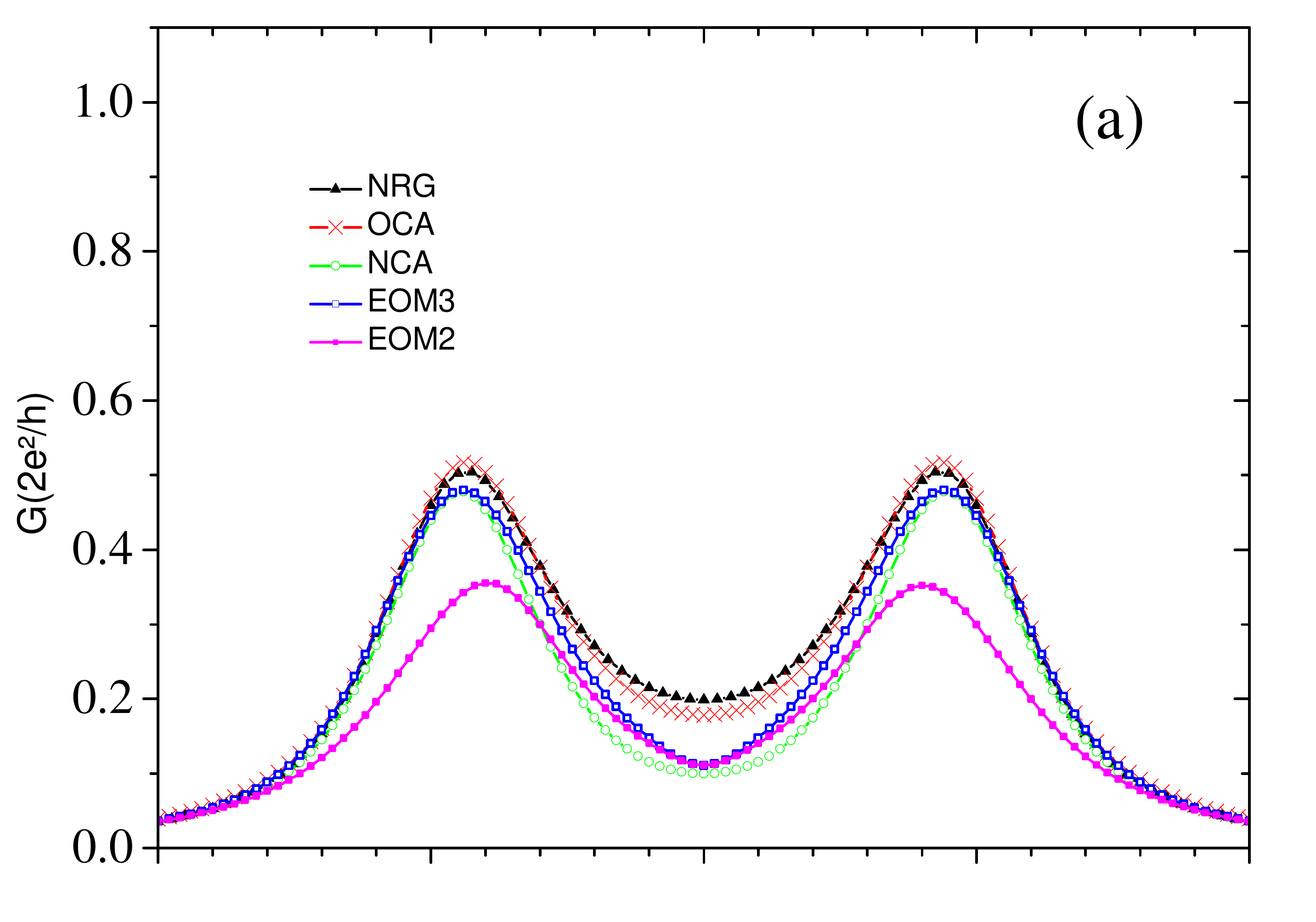}
\includegraphics[width=1.0\columnwidth]{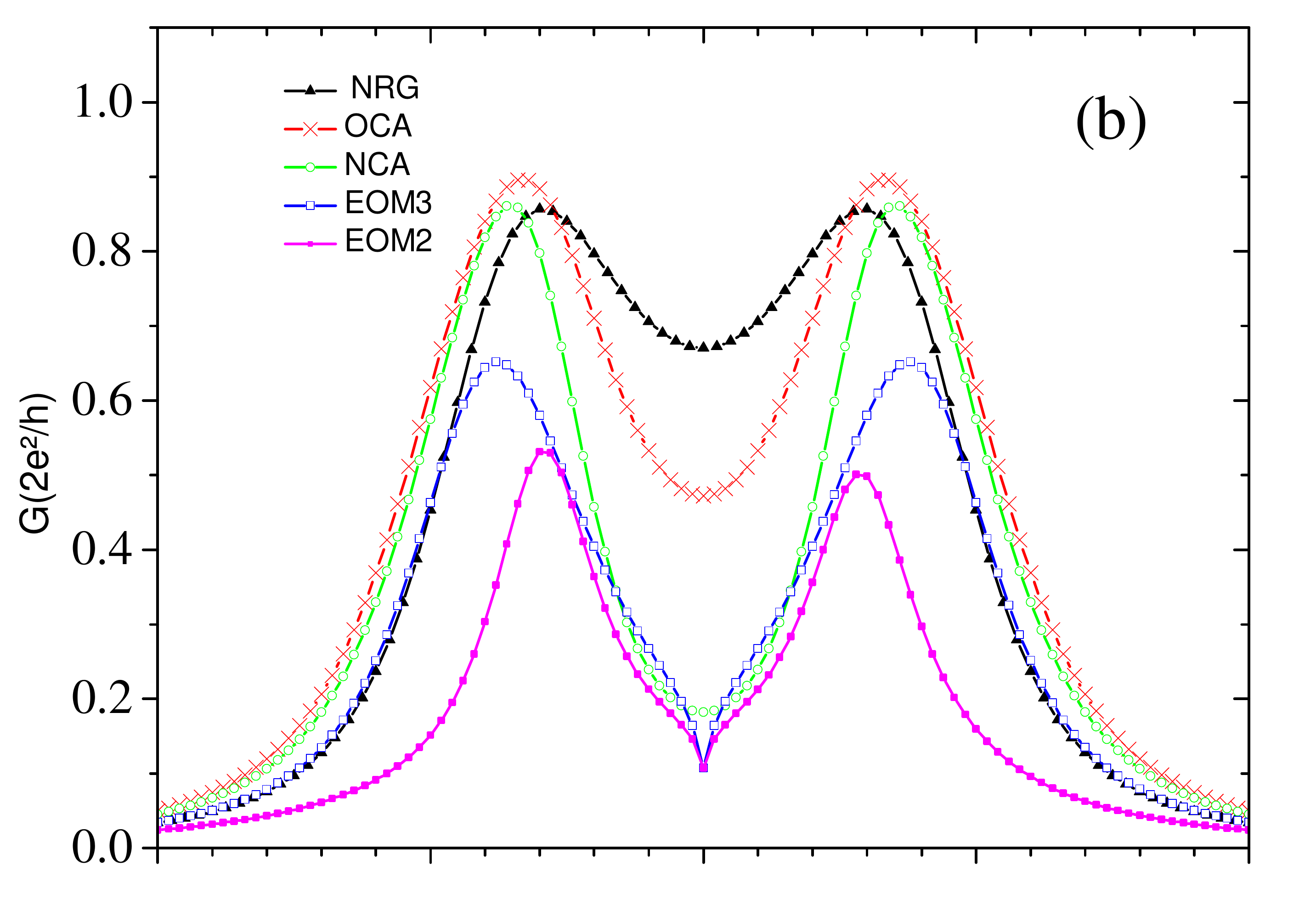}
\includegraphics[width=1.0\columnwidth]{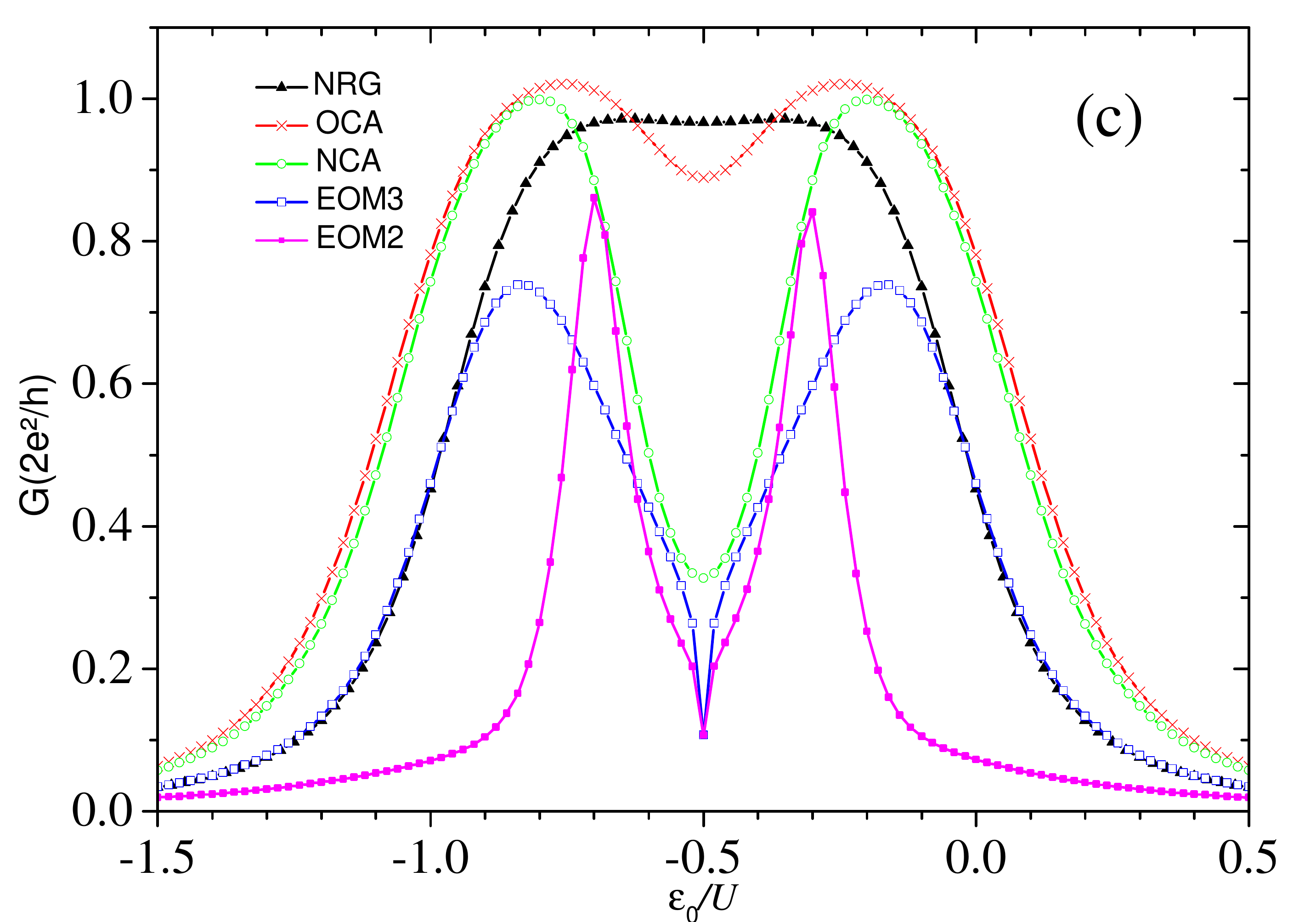}
\vspace{-10pt}
\caption{Conductance ${\cal G}$ in units of $2e^2/h$ as a function of the impurity  energy level $\varepsilon_0/U$ 
for $\mu = 0$, $\Gamma=0.2U$, and 
(a) $T=0.04U$, (b) $T=0.003U$, and  (c) $T=0.0004U$.  
}
\label{fig:conductancexTxe0}
\vspace{-10pt}
\end{figure}

In order to address the Kondo regime, $ T\lesssim T_K$, we increase the coupling strength to $\Gamma=0.2U$ 
such that $T_K$ increases above $\sim 0.002U$ for all $\varepsilon_0$ values in the range of interest. This speeds 
the convergence of the NCA and OCA algorithms, while avoiding the low-$T$ instabilities of these methods, see discussion
at the end of Sec.~\ref{sec:NCA_OCA}.  The results are presented in Fig.\ \ref{fig:conductancexTxe0} . 
In this temperature regime, the limitations of the EOM method in treating correlations become even more manifest, 
particularly at $\varepsilon_0 = -U/2$, which corresponds to the particle-hole symmetry point. 
We emphasize that, as in the previous case, EOM3 is very accurate in describing the  $\varepsilon_0 > 0$ 
and $\varepsilon_0 < - U$ regions.

% ------------------------------ FIGURE 6 --------------------------------------
% --------------------------- END FIGURE 6 -------------------------------------

Taking the NRG conductance as benchmark, we find that the OCA represents a significant improvement over the NCA results.  
This is a direct consequence of the better description of the Kondo peak by the OCA, as discussed in Sec.~\ref{sec:DOS}: 
The Kondo peak in the OCA spectral function is higher and wider than in the NCA one. However, a closer comparison with 
the NRG result shows that the OCA conductance fails to develop a Kondo plateau, as can be seen in 
Fig. \ref{fig:conductancexTxe0}c. 

When the temperature is further decreased, the OCA conductance exhibits a plateau, but it exceeds the maximum conductance, 
${\cal G}_{\rm max} = 2e^2/h$ (see also Ref.~\onlinecite{Melo2019}). This undesired unphysical feature is related to the violation 
of the Friedel sum rule by the NCA/OCA procedure discussed in Sec.~\ref{sec:NCA_OCA}, which results in the overestimation of the height of the Kondo peak \cite{Tosi2011}. 
%\Luis{Ok, I think that does it. A comprehensive analysis of how to enforce the Friedel sum rule is beyond the scope here...}

The NCA and OCA results are also at odds with NRG for $\varepsilon_0 < - U$ and $\varepsilon_0 > 0$, showing an increase 
of ${\cal G}$ as the temperature is decreased. This behavior can be understood by looking at the spectral functions in these 
parameter regions. In the discussion of the results of Fig. \ref{fig:DOS-lowT}c, we noted that the density of states obtained by the NCA approach exhibits a sharp peak at the Fermi level and this peak increases when the temperature decreases.  
As shown in Fig.~\ref{fig:DOS-lowT}c, the $T\!=\!0$ there is no such peak in the NRG data. 
We conclude that the observed increase in the conductance shown in the NCA and OCA data are related to 
this spurious resonance. Intriguingly, we have not been able to determine the numerical origin of this  behavior in the 
NCA/OCA density of states' close to the charge fluctuation points.

Let us now investigate the temperature dependence of the conductance with more detail. For this purpose we select three different configurations, namely, $\varepsilon_0 + U/2=0$, $\varepsilon_0 + U/4=0$, and $\varepsilon_0=0$. The latter cases correspond to the mid-valley, a crossover, and the resonance peak configurations, respectively. 

\begin{figure}[!ht]
\centering
\includegraphics[width=0.9\columnwidth]{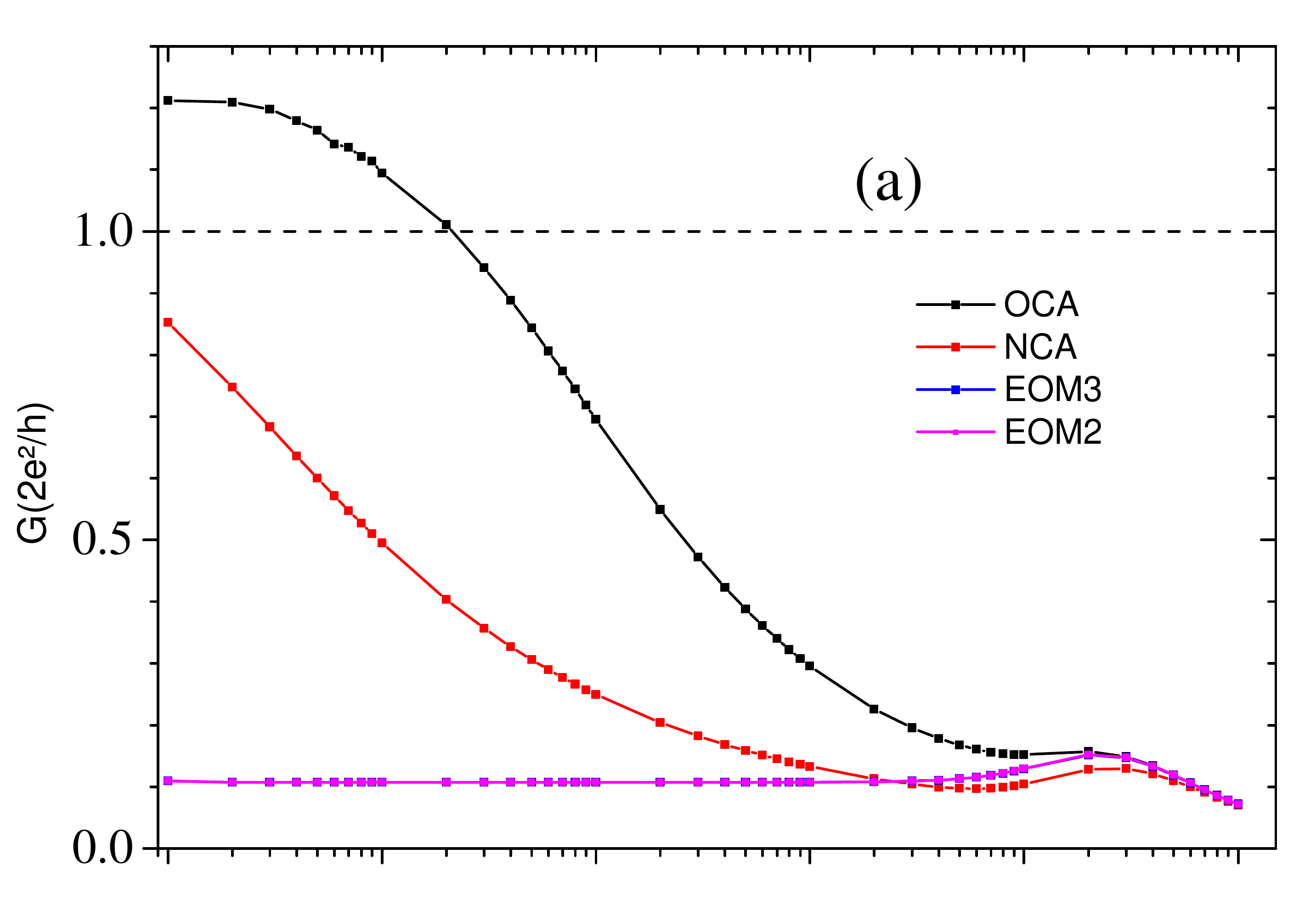}
\includegraphics[width=0.9\columnwidth]{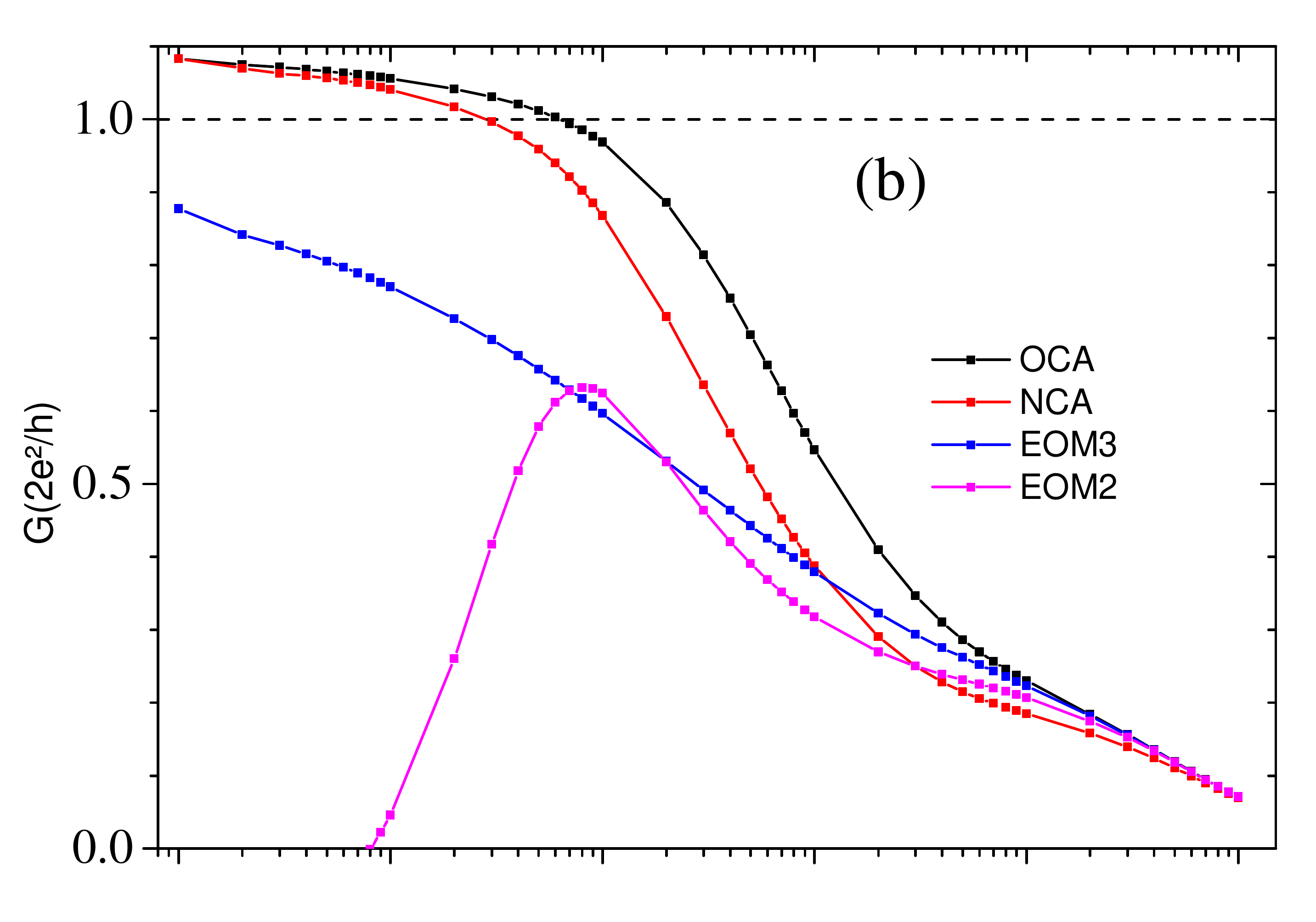}
\includegraphics[width=0.9\columnwidth]{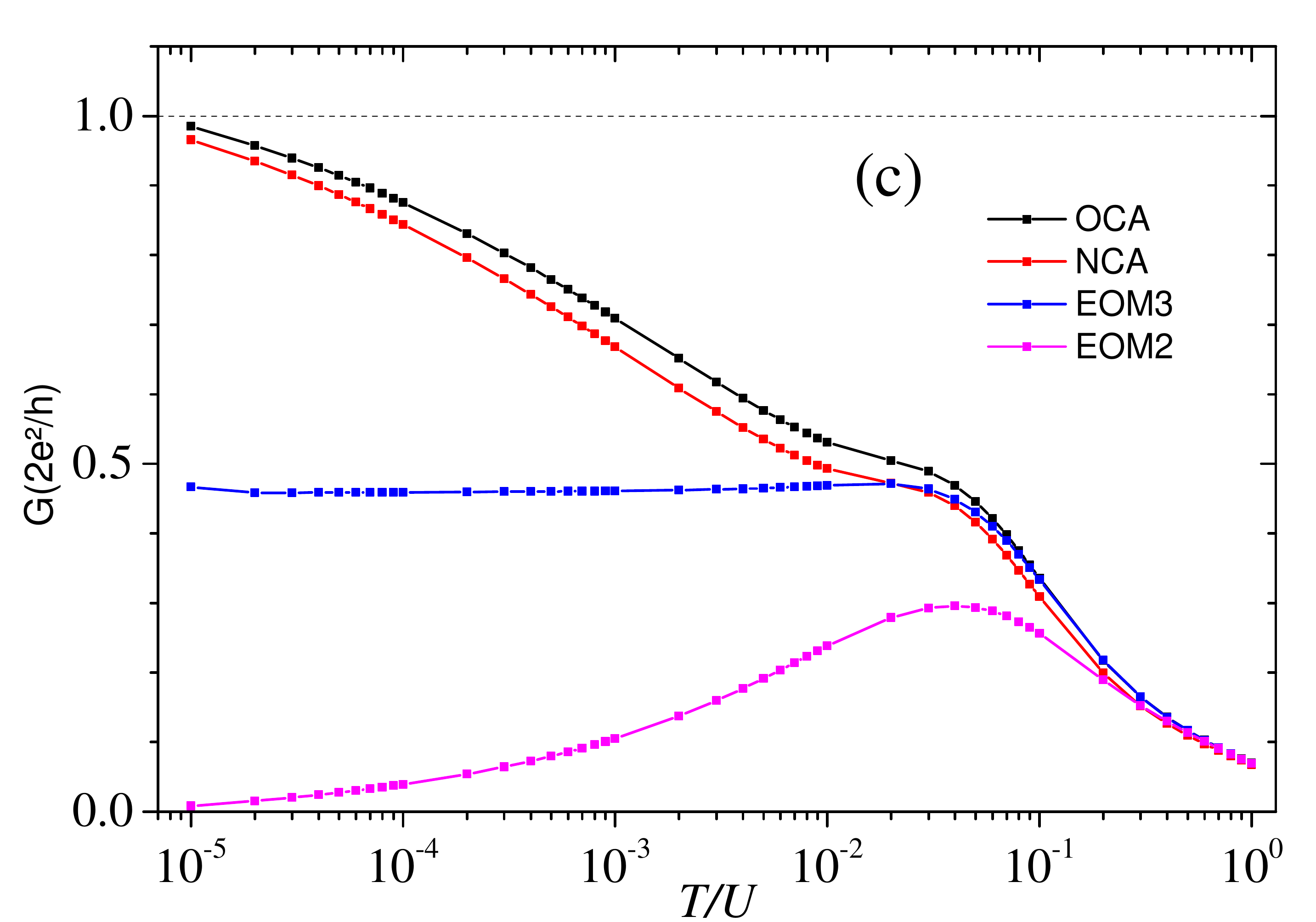}
%\vspace{-10pt}
\caption{Conductance $\mathcal{G}$ in units of $2e^2/h$ as a function of the temperature $T $ for 
(a) $\varepsilon_0 = -U/2$, (b) $\varepsilon_0 = -U/4$, (c) $\varepsilon_0 = 0$ corresponding respectively
 to the mid valley, a crossover, and the resonance peak configuration. The dashed lines indicate the unitary limit. 
 Here $\Gamma=0.2U$.}
\label{fig:conductancexT_e0=-U/2-e0=-U/4-e0=0}
\vspace{0pt}
\end{figure}

Figure \ref{fig:conductancexT_e0=-U/2-e0=-U/4-e0=0} shows that in the high temperature limit all methods agree very well, as it has been already suggested by Fig. \ref{fig:conductancexTxe0_Gamma1}a. In the present analysis, we are not including the NRG results, which we will discuss later when addressing the scaling behavior of the conductance.  

As the temperature decreases, the differences are enhanced. As expected, in the particle-hole symmetric point (Fig. \ref{fig:conductancexT_e0=-U/2-e0=-U/4-e0=0}a), both EOM schemes dramatically fail to capture the increase of the conductance due to the missing Kondo peak in the spectral function. We notice that  these methods render nearly temperature-independent spectral functions. The OCA method gives significantly better results than NCA, capturing the onset of conductance increase with good accuracy. When the temperature is decreased below about $T_K/10$, the OCA conductance saturates, although at an nonphysical value greater than the quantum of conductance $G_{\text{plateau}} > 2e^2/h$. This feature is related to the violation of the Friedel sum rule  in the OCA procedure mentioned above. \cite{Tosi2011,Melo2019}.

% ------------------------------ FIGURE 6 --------------------------------------
% --------------------------- END FIGURE 6 -------------------------------------

For $\varepsilon_0 = - U/4$  (Fig. \ref{fig:conductancexT_e0=-U/2-e0=-U/4-e0=0}b), the EOM2 conductance starts to decrease below a certain temperature. The EOM3 method seems to correct this unexpected behavior to a certain extent, but its already mentioned nonphysical features prevent the method to yield conductance values close to the OCA/NCA ones. The difference between the NCA and OCA results becomes smaller, the NCA curve  being always below the OCA one which is consistent with the idea that NCA yields a smaller Kondo temperature. For $\varepsilon_0 = 0$, Fig.~\ref{fig:conductancexT_e0=-U/2-e0=-U/4-e0=0}c, EOM2 performs well only in the high temperature limit $10^{-1} \lesssim T/U < 1$, while EOM3 gives good results for $T/U \gtrsim 10^{-2}$. Remarkably, the NCA conductance shows results {very similar to those of} the OCA curve.

The analogy between quantum dots and impurity systems extends to the scaling behavior of their properties. Indeed, in Ref. \onlinecite{Kouwenhoven2001} it has been experimentally verified that the conductance of a quantum dot,  in the low temperature limit, depends only on the ratio $T/T_K$.

According to a semi-empirical result obtained by NRG calculations for a spin-$1/2$ system, the conductance of a quantum dot is given by \cite{Costi1994,Goldhaber-Gordon1998PRL,Roura-Bas2010,Tosi2011}
\begin{equation}
\mathcal{G}(T)= \frac{2e^2}{h}{\frac{1}{ [ 1 + (2^{1/s}-1)(T/T_K)^2 ] ^s}},
\end{equation}
with $s=0.22$ chosen to fit the experimental data\cite{Goldhaber-Gordon1998PRL}.

Let us discuss the scaling behavior of $\mathcal G$ obtained by the OCA and NCA methods in comparison with the 
NRG formula above. For that purpose, it is important to note that the Kondo temperature given by the OCA method is larger than the one given by NCA (see discussion in Sec. \ref{sec:DOS}). 

In order to calculate the Kondo temperature we adopt the procedure put forward in Ref. \onlinecite{Tosi2011} which defines $T_K$ as the temperature for which $\mathcal{G}/(2e^2/h)=1/2$. In this way, for $\Gamma = 0.2U$ and $\varepsilon_0 = -U/2$, we obtain $T_K^{\text{OCA}} \approx 0.0028U$ and $T_K^{\text{NCA}} \approx 0.000097U$. (It is interesting to recall that we obtain 
$T_K^{\text{NRG}} \approx 0.0021U$ from the NRG susceptibility analysis.)
We scale the conductance curves for each method by the corresponding Kondo temperature. 

\begin{figure}[!h]
\centering
%\vspace{-5pt}
\includegraphics[width=\columnwidth]{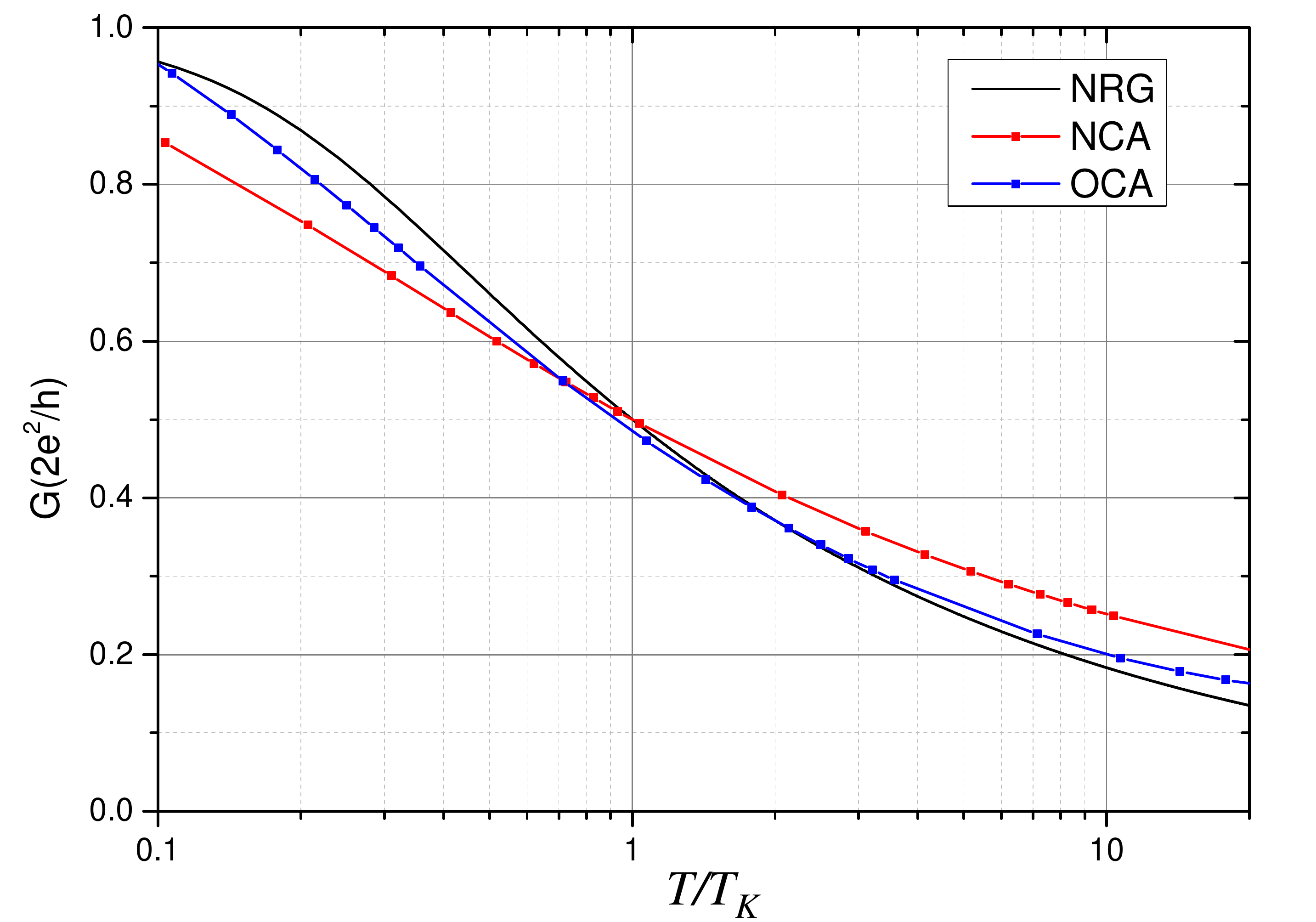}
\label{fig:conductancexT_differents_e0_a}
\vspace{-10pt}
\caption{Conductance $\mathcal{G}$ in units of $2e^2/h$ as a function of the scaled temperature $T/T_K$ 
for $\varepsilon_0 = -U/2$. Here $\Gamma=0.2U$ and  $T_K^{\text{OCA}}\approx 0.0028U$ and 
$T_K^{\text{NCA}}\approx 0.000097U$. The Haldane estimate gives $T_K = 0.009U$.}
\label{fig:conductancexT_scaling}
\vspace{0pt}
\end{figure}

 In Fig. \ref{fig:conductancexT_scaling} we show the conductance as function of $T/T_K$ obtained in NRG, OCA, and NCA methods. Since the OCA method gives unphysical conductance values when $T \lesssim 0.1 T_K$, our analysis of the scaling starts from this temperature \cite{Tosi2011}. Clearly, the scaling obtained using OCA is better than the one yielded by the NCA solver in the sense that the former is closer to the NRG result than the latter. However, as previously reported in Ref. \onlinecite{Tosi2011}, the inclusion of vertex corrections is not enough to recover the NRG prediction.

% ------------------------------ FIGURE 8 --------------------------------------
% --------------------------- END FIGURE 8 -------------------------------------

Let us finish discussing CPU times. For the EOM and SBA methods the CPU times involved in the 
conductance calculations for a given value of $\varepsilon_0/U$ are essentially the same as the 
spectral function ones, discussed in the previous section. Since ${\cal G}$ is calculated by NRG 
using Eq.~\eqref{eq:CondNRG}, its computation is much faster than the one for DOS. The approximate 
clock times are about 20 minutes per calculation using Intel(R) Xeon(R) CPU E5620 @ 2.40GHz 
(one core per job) with 8Gb RAM. 

%%%%%%%%%%%%%%%%%%%%%%%%%%%%%%%%%%%%%%%%%%%
\section{Summary and conclusions}
\label{sec:conclusion}
%%%%%%%%%%%%%%%%%%%%%%%%%%%%%%%%%%%%%%%%%%%

In this work we study the EOM, SBA, and NRG impurity solvers, comparing their results for the spectral function and linear conductance for the SIAM covering all physical relevant regimes.

In summary, as stated in Ref. \onlinecite{Georges1996}, there is no ``universal" impurity solver that can be considered 
the most appropriate choice to describe both the spectral function and transport properties over the all situations of interest. 
We provide a thorough analysis that serves as a guide to choose the most adequate approximation depending on the 
accuracy and computational time for a given regime and/or observable. 

We find that for the high temperature regime most approximations give comparable results for $\rho_{\sigma}(\omega)$ 
and $\mathcal{G}$. Surprisingly the widely used EOM0, EOM1, and EOM2 approaches give poor results for the spectral 
functions at the mixed valence regime, a drawback that is fixed to a large extent by EOM3. As already known, NRG fails 
to give reliable high-energy spectral functions, which is a problem for calculation schemes that combine NRG with DMFT.

At intermediate temperatures, corresponding to the onset of Kondo physics, only SBA and NRG lead to quantitative accurate 
results. Even so, NCA dramatically underestimates $T_K$ and, hence, gives poor results for the mid-valley conductance. 
It is worth to mention that EOM3 gives surprisingly good results at the charge fluctuation regime.

For low temperatures, $T \ll T_K$, the SBA violates the Friedel sum rule and gives ${\cal G}$ values larger than the maximum 
predicted by the unitary limit. Hence, as one approaches $T/T_K \to 0$ the SBA results can only capture the qualitative behavior 
of the physical observables of interest.

In practice, the choice of a impurity solver certainly must take into account the computational cost. 
Table \ref{tab:cpu_times} compares the processing times of the methods we analyze here. 

%%%%%%%%%%%%%%%%%%%%%%%%%%%%%%%%%%%%%%%%%%%
\acknowledgements
We thank Rok Zitko for valuable comments and suggestions in the implementation of the FDM-NRG approach. 
This work was supported by the Brazilian funding agencies 
CNPq (Grant Nos. 308801/2015-6, 308351/2017-7, 423137/2018-2),
FAPERJ (Grant E-26/202.882/2018), 
and FAPESP (Grant Nos. 2016/18495-4, 2016/01343-7, 2017/02317-2).

%%%%%%%%%%%%%%%%%%%%%%%%%%%%%%%%%%%%%%%%%%%

%\bibliographystyle{unsrt}
%\bibliography{tese,GNR}
%\bibliographystyle{prsty}
\bibliography{impurity-solvers}

%Control: production of eprint (-1) disabled
\begin{thebibliography}{84}%
\makeatletter
\providecommand \@ifxundefined [1]{%
 \@ifx{#1\undefined}
}%
\providecommand \@ifnum [1]{%
 \ifnum #1\expandafter \@firstoftwo
 \else \expandafter \@secondoftwo
 \fi
}%
\providecommand \@ifx [1]{%
 \ifx #1\expandafter \@firstoftwo
 \else \expandafter \@secondoftwo
 \fi
}%
\providecommand \natexlab [1]{#1}%
\providecommand \enquote  [1]{``#1''}%
\providecommand \bibnamefont  [1]{#1}%
\providecommand \bibfnamefont [1]{#1}%
\providecommand \citenamefont [1]{#1}%
\providecommand \href@noop [0]{\@secondoftwo}%
\providecommand \href [0]{\begingroup \@sanitize@url \@href}%
\providecommand \@href[1]{\@@startlink{#1}\@@href}%
\providecommand \@@href[1]{\endgroup#1\@@endlink}%
\providecommand \@sanitize@url [0]{\catcode `\\12\catcode `\$12\catcode
  `\&12\catcode `\#12\catcode `\^12\catcode `\_12\catcode `\%12\relax}%
\providecommand \@@startlink[1]{}%
\providecommand \@@endlink[0]{}%
\providecommand \url  [0]{\begingroup\@sanitize@url \@url }%
\providecommand \@url [1]{\endgroup\@href {#1}{\urlprefix }}%
\providecommand \urlprefix  [0]{URL }%
\providecommand \Eprint [0]{\href }%
\providecommand \doibase [0]{http://dx.doi.org/}%
\providecommand \selectlanguage [0]{\@gobble}%
\providecommand \bibinfo  [0]{\@secondoftwo}%
\providecommand \bibfield  [0]{\@secondoftwo}%
\providecommand \translation [1]{[#1]}%
\providecommand \BibitemOpen [0]{}%
\providecommand \bibitemStop [0]{}%
\providecommand \bibitemNoStop [0]{.\EOS\space}%
\providecommand \EOS [0]{\spacefactor3000\relax}%
\providecommand \BibitemShut  [1]{\csname bibitem#1\endcsname}%
\let\auto@bib@innerbib\@empty
%</preamble>
\bibitem [{\citenamefont {Tsvelick}\ and\ \citenamefont
  {Wiegmann}(1983)}]{Tsvelick1983}%
  \BibitemOpen
  \bibfield  {author} {\bibinfo {author} {\bibfnamefont {A.~M.}\ \bibnamefont
  {Tsvelick}}\ and\ \bibinfo {author} {\bibfnamefont {P.~B.}\ \bibnamefont
  {Wiegmann}},\ }\href {\doibase 10.1080/00018738300101581} {\bibfield
  {journal} {\bibinfo  {journal} {Adv. Phys.}\ }\textbf {\bibinfo {volume}
  {32}},\ \bibinfo {pages} {453} (\bibinfo {year} {1983})}\BibitemShut
  {NoStop}%
\bibitem [{\citenamefont {Hewson}(1997)}]{Hewson1997}%
  \BibitemOpen
  \bibfield  {author} {\bibinfo {author} {\bibfnamefont {A.~C.}\ \bibnamefont
  {Hewson}},\ }\href@noop {} {\emph {\bibinfo {title} {{The Kondo Problem to
  Heavy Fermions}}}},\ \bibinfo {edition} {2nd}\ ed.\ (\bibinfo  {publisher}
  {Cambridge University Press},\ \bibinfo {year} {1997})\BibitemShut {NoStop}%
\bibitem [{\citenamefont {Georges}\ \emph {et~al.}(1996)\citenamefont
  {Georges}, \citenamefont {Kotliar}, \citenamefont {Krauth},\ and\
  \citenamefont {Rozenberg}}]{Georges1996}%
  \BibitemOpen
  \bibfield  {author} {\bibinfo {author} {\bibfnamefont {A.}~\bibnamefont
  {Georges}}, \bibinfo {author} {\bibfnamefont {G.}~\bibnamefont {Kotliar}},
  \bibinfo {author} {\bibfnamefont {W.}~\bibnamefont {Krauth}}, \ and\ \bibinfo
  {author} {\bibfnamefont {M.}~\bibnamefont {Rozenberg}},\ }\href {\doibase
  http://dx.doi.org/10.1103/RevModPhys.68.13} {\bibfield  {journal} {\bibinfo
  {journal} {Rev. Mod. Phys.}\ }\textbf {\bibinfo {volume} {68}},\ \bibinfo
  {pages} {13} (\bibinfo {year} {1996})}\BibitemShut {NoStop}%
\bibitem [{\citenamefont {Anderson}(1961)}]{Anderson1961}%
  \BibitemOpen
  \bibfield  {author} {\bibinfo {author} {\bibfnamefont {P.~W.}\ \bibnamefont
  {Anderson}},\ }\href {\doibase 10.1103/PhysRev.124.41} {\bibfield  {journal}
  {\bibinfo  {journal} {Phys. Rev.}\ }\textbf {\bibinfo {volume} {124}},\
  \bibinfo {pages} {41} (\bibinfo {year} {1961})}\BibitemShut {NoStop}%
\bibitem [{\citenamefont {Krishna-murthy}\ \emph {et~al.}(1975)\citenamefont
  {Krishna-murthy}, \citenamefont {Wilson},\ and\ \citenamefont
  {Wilkins}}]{Krishna-Murthy1975}%
  \BibitemOpen
  \bibfield  {author} {\bibinfo {author} {\bibfnamefont {H.~R.}\ \bibnamefont
  {Krishna-murthy}}, \bibinfo {author} {\bibfnamefont {K.~G.}\ \bibnamefont
  {Wilson}}, \ and\ \bibinfo {author} {\bibfnamefont {J.~W.}\ \bibnamefont
  {Wilkins}},\ }\href {\doibase 10.1103/PhysRevLett.35.1101} {\bibfield
  {journal} {\bibinfo  {journal} {Phys. Rev. Lett.}\ }\textbf {\bibinfo
  {volume} {35}},\ \bibinfo {pages} {1101} (\bibinfo {year}
  {1975})}\BibitemShut {NoStop}%
\bibitem [{\citenamefont {Krishna-murthy}\ \emph
  {et~al.}(1980{\natexlab{a}})\citenamefont {Krishna-murthy}, \citenamefont
  {Wilkins},\ and\ \citenamefont {Wilson}}]{Krishna-murthy1980a}%
  \BibitemOpen
  \bibfield  {author} {\bibinfo {author} {\bibfnamefont {H.~R.}\ \bibnamefont
  {Krishna-murthy}}, \bibinfo {author} {\bibfnamefont {J.~W.}\ \bibnamefont
  {Wilkins}}, \ and\ \bibinfo {author} {\bibfnamefont {K.~G.}\ \bibnamefont
  {Wilson}},\ }\href {\doibase 10.1103/PhysRevB.21.1003} {\bibfield  {journal}
  {\bibinfo  {journal} {Phys. Rev. B}\ }\textbf {\bibinfo {volume} {21}},\
  \bibinfo {pages} {1003} (\bibinfo {year} {1980}{\natexlab{a}})}\BibitemShut
  {NoStop}%
\bibitem [{\citenamefont {Krishna-murthy}\ \emph
  {et~al.}(1980{\natexlab{b}})\citenamefont {Krishna-murthy}, \citenamefont
  {Wilkins},\ and\ \citenamefont {Wilson}}]{Krishna-murthy1980b}%
  \BibitemOpen
  \bibfield  {author} {\bibinfo {author} {\bibfnamefont {H.~R.}\ \bibnamefont
  {Krishna-murthy}}, \bibinfo {author} {\bibfnamefont {J.~W.}\ \bibnamefont
  {Wilkins}}, \ and\ \bibinfo {author} {\bibfnamefont {K.~G.}\ \bibnamefont
  {Wilson}},\ }\href {\doibase 10.1103/PhysRevB.21.1044} {\bibfield  {journal}
  {\bibinfo  {journal} {Phys. Rev. B}\ }\textbf {\bibinfo {volume} {21}},\
  \bibinfo {pages} {1044} (\bibinfo {year} {1980}{\natexlab{b}})}\BibitemShut
  {NoStop}%
\bibitem [{\citenamefont {Newns}\ and\ \citenamefont {Read}(1987)}]{Newns1987}%
  \BibitemOpen
  \bibfield  {author} {\bibinfo {author} {\bibfnamefont {D.~M.}\ \bibnamefont
  {Newns}}\ and\ \bibinfo {author} {\bibfnamefont {N.}~\bibnamefont {Read}},\
  }\href {\doibase 10.1080/00018738700101082} {\bibfield  {journal} {\bibinfo
  {journal} {Advances in Physics}\ }\textbf {\bibinfo {volume} {36}},\ \bibinfo
  {pages} {799} (\bibinfo {year} {1987})}\BibitemShut {NoStop}%
\bibitem [{\citenamefont {Coleman}(1984)}]{Coleman1984}%
  \BibitemOpen
  \bibfield  {author} {\bibinfo {author} {\bibfnamefont {P.}~\bibnamefont
  {Coleman}},\ }\href {\doibase 10.1103/PhysRevB.29.3035} {\bibfield  {journal}
  {\bibinfo  {journal} {Phys. Rev. B}\ }\textbf {\bibinfo {volume} {29}},\
  \bibinfo {pages} {3035} (\bibinfo {year} {1984})}\BibitemShut {NoStop}%
\bibitem [{\citenamefont {Meir}\ \emph {et~al.}(1991)\citenamefont {Meir},
  \citenamefont {Wingreen},\ and\ \citenamefont {Lee}}]{Meir1991}%
  \BibitemOpen
  \bibfield  {author} {\bibinfo {author} {\bibfnamefont {Y.}~\bibnamefont
  {Meir}}, \bibinfo {author} {\bibfnamefont {N.~S.}\ \bibnamefont {Wingreen}},
  \ and\ \bibinfo {author} {\bibfnamefont {P.~A.}\ \bibnamefont {Lee}},\ }\href
  {\doibase 10.1103/PhysRevLett.66.3048} {\bibfield  {journal} {\bibinfo
  {journal} {Phys. Rev. Lett.}\ }\textbf {\bibinfo {volume} {66}},\ \bibinfo
  {pages} {3048} (\bibinfo {year} {1991})}\BibitemShut {NoStop}%
\bibitem [{\citenamefont {Pustilnik}\ and\ \citenamefont
  {Glazman}(2004)}]{Pustilnik2004}%
  \BibitemOpen
  \bibfield  {author} {\bibinfo {author} {\bibfnamefont {M.}~\bibnamefont
  {Pustilnik}}\ and\ \bibinfo {author} {\bibfnamefont {L.}~\bibnamefont
  {Glazman}},\ }\href {\doibase 10.1088/0953-8984/16/16/R01} {\bibfield
  {journal} {\bibinfo  {journal} {J. Phys. Condens. Matter}\ }\textbf {\bibinfo
  {volume} {16}},\ \bibinfo {pages} {R513} (\bibinfo {year}
  {2004})}\BibitemShut {NoStop}%
\bibitem [{\citenamefont {Meir}\ \emph {et~al.}(1993)\citenamefont {Meir},
  \citenamefont {Wingreen},\ and\ \citenamefont {Lee}}]{Meir1992a}%
  \BibitemOpen
  \bibfield  {author} {\bibinfo {author} {\bibfnamefont {Y.}~\bibnamefont
  {Meir}}, \bibinfo {author} {\bibfnamefont {N.~S.}\ \bibnamefont {Wingreen}},
  \ and\ \bibinfo {author} {\bibfnamefont {P.~a.}\ \bibnamefont {Lee}},\ }\href
  {\doibase 10.1103/PhysRevLett.70.2601} {\bibfield  {journal} {\bibinfo
  {journal} {Phys. Rev. Lett.}\ }\textbf {\bibinfo {volume} {70}},\ \bibinfo
  {pages} {2601} (\bibinfo {year} {1993})}\BibitemShut {NoStop}%
\bibitem [{\citenamefont {Ng}\ and\ \citenamefont {Lee}(1988)}]{Ng1988}%
  \BibitemOpen
  \bibfield  {author} {\bibinfo {author} {\bibfnamefont {T.~K.}\ \bibnamefont
  {Ng}}\ and\ \bibinfo {author} {\bibfnamefont {P.~A.}\ \bibnamefont {Lee}},\
  }\href {\doibase 10.1103/PhysRevLett.61.1768} {\bibfield  {journal} {\bibinfo
   {journal} {Phys. Rev. Lett.}\ }\textbf {\bibinfo {volume} {61}},\ \bibinfo
  {pages} {1768} (\bibinfo {year} {1988})}\BibitemShut {NoStop}%
\bibitem [{\citenamefont {Costi}\ \emph {et~al.}(1994)\citenamefont {Costi},
  \citenamefont {Hewson},\ and\ \citenamefont {Zlatic}}]{Costi1994}%
  \BibitemOpen
  \bibfield  {author} {\bibinfo {author} {\bibfnamefont {T.~A.}\ \bibnamefont
  {Costi}}, \bibinfo {author} {\bibfnamefont {A.~C.}\ \bibnamefont {Hewson}}, \
  and\ \bibinfo {author} {\bibfnamefont {V.}~\bibnamefont {Zlatic}},\ }\href
  {\doibase 10.1088/0953-8984/6/13/013} {\bibfield  {journal} {\bibinfo
  {journal} {J. Phys.: Condens. Matter}\ }\textbf {\bibinfo {volume} {6}},\
  \bibinfo {pages} {2518} (\bibinfo {year} {1994})}\BibitemShut {NoStop}%
\bibitem [{\citenamefont {Goldhaber-Gordon}\ \emph
  {et~al.}(1998{\natexlab{a}})\citenamefont {Goldhaber-Gordon}, \citenamefont
  {Shtrikman}, \citenamefont {Mahalu}, \citenamefont {Abusch-Magderand},
  \citenamefont {Meirav},\ and\ \citenamefont
  {Kastner}}]{Goldhaber-Gordon1998Nature}%
  \BibitemOpen
  \bibfield  {author} {\bibinfo {author} {\bibfnamefont {D.}~\bibnamefont
  {Goldhaber-Gordon}}, \bibinfo {author} {\bibfnamefont {H.}~\bibnamefont
  {Shtrikman}}, \bibinfo {author} {\bibfnamefont {D.}~\bibnamefont {Mahalu}},
  \bibinfo {author} {\bibfnamefont {D.}~\bibnamefont {Abusch-Magderand}},
  \bibinfo {author} {\bibfnamefont {U.}~\bibnamefont {Meirav}}, \ and\ \bibinfo
  {author} {\bibfnamefont {M.~A.}\ \bibnamefont {Kastner}},\ }\href {\doibase
  10.1038/34373} {\bibfield  {journal} {\bibinfo  {journal} {Nature}\ }\textbf
  {\bibinfo {volume} {391}},\ \bibinfo {pages} {156} (\bibinfo {year}
  {1998}{\natexlab{a}})}\BibitemShut {NoStop}%
\bibitem [{\citenamefont {Goldhaber-Gordon}\ \emph
  {et~al.}(1998{\natexlab{b}})\citenamefont {Goldhaber-Gordon}, \citenamefont
  {G{\"{o}}res}, \citenamefont {Kastner}, \citenamefont {Shtrikman},
  \citenamefont {Mahalu},\ and\ \citenamefont
  {Meirav}}]{Goldhaber-Gordon1998PRL}%
  \BibitemOpen
  \bibfield  {author} {\bibinfo {author} {\bibfnamefont {D.}~\bibnamefont
  {Goldhaber-Gordon}}, \bibinfo {author} {\bibfnamefont {J.}~\bibnamefont
  {G{\"{o}}res}}, \bibinfo {author} {\bibfnamefont {M.~A.}\ \bibnamefont
  {Kastner}}, \bibinfo {author} {\bibfnamefont {H.}~\bibnamefont {Shtrikman}},
  \bibinfo {author} {\bibfnamefont {D.}~\bibnamefont {Mahalu}}, \ and\ \bibinfo
  {author} {\bibfnamefont {U.}~\bibnamefont {Meirav}},\ }\href {\doibase
  10.1103/PhysRevLett.81.5225} {\bibfield  {journal} {\bibinfo  {journal}
  {Phys. Rev. Lett.}\ }\textbf {\bibinfo {volume} {81}},\ \bibinfo {pages}
  {5225} (\bibinfo {year} {1998}{\natexlab{b}})}\BibitemShut {NoStop}%
\bibitem [{\citenamefont {Pustilnik}\ and\ \citenamefont
  {Glazman}(2001)}]{Pustilnik2001}%
  \BibitemOpen
  \bibfield  {author} {\bibinfo {author} {\bibfnamefont {M.}~\bibnamefont
  {Pustilnik}}\ and\ \bibinfo {author} {\bibfnamefont {L.~I.}\ \bibnamefont
  {Glazman}},\ }\href {\doibase 10.1103/PhysRevLett.87.216601} {\bibfield
  {journal} {\bibinfo  {journal} {Phys. Rev. Lett.}\ }\textbf {\bibinfo
  {volume} {87}},\ \bibinfo {pages} {216601} (\bibinfo {year}
  {2001})}\BibitemShut {NoStop}%
\bibitem [{\citenamefont {Glazman}\ and\ \citenamefont
  {Pustilnik}(2003)}]{Glazman2003}%
  \BibitemOpen
  \bibfield  {author} {\bibinfo {author} {\bibfnamefont {L.~I.}\ \bibnamefont
  {Glazman}}\ and\ \bibinfo {author} {\bibfnamefont {M.}~\bibnamefont
  {Pustilnik}},\ }\href {\doibase 10.1007/978-94-007-1021-4_4} {\emph {\bibinfo
  {title} {New Directions in Mesoscopic Physics (Towards Nanoscience)}}},\
  edited by\ \bibinfo {editor} {\bibfnamefont {R.}~\bibnamefont {Fazio}},
  \bibinfo {editor} {\bibfnamefont {V.~F.}\ \bibnamefont {Gantmakher}}, \ and\
  \bibinfo {editor} {\bibfnamefont {Y.}~\bibnamefont {Imry}}\ (\bibinfo
  {publisher} {Springer Netherlands},\ \bibinfo {address} {Dordrecht},\
  \bibinfo {year} {2003})\ Chap.\ \bibinfo {chapter} {{Coulomb blockade and
  Kondo effect in quantum dots}}, pp.\ \bibinfo {pages} {93--115}\BibitemShut
  {NoStop}%
\bibitem [{\citenamefont {Kugler}\ \emph {et~al.}(2019)\citenamefont {Kugler},
  \citenamefont {Zingl}, \citenamefont {Strand}, \citenamefont {Lee},
  \citenamefont {von Delft},\ and\ \citenamefont {Georges}}]{Kugler2019}%
  \BibitemOpen
  \bibfield  {author} {\bibinfo {author} {\bibfnamefont {F.~B.}\ \bibnamefont
  {Kugler}}, \bibinfo {author} {\bibfnamefont {M.}~\bibnamefont {Zingl}},
  \bibinfo {author} {\bibfnamefont {H.~U.~R.}\ \bibnamefont {Strand}}, \bibinfo
  {author} {\bibfnamefont {S.-S.~B.}\ \bibnamefont {Lee}}, \bibinfo {author}
  {\bibfnamefont {J.}~\bibnamefont {von Delft}}, \ and\ \bibinfo {author}
  {\bibfnamefont {A.}~\bibnamefont {Georges}},\ }\href
  {https://arxiv.org/abs/1909.02389} {\bibfield  {journal} {\bibinfo  {journal}
  {arXiv/1909.02389}\ } (\bibinfo {year} {2019})}\BibitemShut {NoStop}%
\bibitem [{\citenamefont {Kotliar}\ \emph {et~al.}(2006)\citenamefont
  {Kotliar}, \citenamefont {Savrasov}, \citenamefont {Haule}, \citenamefont
  {Oudovenko}, \citenamefont {Parcollet},\ and\ \citenamefont
  {Marianetti}}]{Kotliar2006}%
  \BibitemOpen
  \bibfield  {author} {\bibinfo {author} {\bibfnamefont {G.}~\bibnamefont
  {Kotliar}}, \bibinfo {author} {\bibfnamefont {S.~Y.}\ \bibnamefont
  {Savrasov}}, \bibinfo {author} {\bibfnamefont {K.}~\bibnamefont {Haule}},
  \bibinfo {author} {\bibfnamefont {V.~S.}\ \bibnamefont {Oudovenko}}, \bibinfo
  {author} {\bibfnamefont {O.}~\bibnamefont {Parcollet}}, \ and\ \bibinfo
  {author} {\bibfnamefont {C.~A.}\ \bibnamefont {Marianetti}},\ }\href
  {\doibase 10.1103/RevModPhys.78.865} {\bibfield  {journal} {\bibinfo
  {journal} {Rev. Mod. Phys.}\ }\textbf {\bibinfo {volume} {78}},\ \bibinfo
  {pages} {865} (\bibinfo {year} {2006})}\BibitemShut {NoStop}%
\bibitem [{\citenamefont {Liang}\ \emph {et~al.}(2002)\citenamefont {Liang},
  \citenamefont {Shores}, \citenamefont {Bockrath}, \citenamefont {Long},\ and\
  \citenamefont {Park}}]{Liang2002}%
  \BibitemOpen
  \bibfield  {author} {\bibinfo {author} {\bibfnamefont {W.}~\bibnamefont
  {Liang}}, \bibinfo {author} {\bibfnamefont {M.~P.}\ \bibnamefont {Shores}},
  \bibinfo {author} {\bibfnamefont {M.}~\bibnamefont {Bockrath}}, \bibinfo
  {author} {\bibfnamefont {J.~R.}\ \bibnamefont {Long}}, \ and\ \bibinfo
  {author} {\bibfnamefont {H.}~\bibnamefont {Park}},\ }\href {\doibase
  10.1038/nature00790} {\bibfield  {journal} {\bibinfo  {journal} {Nature}\
  }\textbf {\bibinfo {volume} {417}},\ \bibinfo {pages} {725} (\bibinfo {year}
  {2002})}\BibitemShut {NoStop}%
\bibitem [{\citenamefont {Park}\ \emph {et~al.}(2002)\citenamefont {Park},
  \citenamefont {Pasupathy}, \citenamefont {Goldsmith}, \citenamefont {Chang},
  \citenamefont {Yaish}, \citenamefont {Petta}, \citenamefont {Rinkoski},
  \citenamefont {Sethna}, \citenamefont {Abru{\~{n}}a}, \citenamefont
  {McEuen},\ and\ \citenamefont {Ralph}}]{Park2002}%
  \BibitemOpen
  \bibfield  {author} {\bibinfo {author} {\bibfnamefont {J.}~\bibnamefont
  {Park}}, \bibinfo {author} {\bibfnamefont {A.~N.}\ \bibnamefont {Pasupathy}},
  \bibinfo {author} {\bibfnamefont {J.~I.}\ \bibnamefont {Goldsmith}}, \bibinfo
  {author} {\bibfnamefont {C.}~\bibnamefont {Chang}}, \bibinfo {author}
  {\bibfnamefont {Y.}~\bibnamefont {Yaish}}, \bibinfo {author} {\bibfnamefont
  {J.~R.}\ \bibnamefont {Petta}}, \bibinfo {author} {\bibfnamefont
  {M.}~\bibnamefont {Rinkoski}}, \bibinfo {author} {\bibfnamefont {J.~P.}\
  \bibnamefont {Sethna}}, \bibinfo {author} {\bibfnamefont {H.~D.}\
  \bibnamefont {Abru{\~{n}}a}}, \bibinfo {author} {\bibfnamefont {P.~L.}\
  \bibnamefont {McEuen}}, \ and\ \bibinfo {author} {\bibfnamefont {D.~C.}\
  \bibnamefont {Ralph}},\ }\href {\doibase 10.1038/nature00791} {\bibfield
  {journal} {\bibinfo  {journal} {Nature}\ }\textbf {\bibinfo {volume} {417}},\
  \bibinfo {pages} {722} (\bibinfo {year} {2002})}\BibitemShut {NoStop}%
\bibitem [{\citenamefont {Thoss}\ and\ \citenamefont
  {Evers}(2018)}]{Thoss2018}%
  \BibitemOpen
  \bibfield  {author} {\bibinfo {author} {\bibfnamefont {M.}~\bibnamefont
  {Thoss}}\ and\ \bibinfo {author} {\bibfnamefont {F.}~\bibnamefont {Evers}},\
  }\href {\doibase 10.1063/1.5003306} {\bibfield  {journal} {\bibinfo
  {journal} {J. Chem. Phys.}\ }\textbf {\bibinfo {volume} {148}},\ \bibinfo
  {pages} {030901} (\bibinfo {year} {2018})}\BibitemShut {NoStop}%
\bibitem [{\citenamefont {Droghetti}\ and\ \citenamefont
  {Rungger}(2017)}]{Droghetti2017}%
  \BibitemOpen
  \bibfield  {author} {\bibinfo {author} {\bibfnamefont {A.}~\bibnamefont
  {Droghetti}}\ and\ \bibinfo {author} {\bibfnamefont {I.}~\bibnamefont
  {Rungger}},\ }\href {\doibase 10.1103/PhysRevB.95.085131} {\bibfield
  {journal} {\bibinfo  {journal} {Phys. Rev. B}\ }\textbf {\bibinfo {volume}
  {95}},\ \bibinfo {pages} {085131} (\bibinfo {year} {2017})}\BibitemShut
  {NoStop}%
\bibitem [{\citenamefont {Chioncel}\ \emph {et~al.}(2015)\citenamefont
  {Chioncel}, \citenamefont {Morari}, \citenamefont {{\"{O}}stlin},
  \citenamefont {Appelt}, \citenamefont {Droghetti}, \citenamefont
  {Radonji{\'{c}}}, \citenamefont {Rungger}, \citenamefont {Vitos},
  \citenamefont {Eckern},\ and\ \citenamefont {Postnikov}}]{Chioncel2015}%
  \BibitemOpen
  \bibfield  {author} {\bibinfo {author} {\bibfnamefont {L.}~\bibnamefont
  {Chioncel}}, \bibinfo {author} {\bibfnamefont {C.}~\bibnamefont {Morari}},
  \bibinfo {author} {\bibfnamefont {A.}~\bibnamefont {{\"{O}}stlin}}, \bibinfo
  {author} {\bibfnamefont {W.~H.}\ \bibnamefont {Appelt}}, \bibinfo {author}
  {\bibfnamefont {A.}~\bibnamefont {Droghetti}}, \bibinfo {author}
  {\bibfnamefont {M.~M.}\ \bibnamefont {Radonji{\'{c}}}}, \bibinfo {author}
  {\bibfnamefont {I.}~\bibnamefont {Rungger}}, \bibinfo {author} {\bibfnamefont
  {L.}~\bibnamefont {Vitos}}, \bibinfo {author} {\bibfnamefont
  {U.}~\bibnamefont {Eckern}}, \ and\ \bibinfo {author} {\bibfnamefont {A.~V.}\
  \bibnamefont {Postnikov}},\ }\href {\doibase 10.1103/PhysRevB.92.054431}
  {\bibfield  {journal} {\bibinfo  {journal} {Phys. Rev. B}\ }\textbf {\bibinfo
  {volume} {92}},\ \bibinfo {pages} {054431} (\bibinfo {year}
  {2015})}\BibitemShut {NoStop}%
\bibitem [{\citenamefont {Appelt}\ \emph {et~al.}(2018)\citenamefont {Appelt},
  \citenamefont {Droghetti}, \citenamefont {Chioncel}, \citenamefont
  {Radonji{\'{c}}}, \citenamefont {Mu{\~{n}}oz}, \citenamefont {Kirchner},
  \citenamefont {Vollhardt},\ and\ \citenamefont {Rungger}}]{Appelt2018}%
  \BibitemOpen
  \bibfield  {author} {\bibinfo {author} {\bibfnamefont {W.~H.}\ \bibnamefont
  {Appelt}}, \bibinfo {author} {\bibfnamefont {A.}~\bibnamefont {Droghetti}},
  \bibinfo {author} {\bibfnamefont {L.}~\bibnamefont {Chioncel}}, \bibinfo
  {author} {\bibfnamefont {M.~M.}\ \bibnamefont {Radonji{\'{c}}}}, \bibinfo
  {author} {\bibfnamefont {E.}~\bibnamefont {Mu{\~{n}}oz}}, \bibinfo {author}
  {\bibfnamefont {S.}~\bibnamefont {Kirchner}}, \bibinfo {author}
  {\bibfnamefont {D.}~\bibnamefont {Vollhardt}}, \ and\ \bibinfo {author}
  {\bibfnamefont {I.}~\bibnamefont {Rungger}},\ }\href {\doibase
  10.1039/c8nr03991g} {\bibfield  {journal} {\bibinfo  {journal} {Nanoscale}\
  }\textbf {\bibinfo {volume} {10}},\ \bibinfo {pages} {17738} (\bibinfo {year}
  {2018})}\BibitemShut {NoStop}%
\bibitem [{\citenamefont {{David Jacob}}(2015)}]{DavidJacob2015}%
  \BibitemOpen
  \bibfield  {author} {\bibinfo {author} {\bibnamefont {{David Jacob}}},\
  }\href {\doibase http://dx.doi.org/10.1088/0953-8984/27/24/245606} {\bibfield
   {journal} {\bibinfo  {journal} {J. Phys. Condens. Matter}\ }\textbf
  {\bibinfo {volume} {27}},\ \bibinfo {pages} {245606} (\bibinfo {year}
  {2015})}\BibitemShut {NoStop}%
\bibitem [{\citenamefont {Jacob}\ \emph {et~al.}(2009)\citenamefont {Jacob},
  \citenamefont {Haule},\ and\ \citenamefont {Kotliar}}]{Jacob2009}%
  \BibitemOpen
  \bibfield  {author} {\bibinfo {author} {\bibfnamefont {D.}~\bibnamefont
  {Jacob}}, \bibinfo {author} {\bibfnamefont {K.}~\bibnamefont {Haule}}, \ and\
  \bibinfo {author} {\bibfnamefont {G.}~\bibnamefont {Kotliar}},\ }\href
  {\doibase 10.1103/PhysRevLett.103.016803} {\bibfield  {journal} {\bibinfo
  {journal} {Phys. Rev. Lett.}\ }\textbf {\bibinfo {volume} {103}},\ \bibinfo
  {pages} {3} (\bibinfo {year} {2009})}\BibitemShut {NoStop}%
\bibitem [{\citenamefont {Jacob}\ \emph {et~al.}(2010)\citenamefont {Jacob},
  \citenamefont {Haule},\ and\ \citenamefont {Kotliar}}]{Jacob2010}%
  \BibitemOpen
  \bibfield  {author} {\bibinfo {author} {\bibfnamefont {D.}~\bibnamefont
  {Jacob}}, \bibinfo {author} {\bibfnamefont {K.}~\bibnamefont {Haule}}, \ and\
  \bibinfo {author} {\bibfnamefont {G.}~\bibnamefont {Kotliar}},\ }\href
  {\doibase 10.1103/PhysRevB.82.195115} {\bibfield  {journal} {\bibinfo
  {journal} {Phys. Rev. B}\ }\textbf {\bibinfo {volume} {82}},\ \bibinfo
  {pages} {195115} (\bibinfo {year} {2010})}\BibitemShut {NoStop}%
\bibitem [{\citenamefont {Theumann}(1969)}]{Theumann1969}%
  \BibitemOpen
  \bibfield  {author} {\bibinfo {author} {\bibfnamefont {A.}~\bibnamefont
  {Theumann}},\ }\href {\doibase 10.1103/PhysRev.178.978} {\bibfield  {journal}
  {\bibinfo  {journal} {Phys. Rev.}\ }\textbf {\bibinfo {volume} {178}},\
  \bibinfo {pages} {978} (\bibinfo {year} {1969})}\BibitemShut {NoStop}%
\bibitem [{\citenamefont {Lacroix}(1981)}]{Lacroix1981}%
  \BibitemOpen
  \bibfield  {author} {\bibinfo {author} {\bibfnamefont {C.}~\bibnamefont
  {Lacroix}},\ }\href {\doibase 10.1088/0305-4608/11/11/020} {\bibfield
  {journal} {\bibinfo  {journal} {J. Phys. F: Met. Phys.}\ }\textbf {\bibinfo
  {volume} {11}},\ \bibinfo {pages} {2389} (\bibinfo {year}
  {1981})}\BibitemShut {NoStop}%
\bibitem [{\citenamefont {Lacroix}(1982)}]{Lacroix1982}%
  \BibitemOpen
  \bibfield  {author} {\bibinfo {author} {\bibfnamefont {C.}~\bibnamefont
  {Lacroix}},\ }\href {\doibase 10.1063/1.330756} {\bibfield  {journal}
  {\bibinfo  {journal} {J. Appl. Phys.}\ }\textbf {\bibinfo {volume} {53}},\
  \bibinfo {pages} {2131} (\bibinfo {year} {1982})}\BibitemShut {NoStop}%
\bibitem [{\citenamefont {Bickers}(1987)}]{bickers1987}%
  \BibitemOpen
  \bibfield  {author} {\bibinfo {author} {\bibfnamefont {N.~E.}\ \bibnamefont
  {Bickers}},\ }\href {\doibase 10.1103/RevModPhys.59.845} {\bibfield
  {journal} {\bibinfo  {journal} {Rev. Mod. Phys.}\ }\textbf {\bibinfo {volume}
  {59}},\ \bibinfo {pages} {845} (\bibinfo {year} {1987})}\BibitemShut
  {NoStop}%
\bibitem [{\citenamefont {Pruschke}\ and\ \citenamefont
  {Grewe}(1989)}]{Pruschke1989}%
  \BibitemOpen
  \bibfield  {author} {\bibinfo {author} {\bibfnamefont {T.}~\bibnamefont
  {Pruschke}}\ and\ \bibinfo {author} {\bibfnamefont {N.}~\bibnamefont
  {Grewe}},\ }\href {\doibase 10.1007/BF01311391} {\bibfield  {journal}
  {\bibinfo  {journal} {Z. Phys. B}\ }\textbf {\bibinfo {volume} {74}},\
  \bibinfo {pages} {439} (\bibinfo {year} {1989})}\BibitemShut {NoStop}%
\bibitem [{\citenamefont {Hirsch}\ and\ \citenamefont
  {Fye}(1986)}]{Hirsch1986}%
  \BibitemOpen
  \bibfield  {author} {\bibinfo {author} {\bibfnamefont {J.~E.}\ \bibnamefont
  {Hirsch}}\ and\ \bibinfo {author} {\bibfnamefont {R.~M.}\ \bibnamefont
  {Fye}},\ }\href {\doibase 10.1103/PhysRevLett.56.2521} {\bibfield  {journal}
  {\bibinfo  {journal} {Phys. Rev. Lett.}\ }\textbf {\bibinfo {volume} {56}},\
  \bibinfo {pages} {2521} (\bibinfo {year} {1986})}\BibitemShut {NoStop}%
\bibitem [{\citenamefont {Yamada}(1975)}]{Yamada1975}%
  \BibitemOpen
  \bibfield  {author} {\bibinfo {author} {\bibfnamefont {K.}~\bibnamefont
  {Yamada}},\ }\href {\doibase 10.1143/PTP.53.970} {\bibfield  {journal}
  {\bibinfo  {journal} {Prog. Theor. Phys.}\ }\textbf {\bibinfo {volume}
  {53}},\ \bibinfo {pages} {970} (\bibinfo {year} {1975})}\BibitemShut
  {NoStop}%
\bibitem [{\citenamefont {Yosida}\ and\ \citenamefont
  {Yamada}(1970)}]{Yosida1970}%
  \BibitemOpen
  \bibfield  {author} {\bibinfo {author} {\bibfnamefont {K.}~\bibnamefont
  {Yosida}}\ and\ \bibinfo {author} {\bibfnamefont {K.}~\bibnamefont
  {Yamada}},\ }\href {\doibase 10.1143/PTPS.46.244} {\bibfield  {journal}
  {\bibinfo  {journal} {Prog. Theor. Phys. Sup.}\ }\textbf {\bibinfo {volume}
  {46}},\ \bibinfo {pages} {244} (\bibinfo {year} {1970})}\BibitemShut
  {NoStop}%
\bibitem [{\citenamefont {Yosida}\ and\ \citenamefont
  {Yamada}(1975)}]{Yosida1970a}%
  \BibitemOpen
  \bibfield  {author} {\bibinfo {author} {\bibfnamefont {K.}~\bibnamefont
  {Yosida}}\ and\ \bibinfo {author} {\bibfnamefont {K.}~\bibnamefont
  {Yamada}},\ }\href {\doibase 10.1143/PTP.53.1286} {\bibfield  {journal}
  {\bibinfo  {journal} {Prog. Theor. Phys.}\ }\textbf {\bibinfo {volume}
  {53}},\ \bibinfo {pages} {1286} (\bibinfo {year} {1975})}\BibitemShut
  {NoStop}%
\bibitem [{\citenamefont {Kashcheyevs}\ \emph {et~al.}(2006)\citenamefont
  {Kashcheyevs}, \citenamefont {Aharony},\ and\ \citenamefont
  {Entin-Wohlman}}]{Kashcheyevs2006}%
  \BibitemOpen
  \bibfield  {author} {\bibinfo {author} {\bibfnamefont {V.}~\bibnamefont
  {Kashcheyevs}}, \bibinfo {author} {\bibfnamefont {A.}~\bibnamefont
  {Aharony}}, \ and\ \bibinfo {author} {\bibfnamefont {O.}~\bibnamefont
  {Entin-Wohlman}},\ }\href {\doibase 10.1103/PhysRevB.73.125338} {\bibfield
  {journal} {\bibinfo  {journal} {Phys. Rev. B}\ }\textbf {\bibinfo {volume}
  {73}},\ \bibinfo {pages} {125338} (\bibinfo {year} {2006})}\BibitemShut
  {NoStop}%
\bibitem [{\citenamefont {Haule}\ \emph {et~al.}(2010)\citenamefont {Haule},
  \citenamefont {Yee},\ and\ \citenamefont {Kim}}]{Haule2010}%
  \BibitemOpen
  \bibfield  {author} {\bibinfo {author} {\bibfnamefont {K.}~\bibnamefont
  {Haule}}, \bibinfo {author} {\bibfnamefont {C.~H.}\ \bibnamefont {Yee}}, \
  and\ \bibinfo {author} {\bibfnamefont {K.}~\bibnamefont {Kim}},\ }\href
  {\doibase 10.1103/PhysRevB.81.195107} {\bibfield  {journal} {\bibinfo
  {journal} {Phys. Rev. B}\ }\textbf {\bibinfo {volume} {81}},\ \bibinfo
  {pages} {195107} (\bibinfo {year} {2010})}\BibitemShut {NoStop}%
\bibitem [{\citenamefont {Bulla}\ \emph {et~al.}(2008)\citenamefont {Bulla},
  \citenamefont {Costi},\ and\ \citenamefont {Pruschke}}]{Bulla2008}%
  \BibitemOpen
  \bibfield  {author} {\bibinfo {author} {\bibfnamefont {R.}~\bibnamefont
  {Bulla}}, \bibinfo {author} {\bibfnamefont {T.~A.}\ \bibnamefont {Costi}}, \
  and\ \bibinfo {author} {\bibfnamefont {T.}~\bibnamefont {Pruschke}},\ }\href
  {\doibase 10.1103/RevModPhys.80.395} {\bibfield  {journal} {\bibinfo
  {journal} {Rev. Mod. Phys.}\ }\textbf {\bibinfo {volume} {80}},\ \bibinfo
  {pages} {395} (\bibinfo {year} {2008})}\BibitemShut {NoStop}%
\bibitem [{\citenamefont {Meir}\ and\ \citenamefont
  {Wingreen}(1992)}]{MeirWingreen1992}%
  \BibitemOpen
  \bibfield  {author} {\bibinfo {author} {\bibfnamefont {Y.}~\bibnamefont
  {Meir}}\ and\ \bibinfo {author} {\bibfnamefont {N.~S.}\ \bibnamefont
  {Wingreen}},\ }\href {\doibase 10.1103/PhysRevLett.68.2512} {\bibfield
  {journal} {\bibinfo  {journal} {Phys. Rev. Lett.}\ }\textbf {\bibinfo
  {volume} {68}},\ \bibinfo {pages} {2512} (\bibinfo {year}
  {1992})}\BibitemShut {NoStop}%
\bibitem [{\citenamefont {Wingreen}\ and\ \citenamefont
  {Meir}(1994)}]{Wingreen1994}%
  \BibitemOpen
  \bibfield  {author} {\bibinfo {author} {\bibfnamefont {N.~S.}\ \bibnamefont
  {Wingreen}}\ and\ \bibinfo {author} {\bibfnamefont {Y.}~\bibnamefont
  {Meir}},\ }\href {\doibase 10.1103/PhysRevB.49.11040} {\bibfield  {journal}
  {\bibinfo  {journal} {Phys. Rev. B}\ }\textbf {\bibinfo {volume} {49}},\
  \bibinfo {pages} {11040} (\bibinfo {year} {1994})}\BibitemShut {NoStop}%
\bibitem [{Note1()}]{Note1}%
  \BibitemOpen
  \bibinfo {note} {In this paper we use $\Gamma $, the typical scattering
  theory notation used in quantum dots and molecular electronics, instead of
  the hybridization function $\Delta $, more familiar to the strongly
  correlated systems community. Note that $\Gamma = 2 \Delta $.}\BibitemShut
  {Stop}%
\bibitem [{\citenamefont {Haug}\ and\ \citenamefont {Jauho}(2008)}]{Haug2008}%
  \BibitemOpen
  \bibfield  {author} {\bibinfo {author} {\bibfnamefont {H.}~\bibnamefont
  {Haug}}\ and\ \bibinfo {author} {\bibfnamefont {A.-P.}\ \bibnamefont
  {Jauho}},\ }\href@noop {} {\emph {\bibinfo {title} {{Quantum Kinectics in
  Transport and Optics of Semiconductors}}}},\ \bibinfo {edition} {2nd}\ ed.\
  (\bibinfo  {publisher} {Springer},\ \bibinfo {year} {2008})\BibitemShut
  {NoStop}%
\bibitem [{\citenamefont {{Dias da Silva}}\ \emph {et~al.}(2017)\citenamefont
  {{Dias da Silva}}, \citenamefont {Lewenkopf}, \citenamefont {Vernek},
  \citenamefont {Ferreira},\ and\ \citenamefont {Ulloa}}]{DiasdaSilva2017}%
  \BibitemOpen
  \bibfield  {author} {\bibinfo {author} {\bibfnamefont {L.~G. G.~V.}\
  \bibnamefont {{Dias da Silva}}}, \bibinfo {author} {\bibfnamefont
  {C.}~\bibnamefont {Lewenkopf}}, \bibinfo {author} {\bibfnamefont
  {E.}~\bibnamefont {Vernek}}, \bibinfo {author} {\bibfnamefont {G.~J.}\
  \bibnamefont {Ferreira}}, \ and\ \bibinfo {author} {\bibfnamefont {S.~E.}\
  \bibnamefont {Ulloa}},\ }\href {\doibase 10.1103/PhysRevLett.119.116801}
  {\bibfield  {journal} {\bibinfo  {journal} {Phys. Rev. Lett.}\ }\textbf
  {\bibinfo {volume} {119}},\ \bibinfo {pages} {116801} (\bibinfo {year}
  {2017})}\BibitemShut {NoStop}%
\bibitem [{\citenamefont {Hern{\'{a}}ndez}\ \emph {et~al.}(2007)\citenamefont
  {Hern{\'{a}}ndez}, \citenamefont {Apel}, \citenamefont {Pinheiro},\ and\
  \citenamefont {Lewenkopf}}]{Hernandez2007}%
  \BibitemOpen
  \bibfield  {author} {\bibinfo {author} {\bibfnamefont {A.}~\bibnamefont
  {Hern{\'{a}}ndez}}, \bibinfo {author} {\bibfnamefont {V.~M.}\ \bibnamefont
  {Apel}}, \bibinfo {author} {\bibfnamefont {F.~A.}\ \bibnamefont {Pinheiro}},
  \ and\ \bibinfo {author} {\bibfnamefont {C.~H.}\ \bibnamefont {Lewenkopf}},\
  }\href {\doibase 10.1016/j.physa.2007.06.032} {\bibfield  {journal} {\bibinfo
   {journal} {Physica A}\ }\textbf {\bibinfo {volume} {385}},\ \bibinfo {pages}
  {148} (\bibinfo {year} {2007})}\BibitemShut {NoStop}%
\bibitem [{\citenamefont {Hubbard}(1963)}]{Hubbard1963}%
  \BibitemOpen
  \bibfield  {author} {\bibinfo {author} {\bibfnamefont {J.}~\bibnamefont
  {Hubbard}},\ }\href {\doibase 10.1098/rspa.1963.0204} {\bibfield  {journal}
  {\bibinfo  {journal} {Proc. R. Soc. Lond. A}\ }\textbf {\bibinfo {volume}
  {276}},\ \bibinfo {pages} {238} (\bibinfo {year} {1963})}\BibitemShut
  {NoStop}%
\bibitem [{\citenamefont {Gebhard}(1997)}]{Gebhard1997}%
  \BibitemOpen
  \bibfield  {author} {\bibinfo {author} {\bibfnamefont {F.}~\bibnamefont
  {Gebhard}},\ }\href@noop {} {\emph {\bibinfo {title} {{The Mott
  metal-insulator transition: models and methods}}}}\ (\bibinfo  {publisher}
  {Springer},\ \bibinfo {year} {1997})\BibitemShut {NoStop}%
\bibitem [{\citenamefont {Hubbard}(1964)}]{Hubbard1964}%
  \BibitemOpen
  \bibfield  {author} {\bibinfo {author} {\bibfnamefont {J.}~\bibnamefont
  {Hubbard}},\ }\href {\doibase 10.1098/rspa.1964.0019} {\bibfield  {journal}
  {\bibinfo  {journal} {Proc. R. Soc. Lond. A}\ }\textbf {\bibinfo {volume}
  {277}},\ \bibinfo {pages} {237} (\bibinfo {year} {1964})}\BibitemShut
  {NoStop}%
\bibitem [{\citenamefont {Gerace}\ \emph {et~al.}(2002)\citenamefont {Gerace},
  \citenamefont {Pavarini},\ and\ \citenamefont {Andreani}}]{Gerace2002}%
  \BibitemOpen
  \bibfield  {author} {\bibinfo {author} {\bibfnamefont {D.}~\bibnamefont
  {Gerace}}, \bibinfo {author} {\bibfnamefont {E.}~\bibnamefont {Pavarini}}, \
  and\ \bibinfo {author} {\bibfnamefont {L.~C.}\ \bibnamefont {Andreani}},\
  }\href {\doibase 10.1103/PhysRevB.65.155331} {\bibfield  {journal} {\bibinfo
  {journal} {Phys. Rev. B}\ }\textbf {\bibinfo {volume} {65}},\ \bibinfo
  {pages} {155331} (\bibinfo {year} {2002})}\BibitemShut {NoStop}%
\bibitem [{\citenamefont {Aguado}\ and\ \citenamefont
  {Langreth}(2003)}]{Aguado2003}%
  \BibitemOpen
  \bibfield  {author} {\bibinfo {author} {\bibfnamefont {R.}~\bibnamefont
  {Aguado}}\ and\ \bibinfo {author} {\bibfnamefont {D.~C.}\ \bibnamefont
  {Langreth}},\ }\href {\doibase 10.1103/PhysRevB.67.245307} {\bibfield
  {journal} {\bibinfo  {journal} {Phys. Rev. B}\ }\textbf {\bibinfo {volume}
  {67}},\ \bibinfo {pages} {245307} (\bibinfo {year} {2003})}\BibitemShut
  {NoStop}%
\bibitem [{\citenamefont {Haule}\ \emph {et~al.}(2001)\citenamefont {Haule},
  \citenamefont {Kirchner}, \citenamefont {Kroha},\ and\ \citenamefont
  {W{\"{o}}lfle}}]{Haule2001}%
  \BibitemOpen
  \bibfield  {author} {\bibinfo {author} {\bibfnamefont {K.}~\bibnamefont
  {Haule}}, \bibinfo {author} {\bibfnamefont {S.}~\bibnamefont {Kirchner}},
  \bibinfo {author} {\bibfnamefont {J.}~\bibnamefont {Kroha}}, \ and\ \bibinfo
  {author} {\bibfnamefont {P.}~\bibnamefont {W{\"{o}}lfle}},\ }\href {\doibase
  10.1103/PhysRevB.64.155111} {\bibfield  {journal} {\bibinfo  {journal} {Phys.
  Rev. B}\ }\textbf {\bibinfo {volume} {64}},\ \bibinfo {pages} {155111}
  (\bibinfo {year} {2001})}\BibitemShut {NoStop}%
\bibitem [{\citenamefont {Sposetti}\ \emph {et~al.}(2016)\citenamefont
  {Sposetti}, \citenamefont {Manuel},\ and\ \citenamefont
  {Roura-Bas}}]{Sposetti2016}%
  \BibitemOpen
  \bibfield  {author} {\bibinfo {author} {\bibfnamefont {C.~N.}\ \bibnamefont
  {Sposetti}}, \bibinfo {author} {\bibfnamefont {L.~O.}\ \bibnamefont
  {Manuel}}, \ and\ \bibinfo {author} {\bibfnamefont {P.}~\bibnamefont
  {Roura-Bas}},\ }\href {\doibase 10.1103/PhysRevB.94.085139} {\bibfield
  {journal} {\bibinfo  {journal} {Phys. Rev. B}\ }\textbf {\bibinfo {volume}
  {94}},\ \bibinfo {pages} {085139} (\bibinfo {year} {2016})}\BibitemShut
  {NoStop}%
\bibitem [{\citenamefont {Barnes}(1976)}]{Barnes1976}%
  \BibitemOpen
  \bibfield  {author} {\bibinfo {author} {\bibfnamefont {S.~E.}\ \bibnamefont
  {Barnes}},\ }\href {\doibase 10.1088/0305-4608/6/7/018} {\bibfield  {journal}
  {\bibinfo  {journal} {J. Phys. F: Met. Phys.}\ }\textbf {\bibinfo {volume}
  {6}},\ \bibinfo {pages} {1375} (\bibinfo {year} {1976})}\BibitemShut
  {NoStop}%
\bibitem [{\citenamefont {Barnes}(1977)}]{Barnes1977}%
  \BibitemOpen
  \bibfield  {author} {\bibinfo {author} {\bibfnamefont {S.~E.}\ \bibnamefont
  {Barnes}},\ }\href {\doibase 10.1088/0305-4608/7/12/022} {\bibfield
  {journal} {\bibinfo  {journal} {J. Phys. F: Met. Phys.}\ }\textbf {\bibinfo
  {volume} {7}},\ \bibinfo {pages} {2637} (\bibinfo {year} {1977})}\BibitemShut
  {NoStop}%
\bibitem [{\citenamefont {Kroha}\ and\ \citenamefont
  {W{\"{o}}lfle}(1998)}]{Kroha1998}%
  \BibitemOpen
  \bibfield  {author} {\bibinfo {author} {\bibfnamefont {J.}~\bibnamefont
  {Kroha}}\ and\ \bibinfo {author} {\bibfnamefont {P.}~\bibnamefont
  {W{\"{o}}lfle}},\ }\href {\doibase 10.1007/BFb0107485} {\bibfield  {journal}
  {\bibinfo  {journal} {Acta Physica Polonica B}\ }\textbf {\bibinfo {volume}
  {29}},\ \bibinfo {pages} {3781} (\bibinfo {year} {1998})}\BibitemShut
  {NoStop}%
\bibitem [{\citenamefont {Abrikosov}(1965)}]{Abrikosov1965}%
  \BibitemOpen
  \bibfield  {author} {\bibinfo {author} {\bibfnamefont {A.}~\bibnamefont
  {Abrikosov}},\ }\href@noop {} {\bibfield  {journal} {\bibinfo  {journal}
  {Physics World Physique Fizika}\ }\textbf {\bibinfo {volume} {2}},\ \bibinfo
  {pages} {5} (\bibinfo {year} {1965})}\BibitemShut {NoStop}%
\bibitem [{\citenamefont {Hettler}\ \emph {et~al.}(1998)\citenamefont
  {Hettler}, \citenamefont {Kroha},\ and\ \citenamefont
  {Hershfield}}]{Hettler1998}%
  \BibitemOpen
  \bibfield  {author} {\bibinfo {author} {\bibfnamefont {M.~H.}\ \bibnamefont
  {Hettler}}, \bibinfo {author} {\bibfnamefont {J.}~\bibnamefont {Kroha}}, \
  and\ \bibinfo {author} {\bibfnamefont {S.}~\bibnamefont {Hershfield}},\
  }\href {\doibase 10.1103/PhysRevB.58.5649} {\bibfield  {journal} {\bibinfo
  {journal} {Phys. Rev. B}\ }\textbf {\bibinfo {volume} {58}},\ \bibinfo
  {pages} {5649} (\bibinfo {year} {1998})}\BibitemShut {NoStop}%
\bibitem [{\citenamefont {Costi}\ \emph {et~al.}(1996)\citenamefont {Costi},
  \citenamefont {Kroha},\ and\ \citenamefont {W{\"{o}}lfle}}]{Costi1996}%
  \BibitemOpen
  \bibfield  {author} {\bibinfo {author} {\bibfnamefont {T.}~\bibnamefont
  {Costi}}, \bibinfo {author} {\bibfnamefont {J.}~\bibnamefont {Kroha}}, \ and\
  \bibinfo {author} {\bibfnamefont {P.}~\bibnamefont {W{\"{o}}lfle}},\ }\href
  {\doibase 10.1103/PhysRevB.53.1850} {\bibfield  {journal} {\bibinfo
  {journal} {Phys. Rev. B}\ }\textbf {\bibinfo {volume} {53}},\ \bibinfo
  {pages} {1850} (\bibinfo {year} {1996})}\BibitemShut {NoStop}%
\bibitem [{Note2()}]{Note2}%
  \BibitemOpen
  \bibinfo {note} {See, for example, the Appendix of Ref. \protect
  \rev@citealpnum {Sposetti2016}}\BibitemShut {NoStop}%
\bibitem [{Note3()}]{Note3}%
  \BibitemOpen
  \bibinfo {note} {See, for example, the Appendix A of Ref. \protect
  \rev@citealpnum {Hettler1998}}\BibitemShut {NoStop}%
\bibitem [{Note4()}]{Note4}%
  \BibitemOpen
  \bibinfo {note} {The main pseudo-particle is the one with the lowest energy.
  Note that $\varepsilon _b = 0$, $\varepsilon _{s_{\sigma }}=\varepsilon _0$,
  and $\varepsilon _d = 2\varepsilon _0 + U$}\BibitemShut {NoStop}%
\bibitem [{\citenamefont {Tosi}\ \emph {et~al.}(2011)\citenamefont {Tosi},
  \citenamefont {Roura-Bas}, \citenamefont {Llois},\ and\ \citenamefont
  {Manuel}}]{Tosi2011}%
  \BibitemOpen
  \bibfield  {author} {\bibinfo {author} {\bibfnamefont {L.}~\bibnamefont
  {Tosi}}, \bibinfo {author} {\bibfnamefont {P.}~\bibnamefont {Roura-Bas}},
  \bibinfo {author} {\bibfnamefont {A.~M.}\ \bibnamefont {Llois}}, \ and\
  \bibinfo {author} {\bibfnamefont {L.~O.}\ \bibnamefont {Manuel}},\ }\href
  {\doibase 10.1103/PhysRevB.83.073301} {\bibfield  {journal} {\bibinfo
  {journal} {Phys. Rev. B}\ }\textbf {\bibinfo {volume} {83}},\ \bibinfo
  {pages} {73301} (\bibinfo {year} {2011})}\BibitemShut {NoStop}%
\bibitem [{\citenamefont {Grewe}\ \emph {et~al.}(2008)\citenamefont {Grewe},
  \citenamefont {Schmitt}, \citenamefont {Jabben},\ and\ \citenamefont
  {Anders}}]{Grewe2008}%
  \BibitemOpen
  \bibfield  {author} {\bibinfo {author} {\bibfnamefont {N.}~\bibnamefont
  {Grewe}}, \bibinfo {author} {\bibfnamefont {S.}~\bibnamefont {Schmitt}},
  \bibinfo {author} {\bibfnamefont {T.}~\bibnamefont {Jabben}}, \ and\ \bibinfo
  {author} {\bibfnamefont {F.~B.}\ \bibnamefont {Anders}},\ }\href {\doibase
  10.1088/0953-8984/20/36/365217} {\bibfield  {journal} {\bibinfo  {journal}
  {J. Phys. Condens. Matter}\ }\textbf {\bibinfo {volume} {20}},\ \bibinfo
  {pages} {365217} (\bibinfo {year} {2008})}\BibitemShut {NoStop}%
\bibitem [{\citenamefont {Anders}(1995)}]{Anders1995}%
  \BibitemOpen
  \bibfield  {author} {\bibinfo {author} {\bibfnamefont {F.~B.}\ \bibnamefont
  {Anders}},\ }\href {\doibase 10.1088/0953-8984/7/14/018} {\bibfield
  {journal} {\bibinfo  {journal} {J. Phys. Condens. Matter}\ }\textbf {\bibinfo
  {volume} {7}},\ \bibinfo {pages} {2801} (\bibinfo {year} {1995})}\BibitemShut
  {NoStop}%
\bibitem [{\citenamefont {Schmitt}\ \emph {et~al.}(2009)\citenamefont
  {Schmitt}, \citenamefont {Jabben},\ and\ \citenamefont
  {Grewe}}]{Schmitt2009}%
  \BibitemOpen
  \bibfield  {author} {\bibinfo {author} {\bibfnamefont {S.}~\bibnamefont
  {Schmitt}}, \bibinfo {author} {\bibfnamefont {T.}~\bibnamefont {Jabben}}, \
  and\ \bibinfo {author} {\bibfnamefont {N.}~\bibnamefont {Grewe}},\ }\href
  {\doibase 10.1103/PhysRevB.80.235130} {\bibfield  {journal} {\bibinfo
  {journal} {Phys. Rev. B}\ }\textbf {\bibinfo {volume} {80}},\ \bibinfo
  {pages} {1} (\bibinfo {year} {2009})}\BibitemShut {NoStop}%
\bibitem [{\citenamefont {Vildosola}\ \emph {et~al.}(2015)\citenamefont
  {Vildosola}, \citenamefont {Pourovskii}, \citenamefont {Manuel},\ and\
  \citenamefont {Roura-Bas}}]{Vildosola2015}%
  \BibitemOpen
  \bibfield  {author} {\bibinfo {author} {\bibfnamefont {V.}~\bibnamefont
  {Vildosola}}, \bibinfo {author} {\bibfnamefont {L.~V.}\ \bibnamefont
  {Pourovskii}}, \bibinfo {author} {\bibfnamefont {L.~O.}\ \bibnamefont
  {Manuel}}, \ and\ \bibinfo {author} {\bibfnamefont {P.}~\bibnamefont
  {Roura-Bas}},\ }\href {\doibase 10.1088/0953-8984/27/48/485602} {\bibfield
  {journal} {\bibinfo  {journal} {J. Phys. Condens. Matter}\ }\textbf {\bibinfo
  {volume} {27}},\ \bibinfo {pages} {485602} (\bibinfo {year}
  {2015})}\BibitemShut {NoStop}%
\bibitem [{\citenamefont {R{\"{u}}egg}\ \emph {et~al.}(2013)\citenamefont
  {R{\"{u}}egg}, \citenamefont {Gull}, \citenamefont {Fiete},\ and\
  \citenamefont {Millis}}]{Ruegg2013}%
  \BibitemOpen
  \bibfield  {author} {\bibinfo {author} {\bibfnamefont {A.}~\bibnamefont
  {R{\"{u}}egg}}, \bibinfo {author} {\bibfnamefont {E.}~\bibnamefont {Gull}},
  \bibinfo {author} {\bibfnamefont {G.~A.}\ \bibnamefont {Fiete}}, \ and\
  \bibinfo {author} {\bibfnamefont {A.~J.}\ \bibnamefont {Millis}},\ }\href
  {\doibase 10.1103/PhysRevB.87.075124} {\bibfield  {journal} {\bibinfo
  {journal} {Phys. Rev. B}\ }\textbf {\bibinfo {volume} {87}},\ \bibinfo
  {pages} {1} (\bibinfo {year} {2013})}\BibitemShut {NoStop}%
\bibitem [{\citenamefont {\ifmmode~\check{Z}\else
  \v{Z}\fi{}itko}(2011)}]{Zitko:Phys.Rev.B:085142:2011}%
  \BibitemOpen
  \bibfield  {author} {\bibinfo {author} {\bibfnamefont {R.}~\bibnamefont
  {\ifmmode~\check{Z}\else \v{Z}\fi{}itko}},\ }\href {\doibase
  10.1103/PhysRevB.84.085142} {\bibfield  {journal} {\bibinfo  {journal} {Phys.
  Rev. B}\ }\textbf {\bibinfo {volume} {84}},\ \bibinfo {pages} {085142}
  (\bibinfo {year} {2011})}\BibitemShut {NoStop}%
\bibitem [{\citenamefont {Peters}\ \emph {et~al.}(2006)\citenamefont {Peters},
  \citenamefont {Pruschke},\ and\ \citenamefont {Anders}}]{Peters2006}%
  \BibitemOpen
  \bibfield  {author} {\bibinfo {author} {\bibfnamefont {R.}~\bibnamefont
  {Peters}}, \bibinfo {author} {\bibfnamefont {T.}~\bibnamefont {Pruschke}}, \
  and\ \bibinfo {author} {\bibfnamefont {F.~B.}\ \bibnamefont {Anders}},\
  }\href {\doibase 10.1103/PhysRevB.74.245114} {\bibfield  {journal} {\bibinfo
  {journal} {Phys. Rev. B}\ }\textbf {\bibinfo {volume} {74}},\ \bibinfo
  {pages} {1} (\bibinfo {year} {2006})}\BibitemShut {NoStop}%
\bibitem [{\citenamefont {Weichselbaum}\ and\ \citenamefont {{Von
  Delft}}(2007)}]{Weichselbaum2007}%
  \BibitemOpen
  \bibfield  {author} {\bibinfo {author} {\bibfnamefont {A.}~\bibnamefont
  {Weichselbaum}}\ and\ \bibinfo {author} {\bibfnamefont {J.}~\bibnamefont
  {{Von Delft}}},\ }\href {\doibase 10.1103/PhysRevLett.99.076402} {\bibfield
  {journal} {\bibinfo  {journal} {Phys. Rev. Lett.}\ }\textbf {\bibinfo
  {volume} {99}},\ \bibinfo {pages} {076402} (\bibinfo {year}
  {2007})}\BibitemShut {NoStop}%
\bibitem [{\citenamefont {Yoshida}\ \emph {et~al.}(1990)\citenamefont
  {Yoshida}, \citenamefont {Whitaker},\ and\ \citenamefont
  {Oliveira}}]{Yoshida:Phys.Rev.B:41:9403:1990}%
  \BibitemOpen
  \bibfield  {author} {\bibinfo {author} {\bibfnamefont {M.}~\bibnamefont
  {Yoshida}}, \bibinfo {author} {\bibfnamefont {M.~A.}\ \bibnamefont
  {Whitaker}}, \ and\ \bibinfo {author} {\bibfnamefont {L.~N.}\ \bibnamefont
  {Oliveira}},\ }\href {\doibase 10.1103/PhysRevB.41.9403} {\bibfield
  {journal} {\bibinfo  {journal} {Phys. Rev. B}\ }\textbf {\bibinfo {volume}
  {41}},\ \bibinfo {pages} {9403} (\bibinfo {year} {1990})}\BibitemShut
  {NoStop}%
\bibitem [{\citenamefont {Zawadzki}\ and\ \citenamefont
  {Oliveira}(2018)}]{Zawadzki2018}%
  \BibitemOpen
  \bibfield  {author} {\bibinfo {author} {\bibfnamefont {K.}~\bibnamefont
  {Zawadzki}}\ and\ \bibinfo {author} {\bibfnamefont {L.}~\bibnamefont
  {Oliveira}},\ }\href {\doibase doi.org/10.1140/epjb/e2018-90164-y} {\bibfield
   {journal} {\bibinfo  {journal} {Eur. Phys. J. B}\ }\textbf {\bibinfo
  {volume} {91}},\ \bibinfo {pages} {136} (\bibinfo {year} {2018})}\BibitemShut
  {NoStop}%
\bibitem [{\citenamefont {Haldane}(1978)}]{Haldane1978}%
  \BibitemOpen
  \bibfield  {author} {\bibinfo {author} {\bibfnamefont {F.~D.~M.}\
  \bibnamefont {Haldane}},\ }\href {\doibase 10.1103/PhysRevLett.40.416}
  {\bibfield  {journal} {\bibinfo  {journal} {Phys. Rev. Lett.}\ }\textbf
  {\bibinfo {volume} {40}},\ \bibinfo {pages} {416} (\bibinfo {year}
  {1978})}\BibitemShut {NoStop}%
\bibitem [{\citenamefont {Van~Roermund}\ \emph {et~al.}(2010)\citenamefont
  {Van~Roermund}, \citenamefont {Shiau},\ and\ \citenamefont
  {Lavagna}}]{VanRoermund2010}%
  \BibitemOpen
  \bibfield  {author} {\bibinfo {author} {\bibfnamefont {R.}~\bibnamefont
  {Van~Roermund}}, \bibinfo {author} {\bibfnamefont {S.-y.}\ \bibnamefont
  {Shiau}}, \ and\ \bibinfo {author} {\bibfnamefont {M.}~\bibnamefont
  {Lavagna}},\ }\href {\doibase 10.1103/PhysRevB.81.165115} {\bibfield
  {journal} {\bibinfo  {journal} {Phys. Rev. B}\ }\textbf {\bibinfo {volume}
  {81}},\ \bibinfo {pages} {165115} (\bibinfo {year} {2010})}\BibitemShut
  {NoStop}%
\bibitem [{\citenamefont {Lavagna}(2015)}]{Lavagna2015}%
  \BibitemOpen
  \bibfield  {author} {\bibinfo {author} {\bibfnamefont {M.}~\bibnamefont
  {Lavagna}},\ }\href {\doibase 10.1088/1742-6596/592/1/012141} {\bibfield
  {journal} {\bibinfo  {journal} {J. Phys.: Conf. Ser.}\ }\textbf {\bibinfo
  {volume} {592}},\ \bibinfo {pages} {012141} (\bibinfo {year}
  {2015})}\BibitemShut {NoStop}%
\bibitem [{\citenamefont {Czycholl}(1985)}]{Czycholl1985}%
  \BibitemOpen
  \bibfield  {author} {\bibinfo {author} {\bibfnamefont {G.}~\bibnamefont
  {Czycholl}},\ }\href {\doibase 10.1103/PhysRevB.31.2867} {\bibfield
  {journal} {\bibinfo  {journal} {Phys. Rev. B}\ }\textbf {\bibinfo {volume}
  {31}},\ \bibinfo {pages} {2867} (\bibinfo {year} {1985})}\BibitemShut
  {NoStop}%
\bibitem [{\citenamefont {Aleiner}\ \emph {et~al.}(2002)\citenamefont
  {Aleiner}, \citenamefont {Brouwer},\ and\ \citenamefont
  {Glazman}}]{Aleiner2002}%
  \BibitemOpen
  \bibfield  {author} {\bibinfo {author} {\bibfnamefont {I.~L.}\ \bibnamefont
  {Aleiner}}, \bibinfo {author} {\bibfnamefont {P.~W.}\ \bibnamefont
  {Brouwer}}, \ and\ \bibinfo {author} {\bibfnamefont {L.~I.}\ \bibnamefont
  {Glazman}},\ }\href {\doibase 10.1016/S0370-1573(01)00063-1} {\bibfield
  {journal} {\bibinfo  {journal} {Phys. Rep.}\ }\textbf {\bibinfo {volume}
  {358}},\ \bibinfo {pages} {309} (\bibinfo {year} {2002})}\BibitemShut
  {NoStop}%
\bibitem [{\citenamefont {Foa~Torres}\ \emph {et~al.}(2003)\citenamefont
  {Foa~Torres}, \citenamefont {Lewenkopf},\ and\ \citenamefont
  {Pastawski}}]{FoaTorres2003}%
  \BibitemOpen
  \bibfield  {author} {\bibinfo {author} {\bibfnamefont {L.~E.~F.}\
  \bibnamefont {Foa~Torres}}, \bibinfo {author} {\bibfnamefont {C.~H.}\
  \bibnamefont {Lewenkopf}}, \ and\ \bibinfo {author} {\bibfnamefont {H.~M.}\
  \bibnamefont {Pastawski}},\ }\href {\doibase 10.1103/PhysRevLett.91.116801}
  {\bibfield  {journal} {\bibinfo  {journal} {Phys. Rev. Lett.}\ }\textbf
  {\bibinfo {volume} {91}},\ \bibinfo {pages} {116801} (\bibinfo {year}
  {2003})}\BibitemShut {NoStop}%
\bibitem [{\citenamefont {Schoeller}\ and\ \citenamefont
  {Sch\"on}(1994)}]{Schoeller1994}%
  \BibitemOpen
  \bibfield  {author} {\bibinfo {author} {\bibfnamefont {H.}~\bibnamefont
  {Schoeller}}\ and\ \bibinfo {author} {\bibfnamefont {G.}~\bibnamefont
  {Sch\"on}},\ }\href {\doibase 10.1103/PhysRevB.50.18436} {\bibfield
  {journal} {\bibinfo  {journal} {Phys. Rev. B}\ }\textbf {\bibinfo {volume}
  {50}},\ \bibinfo {pages} {18436} (\bibinfo {year} {1994})}\BibitemShut
  {NoStop}%
\bibitem [{\citenamefont {Melo}(2019)}]{Melo2019}%
  \BibitemOpen
  \bibfield  {author} {\bibinfo {author} {\bibfnamefont {B.}~\bibnamefont
  {Melo}},\ }\emph {\bibinfo {title} {{Impurity solvers for the single impurity
  Anderson model: comparison and applications}}},\ \href@noop {} {Ph.D.
  thesis},\ \bibinfo  {school} {Universidade Federal Fluminense} (\bibinfo
  {year} {2019})\BibitemShut {NoStop}%
\bibitem [{\citenamefont {Kouwenhoven}\ and\ \citenamefont
  {Glazman}(2001)}]{Kouwenhoven2001}%
  \BibitemOpen
  \bibfield  {author} {\bibinfo {author} {\bibfnamefont {L.}~\bibnamefont
  {Kouwenhoven}}\ and\ \bibinfo {author} {\bibfnamefont {L.}~\bibnamefont
  {Glazman}},\ }\href {\doibase cond-mat/0104100} {\bibfield  {journal}
  {\bibinfo  {journal} {Physics World}\ }\textbf {\bibinfo {volume} {14}},\
  \bibinfo {pages} {33} (\bibinfo {year} {2001})}\BibitemShut {NoStop}%
\bibitem [{\citenamefont {Roura-Bas}(2010)}]{Roura-Bas2010}%
  \BibitemOpen
  \bibfield  {author} {\bibinfo {author} {\bibfnamefont {P.}~\bibnamefont
  {Roura-Bas}},\ }\href {\doibase 10.1103/PhysRevB.81.155327} {\bibfield
  {journal} {\bibinfo  {journal} {Phys. Rev. B}\ }\textbf {\bibinfo {volume}
  {81}},\ \bibinfo {pages} {1} (\bibinfo {year} {2010})}\BibitemShut {NoStop}%
\end{thebibliography}%

\end{document}